\documentclass[secnumarabic,amssymb,amsfonts,amsnofootinbib,nobibnotes,aps,prd,10pt,twocolumn]{revtex4-2}

\usepackage[utf8]{inputenc}
\usepackage[makeroom]{cancel}
\usepackage[titletoc]{appendix}
\usepackage[hidelinks,breaklinks=true]{hyperref}
\usepackage{feynmf}
\usepackage{feyn}
\usepackage{csquotes}
\usepackage{slashed}
\usepackage{adjustbox}
\usepackage{xcolor}
\usepackage{mathtools}
\usepackage{amsmath,tikz}
\usepackage{dcolumn}
\usepackage{amsfonts}
\usepackage{indentfirst}
\usepackage[top=1in, bottom=0.4in, left=0.55in, right=0.55in,includefoot]{geometry}
\usepackage[margin=15pt,font={sf,it},labelfont={bf,sf},format=plain,justification=centerlast,width=1\linewidth,skip=8pt]{caption}
\usepackage{afterpage}
\usepackage{graphicx,rotating,colortbl}
\usepackage{amssymb}
\usepackage{braket}
\usepackage{mathrsfs}
\usepackage{subcaption}
\usepackage{empheq}
\usepackage{appendix}
\usepackage{natbib}
\usepackage{float}
\usepackage{color,soul}
\usepackage{mathtools}
\usepackage{microtype}
\usepackage{enumerate}

\newcommand{\beq}{\begin{equation}}
\newcommand{\eeq}{\end{equation}}
\newcommand{\bea}{\begin{eqnarray}}
\newcommand{\eea}{\end{eqnarray}}

\newcommand{\alp}{\alpha}

\newcommand{\del}{\delta}
\newcommand{\bet}{\beta}
\newcommand{\lam}{\lambda}

\newcommand{\dd}{\text{d}}
\newcommand{\comm}[1]{}
\newcommand{\rhoc}{\rho_{\mathrm{c}}}
\newcommand{\Ri}{\mathcal{R}}

\definecolor{DarkViolet}{RGB}{148,0,211}
\definecolor{colorForTitle}{RGB}{0,153,0}

\newcommand{\mycolor}{DarkViolet}

\hypersetup{colorlinks=true,bookmarksopen,bookmarksnumbered,citecolor={\mycolor},linkcolor={\mycolor},urlcolor={\mycolor},pdfstartview=FitH,anchorcolor={\mycolor}}

\usepackage{sfmath}

\usepackage{scalerel}
\usepackage{tikz}
\usetikzlibrary{svg.path}
\definecolor{orcidlogocol}{HTML}{A6CE39}
\tikzset{
  orcidlogo/.pic={
    \fill[orcidlogocol] svg{M256,128c0,70.7-57.3,128-128,128C57.3,256,0,198.7,0,128C0,57.3,57.3,0,128,0C198.7,0,256,57.3,256,128z};
    \fill[white] svg{M86.3,186.2H70.9V79.1h15.4v48.4V186.2z}
                 svg{M108.9,79.1h41.6c39.6,0,57,28.3,57,53.6c0,27.5-21.5,53.6-56.8,53.6h-41.8V79.1z M124.3,172.4h24.5c34.9,0,42.9-26.5,42.9-39.7c0-21.5-13.7-39.7-43.7-39.7h-23.7V172.4z}
                 svg{M88.7,56.8c0,5.5-4.5,10.1-10.1,10.1c-5.6,0-10.1-4.6-10.1-10.1c0-5.6,4.5-10.1,10.1-10.1C84.2,46.7,88.7,51.3,88.7,56.8z};}}
\newcommand\orcid[1]{\href{https://orcid.org/#1}{\mbox{\scalerel*{
\begin{tikzpicture}[yscale=-1,transform shape]
\pic{orcidlogo};
\end{tikzpicture}
}{|}}}}

\usetikzlibrary{matrix,calc,decorations.markings}

\begin{document}

\title{\Large{Extended Tolman III and VII solutions in $f(\Ri,T)$ gravity: Models for neutron stars and supermassive stars}}

\author{Thomas D.~Pappas~\orcid{0000-0003-2186-357X}}
\email[]{thomas.pappas@physics.slu.cz}

\author{Camilo Posada~\orcid{0000-0001-8826-1974}}
\email{camilo.posada@physics.slu.cz}

\author{Zden\v{e}k Stuchl\'ik}
\email{zdenek.stuchlik@physics.slu.cz}

\affiliation{Research Centre for Theoretical Physics and Astrophysics, Institute of Physics, Silesian University in Opava, Bezručovo náměstí 13, CZ-746 01 Opava, Czech Republic}

\vspace{0.5cm}\date{\today}

\begin{abstract}
\noindent\textbf{Abstract.} In the context of linear $f(\Ri,T)=\Ri+\chi T$ gravity, where $\Ri$ is the Ricci scalar, $T$ is the trace of the energy-momentum tensor, and $\chi$ is a dimensionless parameter, we have obtained exact analytical and numerical solutions for isotropic perfect-fluid spheres in hydrostatic equilibrium. Our solutions correspond to two-parametric extensions of the Tolman III (T-III) and Tolman VII (T-VII) models, in terms of the compactness $\beta$ and $\chi$. By requiring configurations that exhibit monotonically decreasing radial profiles for both the energy density and pressure, compliance with the energy conditions, as well as subluminal speed of sound, we have constrained the parametric space of our solutions. We have also obtained analytically a parametric deformation of the T-VII solution that continuously interpolates between the T-III and T-VII models for any $\chi$, and in the appropriate limits, provides an analytic approximation for the uniform density configuration in linear $f(\Ri,T)$ gravity. Finally, by integrating numerically the TOV equations, we have obtained a numerical solution for the uniform-density configuration and subsequently, using the mass-radius relations, we have obtained the maximum mass that can be supported by such configurations. We have found that in the appropriate parametric regime our solution is in very good agreement with the observational bounds for the masses and radii of neutron stars.
\end{abstract}

\maketitle

\section{Introduction}
\label{Sec:Introduction}

In general relativity (GR), there is a large number of known solutions to the field equations for spherically symmetric fluid spheres~\cite{Stephani:2003tm}. However, it was found that only 9 (out of 127) satisfy the most fundamental requirements for physical plausibility~\cite{Delgaty:1998uy}. Among those, the Buchdahl~\cite{Buchdahl:1967ApJ}, Nariai IV~\cite{Nariai:1950srt}, and the Tolman VII (T-VII)~\cite{Tolman:1939jz} solutions are usually employed in the modeling of realistic neutron stars (NSs). The T-VII solution is characterized by an energy density which varies quadratically with the radial coordinate. This behavior turns out to be in a relatively good agreement with the equations of state for realistic NSs~\cite{Lattimer:2000nx}. As a consequence, the T-VII model has been the object of intensive study in the literature, and many interesting properties have been reported~\cite{Negi:2001ApS,Neary:2001ai,Postnikov:2010yn,Sham:2014kea,Moustakidis:2016ndw,Raghoonundun:2015wga, Sotani:2018aiz,Jiang:2019vmf,Posada:2021zxk,Posada:2022lij,Stuchlik:2021coc, Stuchlik:2021vdd}.

It is remarkable that in the context of GR, not overly complicated solutions for perfect fluid spheres are available for the modeling of realistic stars. Nevertheless, GR, despite withstanding numerous and increasingly more refined and accurate tests, ranging from astrophysical to cosmological scales for more than a century~\cite{Turyshev:2008dr,LIGOScientific:2016aoc,EventHorizonTelescope:2019dse}, it fails to provide a satisfactory description for all of the observed gravitational phenomena, such as the flatness of the rotation curves of galaxies and the early- and late-time accelerated phases of the Universe~\cite{Dodelson:2003ft}. Also, in the strong gravity regime, the well-known paradoxes posed by mathematical black holes (BHs), e.g., the loss of information~\cite{Unruh:2017uaw}, and the unavoidable development of geodesically incomplete paths in the interior of BHs~\cite{Penrose:1964wq,Senovilla:2014gza}, suggest that Einstein's theory may only correspond to an approximate description of a more fundamental theory of gravity. 
To this end, a plethora of extensions of the Einstein-Hilbert (EH) gravitational Lagrangian density have been considered as an alternative to provide answers for these open questions~\cite{Nojiri:2017ncd}. Famously, the first attempt toward modifying the EH Lagrangian can be traced back to Einstein himself when he introduced the cosmological constant \footnote{The effect of repulsive cosmological constant in the context of Cosmology was summarized in~\cite{Carroll:2000fy}, for the accretion phenomena in~\cite{Stuchlik:2020rls}, for compact objects and dark matter halos in~\cite{Boehmer:2004nu,Stuchlik:2016xiq,Novotny:2021zlq}, and for the relative motion of galaxies in\cite{Stuchlik:2011zz}.} in order to allow his field equations to admit a static Universe solution~\cite{Einstein:1917ce,ORaifeartaigh:2017uct}. The latter modification of the gravity sector can be thought of as promoting the EH term to a linear function of the Ricci scalar $\Ri$. In the spirit of such a generalization, one of the most straightforward and successful extensions of GR corresponds to replacing the EH term with an arbitrary smooth function of $\Ri$ in the so-called $f(\Ri)$ theory of gravity~\cite{Buchdahl:1970ynr,Sotiriou:2008rp}, in the context of which the accelerated phases of the Universe can be accommodated~\cite{Copeland:2006wr,DeFelice:2010aj,Nojiri:2010wj}. Furthermore, solutions of compact objects in this theory have been obtained, see e.g.~\cite{DeFelice:2010aj} and references therein.

A further generalization of $f(\Ri)$ gravity that has attracted a lot of attention over the last years corresponds to the so-called $f(\Ri,T)$ gravity~\cite{Harko:2011kv} where the trace of the energy-momentum tensor (EMT) $T \equiv g^{\mu\nu}T_{\mu\nu}$ is also considered in constructing the modified gravity sector. The $f(\Ri,T)$ theory of gravity belongs to the class of the so-called generalized curvature-matter coupling theories which lead to interesting phenomenology when applied in cosmological contexts~\cite{Harko:2014gwa,Goncalves:2021vci,Goncalves:2022ggq}. On the other hand, solutions to relativistic compact stars have been also investigated in $f(\Ri,T)$ gravity~\cite{Hansraj:2018jzb, Bhar:2021uqr, Kumar:2021vqa,Feng:2022bvk}.

In this work, we concentrate on two models which have been shown to be relevant for the study of the interior structure of NSs, namely, the Tolman III (T-III)~\cite{Tolman:1939jz}, also known as Schwarzschild's constant density interior solution~\cite{Schwarzschild:1916inc}, and the T-VII solution. Although the T-III solution cannot be considered as a realistic model for NSs due to exhibiting a constant density profile, it nevertheless provides a simple and instructive analytical solution to Einstein's equations \footnote{The T-III solution can be considered as the polytrope with polytropic index $n=0$; for the solutions with non-zero cosmological constant see e.g.~\cite{stuchlik2000,Boehmer:2003uz}}. For these two models, we obtain various analytical extensions in linear $f(\Ri,T)$ gravity, and investigate the effect of the parameter $\chi$ on their interior structure. Additionally, since the analytic extensions of the T-III model do not correspond to exact uniform density configurations, we solve numerically the modified Tolman-Oppenheimer-Volkoff (TOV) system of equations for a uniform density profile in order to obtain this solution.

This article is organized as follows: In Sec.~\ref{Sec:Overview_of_f(R,T)} we present a brief overview of the formalism of $f(\Ri,T)$ gravity, where we discuss the generalized field equations and the hydrostatic equilibrium conditions for perfect-fluid spheres. Subsequently, following Tolman's approach~\cite{Tolman:1939jz}, in Sec.~\ref{Sec:T_III_analytic_extension} we obtain an exact analytic solution to the field equations of linear $f(\Ri,T)$ gravity that corresponds to an extension of the original T-III model, while in Sec.~\ref{Sec:T_VII_analytic_extension}, we obtain analytically the corresponding extension for the T-VII model. By performing a parametric deformation of our T-VII extended solution, in Sec.~\ref{Sec:rho_s_extended_T_VII} we obtain another exact analytic solution that continuously interpolates between the T-III and T-VII models for any value of the parameter $\chi$, and it also provides approximate analytic expressions for the uniform-density configuration in linear $f(\Ri,T)$. In Sec.~\ref{Sec:General_TVII_numer_sol}, we solve numerically the modified TOV system, for hydrostatic equilibrium, for a configuration with uniform energy density. In Sec.~\ref{Sec:Conclusions} we draw our final conclusions.

\vspace{0.15cm}\noindent\textbf{Conventions and notation:} We work in geometrized units ($G=c=1$) and we adopt the signature $(-,+,+,+)$ for the metric. Unless otherwise stated, primes denote derivatives with respect to the radial coordinate.

\section{Perfect fluid spheres and hydrostatic equilibrium in $f(\Ri,T)$ gravity}\label{Sec:Overview_of_f(R,T)}

In this section, we give a brief overview of $f(\Ri,T)$ gravity~\cite{Harko:2011kv} in the metric formalism where the connection is the Levi-Civita connection of the metric. For the alternative Palatini formulation of the theory where the connection is treated as an independent variable for the variation of the action see e.g.~\cite{Wu:2018idg}. We present the field equations for static, spherically-symmetric geometries sourced by isotropic perfect fluids. The action of the theory is given by
\beq
S=\int{\dd^4x\sqrt{-g}}\left[ \frac{f(\Ri,T)}{16\pi}+\mathcal{L}_m \right]\,,
\label{f(R,T) action}
\eeq
where the gravity sector $f(\Ri,T)$ is an arbitrary smooth function of both, the Ricci scalar $\Ri$ and the trace of the energy-momentum tensor (EMT) $T$, while $\mathcal{L}_m$ is the matter Lagrangian density. Assuming that $\mathcal{L}_m$ depends on the metric, but not on its derivatives,  the EMT of matter is defined in the usual way
\beq
T_{\mu\nu}=\frac{-2}{\sqrt{-g}}\frac{\del \left(\sqrt{-g} \mathcal{L}_m \right)}{\del g^{\mu\nu}}\,.
\eeq
We model the interior of the star as an isotropic perfect fluid, with EMT given by
\beq
T_{\mu\nu}=\left(\rho+p \right)u_{\mu}u_{\nu}+p\, g_{\mu\nu}\,,
\label{EMT}
\eeq
where $u^{\mu}$ is the four-velocity of the fluid, with $u^{\mu}u_{\mu}=-1$, and $u^{\mu}\nabla_{\nu} u_{\mu}=0$, while $\rho(r)$ and $p(r)$ are the fluid's energy density and isotropic pressure respectively. Upon variation of~\eqref{f(R,T) action} with respect to the metric one obtains the modified Einstein equations~\cite{Harko:2011kv}
\beq
f_\Ri \Ri_{\mu\nu}-\frac{1}{2}f\, g_{\mu\nu}+D_{\mu\nu}f_\Ri=8\pi T_{\mu\nu}-f_T\left(T_{\mu\nu}+ \Theta_{\mu\nu} \right)\equiv \widetilde{T}_{\mu\nu}\,,
\label{f(R,T) EOM}
\eeq
where we have used the shorthand notations $f_\Ri \equiv \partial_\Ri f(\Ri,T)$, $f_T \equiv \partial_T f(\Ri,T)$, $D_{\mu\nu} \equiv \left(g_{\mu\nu} \Box-\nabla_{\mu}\nabla_{\nu} \right)$ and
\beq
\Theta_{\mu\nu} \equiv g^{\alp\bet}\del T_{\alp\bet}/\del g^{\mu\nu}=-2T_{\mu\nu}+g_{\mu\nu} \mathcal{L}_m-2g^{\alp\bet}\frac{\partial^2 \mathcal{L}_m}{\partial g^{\mu\nu} \partial g^{\alp \bet}}\,.
\eeq
It is a well-known fact that for perfect fluids there is no unique definition for $\mathcal{L}_m$ (see e.g.~\cite{Faraoni:2009rk,Harko:2011kv,Avelino:2018rsb}) and so we have to make an assumption about it in order to proceed any further. Following Harko \emph{et al.}~\cite{Harko:2011kv}, we choose $\mathcal{L}_m=p$, (note that in~\cite{Harko:2011kv}, $\mathcal{L}_m=-p$, however we use the opposite signature for the metric) and with this choice, we have that $\Theta_{\mu\nu}=-2 T_{\mu\nu}+pg_{\mu\nu}$. Upon taking the trace of Eq.~\eqref{f(R,T) EOM} one obtains the dynamical equation for the Ricci scalar
\beq
f_\Ri \Ri-2 f+3 \Box f_\Ri=8\pi \widetilde{T}\,,
\label{eq:trace_eom}
\eeq
which, for $f(\Ri,T)$ models that are nonlinear in $\Ri$, corresponds to a second-order differential equation, rather than an algebraic one, as it is the case with pure GR. Consequently, nonlinear models admit a nonvanishing Ricci scalar even in the exterior region of a star. Due to the presence of the trace of the EMT in the Lagrangian of the theory, $f(\Ri,T)$ gravity is characterized by a nonminimal coupling between matter and curvature; as a consequence, the four-divergence of the EMT will be in general nonzero and given by~\cite{BarrientosO:2014mys}
\beq
\nabla^{\mu}T_{\mu\nu}=\frac{f_T}{ 8\pi-f_T} \left[ \left(T_{\mu\nu}+ \Theta_{\mu\nu} \right) \nabla^{\mu} \ln{f_T} + \nabla^{\mu}\Theta_{\mu\nu}-\frac{1}{2}\nabla_{\nu}T \right]\,.
\label{Tmn divergence}
\eeq
The fact that the EMT of matter is not covariantly conserved for general $f(\Ri,T)$ gravity models, entails the existence of an extra force perpendicular to the velocity of test particles causing their trajectories to deviate from geodesics. Such a force may induce gravitational effects beyond GR that can be relevant at the solar-system and galactic scales~\cite{Harko:2011kv}. The conservation of the EMT is restored in the limit of $f(\Ri)$ gravity i.e. when $f_T=0$ and for particular models that impose the conservation of $T_{\mu\nu}$ as an a priory requirement of the theory, and reconstruct from this condition the form of $f(\Ri,T)$ (see e.g.~\cite{Pretel:2021kgl}). However, the corresponding field equations can then only be integrated numerically and since in this work we are mainly interested in analytic solutions we will not consider such special models of $f(\Ri,T)$ gravity.

It is convenient to rewrite the field equations~\eqref{f(R,T) EOM} in their equivalent effective GR form, where the Einstein tensor appears on the left-hand side (lhs) while the right-hand side (rhs) corresponds to an effective EMT that involves contributions from both, matter fields, and curvature terms as follows
\beq
G_{\mu\nu}=\frac{1}{f_\Ri}\left[\widetilde{T}_{\mu\nu} +\frac{f-\Ri\,f_\Ri}{2}g_{\mu\nu}-D_{\mu\nu}f_\Ri \right] \equiv T^{(\text{eff})}_{\mu\nu}\,.
\label{Gmn_Teff}
\eeq
In principle, for an arbitrary $f(\Ri,T)$ theory, the expressions for the components of $T^{(\text{eff})}_{\mu\nu}$ will be very complicated and consequently obtaining exact analytic solutions in this general case is a very difficult task. Notice however, that for the special class of models
\beq
f(\Ri,T)=\Ri+h(T)\,,
\label{f(R,T) separable}
\eeq
which are linear in the Ricci scalar and do not involve mixing terms between $\Ri$ and $T$, the contributions of the curvature fluids in the effective EMT simplify significantly such that the structure of the field equations becomes
\beq
G_{\mu\nu} = 8\pi T_{\mu\nu}+\frac{h}{2}g_{\mu\nu}+h_T \left(T_{\mu\nu}-p\,g_{\mu\nu} \right)\,.
\label{GR+T effects EOM}
\eeq
In this work, in order to investigate the minimal extension of GR in the regime of $f(\Ri,T)$ gravity, and in an effort to avoid overly complicated field equations that would otherwise hinder the search for exact analytic solutions, we restrict our analysis to the case of linear $f(\Ri,T)$ gravity in the form
\beq
f(\Ri,T)=\Ri+2\chi T\,,
\label{eq:linear_f(R,T)}
\eeq
where $\chi$ is a dimensionless constant free parameter of the theory.
For such linear models, analytic solutions can be readily obtained for the field equations (see e.g.~\cite{Hansraj:2018jzb,Bhar:2021uqr,Pretel:2020oae,Pretel:2021kgl}). Notice that for more complicated nonlinear models of $f(\Ri,T)$ gravity, one may in principle, deal with the complexity of the field equations via dynamically equivalent scalar-tensor representations of the theory and by utilizing the Palatini formalism, see e.g.~\cite{Rosa:2021teg,Rosa:2022cen}.

The most general static and spherically symmetric geometry can be described in curvature coordinates by the line element
\beq
ds^2=-e^{\nu(r)}dt^2+e^{\lambda(r)}dr^2+r^2d\Omega^2_2\,,
\label{line element}
\eeq
where $\nu(r)$ and $\lambda(r)$ are functions of the radial coordinate and $d\Omega^2_2$ is the surface element of the unit 2-sphere. Since Eq.~\eqref{eq:linear_f(R,T)} is a special case of the theories described by Eq.~\eqref{f(R,T) separable}, the substitution of Eqs.~\eqref{EMT} and~\eqref{line element} into Eq.~\eqref{GR+T effects EOM} yields the system of equations
\beq
G^0_{\,\,\,0} =\frac{1}{r^2}\frac{d}{dr}\left(re^{-\lambda}\right)-\frac{1}{r^2} = 8\pi\rho + \chi(3\rho-p) \equiv 8 \pi \rho_{\text{eff}}\,,\label{EinsG00}
\eeq
\vspace{-0.4cm}
\beq
G^1_{\,\,\,1}=  e^{-\lam} \left( \frac{\nu'}{r}+\frac{1}{r^2}\right)-\frac{1}{r^2}
= 8\pi p + \chi(3p-\rho) \equiv 8 \pi p_{\text{eff}}\,, \label{EinsG11}
\eeq
along with the pressure isotropy condition ($G^1_{\,\,\,1}= G^2_{\,\,\,2}$) that gives the following constraint equation between the metric functions
\beq
r^2 \left( 2 \nu''+\nu'^2-\nu' \lam' \right)-2r \left( \nu'+\lam' \right)+4\left(e^{\lam}-1 \right)=0\,.
\label{p isotropy Eq.}
\eeq
From the form of the above equations, it is evident that the linear $f(\Ri,T)$ model naturally yields field equations that are identical to the ones of standard GR with respect to the metric components, and all the beyond-GR effects of the theory are encoded solely in the modified EMT. It is important to point out the absence of curvature terms (with ambiguous physical interpretation) in the effective EMT which would otherwise be present had the theory been chosen to be nonlinear in $\Ri$, or involving mixing terms with $T$. As it can be seen from the form of the effective energy density~\eqref{EinsG00} and pressure~\eqref{EinsG11}, the overall effect of modified gravity on the system is to introduce a mixing between the energy density and pressure of matter on the rhs of the Einstein equations.

The four-divergence of the EMT~\eqref{Tmn divergence} yields the following equation
\beq
p'+\left( \rho +p \right) \frac{\nu'}{2}=\frac{\chi}{8 \pi +2 \chi} \left(\rho'-p'\right)\,,
\label{DivEMT_mod}
\eeq
which reduces to the usual GR conservation equation when $\chi=0$. Upon introducing the mass function
\beq\label{mass}
e^{-\lambda(r)}\equiv 1-\frac{2m(r)}{r}, 
\eeq
\noindent the $tt$ component of the Einstein equations yields the expression for the mass of the system enclosed within a radius $r$
\bea\label{mass_int}
m(r)&=&4\pi \int_0^r{\tilde{r}^2 \rho_{\text{eff}}(\tilde{r}) }d\tilde{r}\nonumber\\
&=&\frac{\left(8\pi+3 \chi \right)}{2} \int_0^r{\tilde{r}^2 \rho(\tilde{r})}d\tilde{r} -\frac{\chi}{2}\int_0^r{\tilde{r}^2 p(\tilde{r}) }d\tilde{r}\,.
\eea
In contrast to GR, where $m(r)$ is determined only by the energy-density of matter, in linear $f(\Ri,T)$ both the energy density and pressure contribute to the mass. In differential form, Eq.~\eqref{mass_int} can be rewritten as
\beq\label{mass_Mod}
\frac{dm}{dr}=4\pi r^2 \rho(r) + \frac{\chi}{2}(3\rho-p)r^2.
\eeq
The boundary conditions are determined as follows: the mass at the origin vanishes $m(0)=0$, and at the surface $m(R)=M$ we obtain the total mass, where $R$ is the radius of the configuration.

In the following, we assume the configuration to be a perfect fluid in hydrostatic equilibrium. Thus, the EMT is given by Eq.~\eqref{EMT}, where the energy density $\rho$ and pressure $p$ are time-independent. Since we are considering a static spacetime, the four-velocity of the fluid $u^{\mu}=(e^{-\nu/2},0,0,0)$, which is normalized. This implies that the nonvanishing components of the EMT are given by
\beq
T_{0}^{\,0} = -\rho,\quad\quad T_{1}^{\,1}=T_{2}^{\,2}=T_{3}^{\,3}=p.
\eeq
\noindent Substituting Eqs.~\eqref{EinsG11} and~\eqref{mass} into Eq.~\eqref{DivEMT_mod}, we obtain the TOV equation in modified linear $f(\Ri,T)$ gravity~\cite{Pretel:2020oae} 
\bea
\frac{dp}{dr}&=&-\frac{(\rho+p)}{(1+\sigma)}\frac{\left[m(r)+4\pi\,p\,r^3 - \frac{1}{2}\chi\,r^3\,(\rho-3p)\right]}{r\left[r-2m(r)\right]} +\nonumber\\
&&\frac{\sigma}{(1+\sigma)}\frac{d\rho}{dr}\,,
\label{modTOV}
\eea
\noindent where $\sigma\equiv \chi/(8\pi + 2\chi)$. Note that for $\chi=0$, Eq.~\eqref{modTOV} reduces to the standard TOV equation in GR~\cite{Schaffner-Bielich:2020psc}. In order to determine the structure of a relativistic spherical mass, in hydrostatic equilibrium, one must solve the TOV system subject to the conditions, $m(0)=0$ and $p(0)=p_{\text{c}}$, with $p_{\text{c}}$ denoting the pressure at the center. Thus, one can obtain profiles for pressure, total mass, and mass-radius relations. We will discuss further the solutions to the TOV system in Sec.~\ref{Sec:General_TVII_numer_sol}.

\section{Tolman III extended model in $f(\Ri,T)$ gravity: The quasiconstant density configuration}
\label{Sec:T_III_analytic_extension}

As a first example, we consider Schwarzschild's constant-density interior solution of GR (see e.g.~\cite{Schaffner-Bielich:2020psc}), also known as Tolman III (T-III) model and obtain its generalization to linear $f(\Ri,T)$ gravity. Although the uniform density assumption is not a physically realistic approximation for a stellar configuration, it nevertheless serves as an instructive toy model that deserves to be further investigated in the framework of modified gravity.

In order to obtain exact analytic solutions for the Einstein field equations of GR, Tolman~\cite{Tolman:1939jz} followed an approach that is based on mathematical arguments rather than physical ones. Instead of choosing a physically sensible energy density, and/or pressure, and subsequently solving the system of equations for the metric functions, he carefully tailored the functional form for the metric functions to simplify the field equations in order to be able to solve them. Thus, he obtained a number of exact analytic solutions some of which are physically sensible. We follow the same approach in this section. Notice that since we are working in the context of linear $f(\Ri,T)$ gravity~\eqref{eq:linear_f(R,T)}, as we have already discussed in the introduction, all the beyond-GR modifications are encoded in the effective energy-density and pressure~\eqref{EinsG00}-\eqref{EinsG11} without introducing further dependence of the field equations on the metric functions or their derivatives. As a consequence, any ansatz for a metric function made on the basis of simplifying the GR field equations will also simplify the field equations in linear $f(\Ri,T)$ gravity. Thus, we fix the functional form for the $g_{rr}(r)$ metric function to be identical to the ansatz introduced by Tolman, and then we obtain the corresponding expressions for the remaining functions $g_{tt}(r)$, $\rho(r)$ and $p(r)$ via the modified Einstein equations~\eqref{EinsG00}-\eqref{p isotropy Eq.}. Following~\cite{Tolman:1939jz} we choose
\beq
g_{rr}(r)=\left(1-\frac{r^2}{A^2} \right)^{-1}\,,
\label{eq:grr Sch.}
\eeq
where $A$ is a constant with units of length. The system of field equations~\eqref{EinsG00}-\eqref{p isotropy Eq.}, with the ansatz~\eqref{eq:grr Sch.}, requires three boundary conditions in total for the determination of the constants of integration. These are imposed at the surface of the star $r=R$ where the radial pressure vanishes and the interior metric matches with the exterior Schwarzschild spacetime. Thus, the boundary conditions read explicitly
\bea
g_{tt}(R)&=&g_{rr}^{-1}(R)=1-\frac{2M}{R} \label{bc1}\,,\\
p(R)&=&0\,, \label{bc2}
\eea
where $M$ denotes the total mass of the star. Upon implementation of the boundary conditions, the system of field equations~\eqref{EinsG00}-\eqref{p isotropy Eq.} admits a two-parametric class of solutions which we write here in terms of the compact radial coordinate $x \equiv r/R\,,\,\, x \in [0,1]$,
\begin{widetext}
\bea
g_{tt}(x)&=&\frac{16 \pi ^2}{(3 \chi +8 \pi )^2} \left[3\sqrt{1-2 \beta}  -\sqrt{1-2\beta x^2} + \left(\frac{3\chi}{4\pi}\right)\sqrt{1-2\beta}\right]^2 \,, \label{gttFSs}\\
g_{rr}(x)&=&\left(1-2 \bet x^2 \right)^{-1} \,,\\
\rho(x)&=& \frac{6\beta\left[\pi\right(\sqrt{1-2\beta x^2}-3\sqrt{1-2\beta}\left)-\chi\sqrt{1-2\beta}\right]}{(2\pi+\chi)\left[4\pi\left(\sqrt{1-2\beta x^2}-3\sqrt{1-2\beta}\right)-3\chi\sqrt{1-2\beta}\right]}\,, \label{rhoFSs}\\
p(x)&=& \frac{3\beta\left(\sqrt{1-2\beta x^2}-\sqrt{1-2\beta}\right)}{2(2\pi +\chi)\left[3\sqrt{1-2\beta}-\sqrt{1-2\beta x^2}+ \left(\frac{3\chi}{4\pi}\right) \sqrt{1-2\beta}\right]} \,, \label{pFSs}
\eea
\end{widetext}
where $\beta \equiv M/R$ is the compactness of the configuration. Reference~\cite{Hansraj:2018jzb} studied the generalization of the Tolman solutions in linear $f(\Ri,T)$ gravity, including the T-III model. However, the implementation of the boundary conditions was not taken into consideration and the constants of integration were left as free parameters, which were assigned arbitrary values after “arduous fine tuning." Furthermore, the integration constant $A$ of the $g_{rr}$ metric function (see Eq.~(25) in~\cite{Hansraj:2018jzb}) was erroneously identified with the radius of the star $R$. These assumptions produced erratic behaviors for the density and pressure profiles. Moreover, this approach does not allow to perform a clear systematic analysis of the properties of the solution due to the presence of a large number of free parameters. As we have shown above, and it was also discussed in~\cite{Bhar:2021uqr} for the Tolman IV fluid spheres, solutions to the field equations are subjected to the boundary conditions~\eqref{bc1} and~\eqref{bc2} which uniquely specify the corresponding values of the integration constants in terms of the only free parameters $\beta$ and $\chi$.

We now turn to the analysis of the properties of our solution. First of all, for $\chi=0$ Eqs.~\eqref{gttFSs}-\eqref{pFSs} reduce to the well-known Schwarzschild's interior solution with constant energy density, or \emph{Schwarzschild star} (see e.g.~\cite{Schaffner-Bielich:2020psc}). Also notice that due to the fact that Eq.~\eqref{p isotropy Eq.} remains invariant from GR, and since we have chosen the $g_{rr}$ metric function to have the same functional form as in the GR version of the model, the $g_{tt}$ metric function also shares the same functional GR form, up to integration constants. The first important difference between the linear-$f(\Ri,T)$ and GR variants of the model, is related to the speed of sound where, using Eqs.~\eqref{rhoFSs} and~\eqref{pFSs}, we obtain
\beq
c_s^2=\frac{dp}{d\rho}=3+\frac{8\pi}{\chi}\,,
\label{eq: Sch speed of sound}
\eeq
which agrees with the value reported in~\cite{Hansraj:2018jzb}. Note that the modified gravity theory (MGT) effects induce a radial dependence on the energy density [Eq.~\eqref{rhoFSs}] which allows the speed of sound to be well-defined, in contrast with the GR case. Causality considerations require the speed of sound to be subluminal, i.e. $0< c_s^2 <1$, which restricts the MGT parameter in the range $\chi \in (-4\pi, -8 \pi /3)$. Furthermore, we point out that the signs of the first derivatives of the energy density~\eqref{rhoFSs} and pressure~\eqref{pFSs} with respect to the compact radial coordinate are given in terms of the following functions of the free parameter $\chi$
\bea
\text{sgn} \rho'(x)&=&\text{sgn} \left(-\frac{\chi}{\chi+2\pi}\right)\,,\label{eq:Sgn drho} \\
\,\,\,\text{sgn} p'(x)&=&\text{sgn} \left(-\frac{8\pi+3\chi}{\chi+2\pi}\right)\,.\label{eq:Sgn dp}
\eea
From Eq.~\eqref{eq:Sgn drho}, we observe that values of $\chi \in (-2\pi,0)$ result in an energy density that is a monotonically increasing function of the radius which is unphysical. The rhs of Eq.~\eqref{eq:Sgn drho} vanishes in the limit of $\chi \to 0$, such that the constant density interior solution of GR is recovered. On the other hand, from Eq.~\eqref{eq:Sgn dp} we have that the first derivative of the pressure with respect to $x$ is positive for any $-8\pi/3<\chi$ and it vanishes when $\chi=-8\pi/3$. Consequently, even though when viewed naively, Eqs.~\eqref{rhoFSs} and~\eqref{eq: Sch speed of sound} seem to infuse the linear $f(\Ri,T)$ variant of the model with the more realistic property of a nonconstant and monotonically decreasing energy density profile (at least for some range of values of $\chi$), a more careful inspection reveals that the model cannot be considered as a more physically realistic one than the GR Schwarzschild star, due to causality violation (see Table~\ref{table:Sch_viable_chis}). However, just as the Schwarzschild star serves as a useful toy model in GR, despite the fact that it has an undefined speed of sound, we will also study the properties of our solution in order to examine the MGT effects.

\begin{figure}[ht!]
\includegraphics[width=\linewidth]{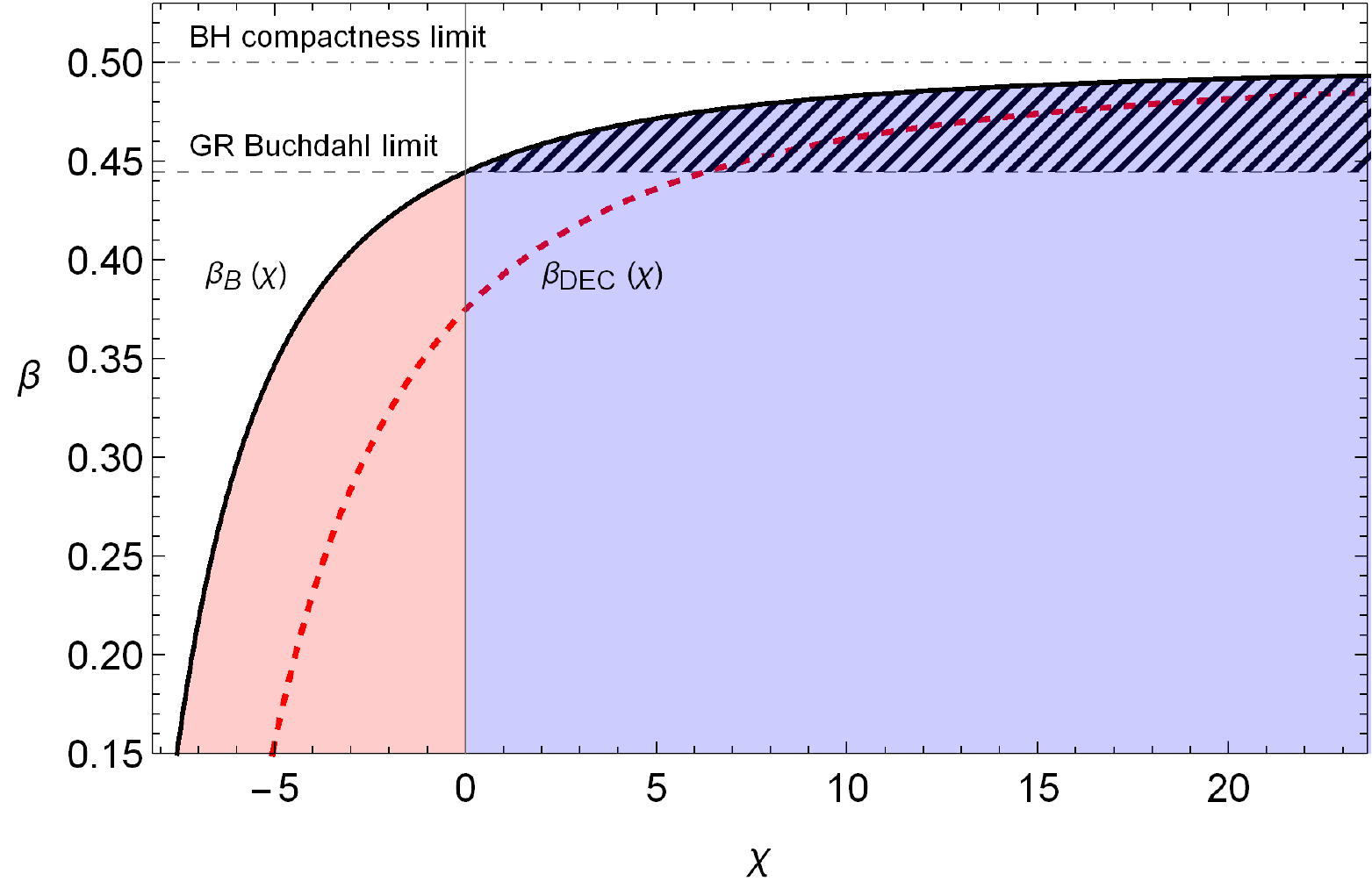}
\caption{The $\chi-\beta$ parametric plane for the extended T-III solution~\eqref{gttFSs}-\eqref{pFSs}. The solid black curve corresponds to the generalized Buchdahl limit $\beta_{\text{B}}(\chi)$ [Eq.~\eqref{eq:Buch_comp_Sch}]. The GR value $(\beta_{\text{B}}=4/9)$ is recovered when $\chi=0$. For large positive values of $\chi$, $\beta_{\text{B}}$ asymptotes to the BH compactness limit ($1/2$), while for negative $\chi$, $\beta_{\text{B}}$ becomes smaller than the typical NS compactness, i.e., $\beta \simeq 0.18-0.2$ for $\chi \lesssim -7.2$. The red dashed curve corresponds to the maximum compactness $\bet_{\text{DEC}}(\chi)$ [Eq.~\eqref{eq:beta_DEC}] for which the dominant energy condition (DEC) is satisfied. For $\chi=0$ we recover the GR value $\bet_{\text{DEC}}=3/8$, while for $\chi \gg 1$, $\bet_{\text{DEC}}(\chi)\to 1/2$.   Configurations in the blue region exhibit monotonically decreasing profiles for both $\rho$ and $p$, while configurations in the red region do not. The shaded region corresponds to configurations with compactness above the Buchdahl limit of GR which, for sufficiently large values of $\chi$, correspond to BH-mimicker states that do not exhibit pressure divergence and do not violate the DEC.}
\label{fig:fig01}
\end{figure}

In GR, the central pressure of the configuration diverges when the compactness $\beta$ reaches the Buchdahl limit $\beta_{\text{B}}=4/9$~\cite{Buchdahl:1959zz}. For the linear $f(\Ri,T)$ variant of the model, we find that the central value of the pressure~\eqref{pFSs}, given by
\beq
p(0)=\frac{6\pi\beta \left(1-\sqrt{1-2\beta}\right)}{(2\pi+\chi) \left[4\pi \left(3\sqrt{1-2\beta}-1\right)+3\chi\sqrt{1-2\beta} \right]}\,,
\label{eq:QCD_p_central}
\eeq
diverges when the star reaches the generalized Buchdahl compactness limit
\beq
\beta_{\text{B}} \left(\chi \right)=\frac{9\chi^2+72\pi\chi+128\pi^2}{18(\chi+4\pi)^2}\,,
\label{eq:Buch_comp_Sch}
\eeq
which is a monotonically increasing function of the parameter $\chi$ (see Fig.~\ref{fig:fig01}). The generalization of the Buchdahl limit is not a unique feature of linear $f(R,T)$ gravity and in fact, various generalizations have been reported in the literature, see e.g. \cite{Goswami:2015dma,EslamPanah:2017ugi,Sharma:2020ooh}. Note that $\beta_{\text{B}}$ reduces to the GR value $4/9$ when $\chi=0$. For large values of $\chi$, $\beta_{\text{B}}$ approaches asymptotically the BH compactness limit $1/2$. For negative values of $\chi$, the function $\beta_B(\chi)$ vanishes at $\chi=-8\pi/3$, however it becomes less than the typical compactness of neutron stars (NSs) $\beta \simeq 0.18-0.2$ already for $\chi \simeq -7.2$. In any case, small deviations from GR imply $|\chi| \ll 1$ and in that parametric regime, we have that $\beta_{\text{B}}$ depends linearly on $\chi$ according to
\beq
\beta_B \left(\chi \right)=\frac{4}{9}+\frac{\chi}{36\pi}+\mathcal{O}\left(\chi^2\right)\,.
\eeq

In Table~\ref{table:Sch_viable_chis}, we collect the various conditions for physically acceptable stellar configurations and their corresponding bounds on the free parameter $\chi$. Only for non-negative values of $\chi$, both the energy density and pressure, exhibit monotonically decreasing radial profiles and the Buchdahl limit is non-negative. Therefore, for the remaining of the analysis in this solution, we consider only non-negative values of $\chi$, keeping in mind that causality is violated in this model.

The surface value of the energy density, for any $\beta$ and $\chi$, can be obtained from Eq.~\eqref{rhoFSs} upon setting $x=1$ and reads
\beq
\rho_s\equiv\rho(1)=\frac{6\bet}{8\pi+3\chi}\,,
\label{eq:T_III_rho_s}
\eeq
while the surface pressure $p(1)=0$ has been fixed to zero by the boundary condition~\eqref{bc2}. Thus, in the constrained parametric regime $\chi \geqslant 0$ and $\beta <\beta_{\text{B}}$, we have that $\rho(x) \geqslant 0$ and $p(x) \geqslant 0$ for any $x$, given the monotonically decreasing radial profile for both the energy density and pressure functions.

\begin{table}[H]
\center{
\caption{Various conditions for physically reasonable configurations and their corresponding bounds on the free parameter $\chi$. It is evident that only for non-negative values of $\chi$ one has physically acceptable radial profiles for both $\rho(r)$ and $p(r)$ and at the same time positive values of the generalized Buchdahl compactness limit. This however entails that the model necessarily violates causality which requires negative values of $\chi$.}
\begin{tabular}{|c|c|}
 \hline
 $\text{condition}$
  &  $\text{allowed values of}\,\, \chi$  \\
 \hline\hline
  $c_s^2 \in (0,1)$ & $\chi \in (-4\pi,-8\pi/3)$   \\
 \hline
  $d\rho(x)/dx \leqslant 0 $ & $\chi \in (-\infty,-2 \pi) \cup [0,+\infty)$ \\
 \hline
   $dp(x)/dx \leqslant 0 $ & $\chi \in (-\infty,-8\pi/3] \cup (-2 \pi,+\infty)$  \\
    \hline
  $\beta_\text{B} >0$ & $\chi \in (-8\pi/3,+\infty)$ \\
 \hline
\end{tabular}
\label{table:Sch_viable_chis}
}
\end{table}

Let us finally investigate the compliance of our solution with respect to the energy conditions. In particular, the ones to be tested are:
\begin{itemize}
\item{Null energy condition (NEC) $\rho+p \geqslant 0$}
\item{Weak energy condition (WEC) $\rho+p \geqslant 0$, $\rho \geqslant 0$}
\item{Strong energy condition (SEC) $\rho+p \geqslant 0$, $\rho + 3p \geqslant 0$}
\item{Dominant energy condition (DEC) $\rho \geqslant \left\vert p\right\vert$}
\end{itemize}
Given that for $\chi \geqslant 0$ and $\bet < \bet_{\text{B}}$, both the energy density and pressure are non-negative functions everywhere in the interior of the configuration, we conclude that the conditions $\rho+p \geqslant 0$, $\rho+3p \geqslant 0$ and consequently the NEC, WEC and SEC are automatically satisfied for our solution in the parametric regime that describes configurations with monotonically decreasing radial profiles, for both the energy density and the pressure.

\begin{figure*}[ht!]
\includegraphics[width=0.495\linewidth]{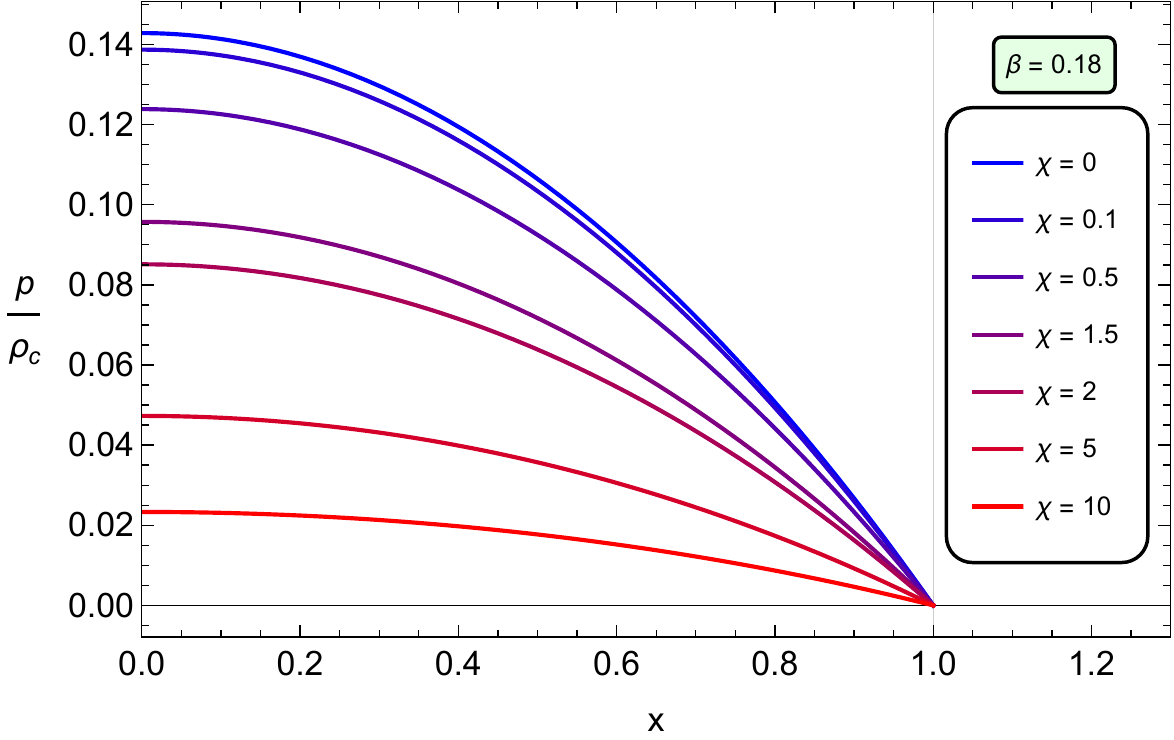}
\includegraphics[width=0.49\linewidth]{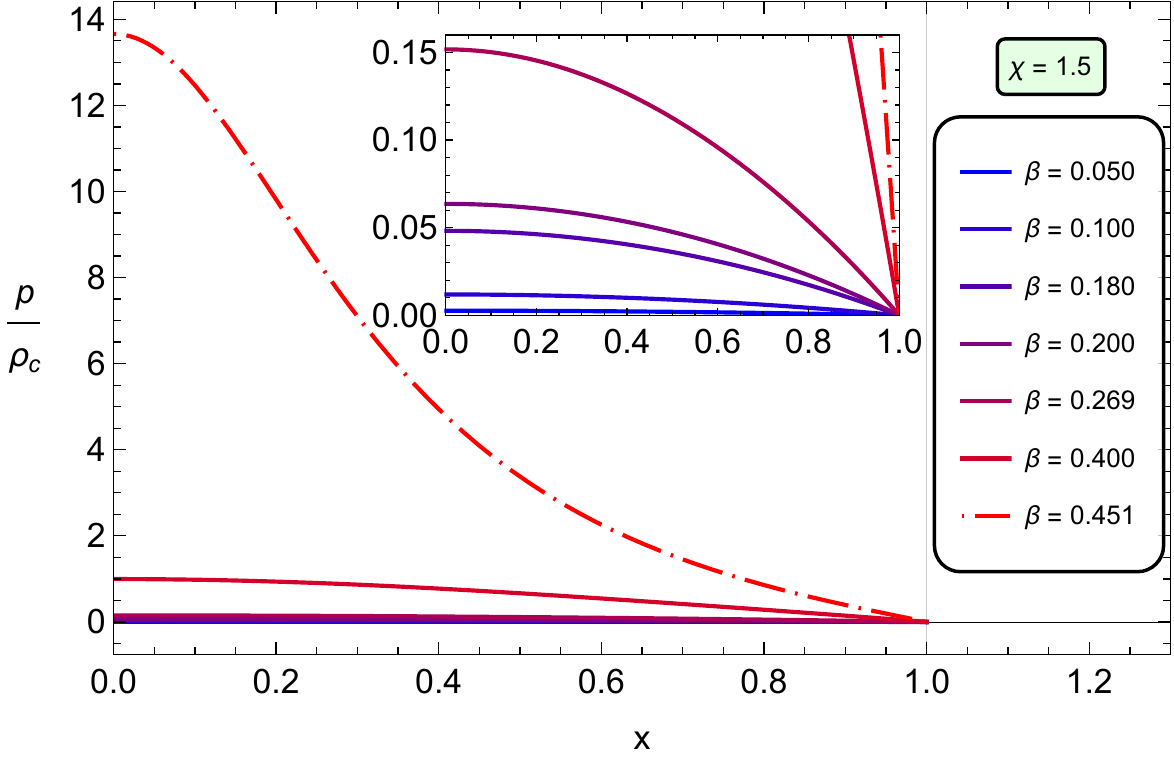}
\caption{The radial profile of the normalized pressure~\eqref{pFSs} for the extended T-III model, in the compact range $x \in [0,1]$. \textbf{Left panel:} for an indicative fixed value of the compactness $\beta=0.18$, and for various values of $\chi$. For the normalization we have used the central density~\eqref{eq:rho0} for $\chi=0$. The overall effect of $\chi$ is to suppress the pressure in the interior of the configuration. \textbf{Right panel:} for an indicative fixed value of the constant $\chi=1.5$, and for various values of compactness. For the normalization we have used the central density~\eqref{eq:rho0} for $\bet=\bet_{\text{DEC}}(1.5)\simeq0.4$. The pressure increases with $\beta$ and as the Buchdahl limit~\eqref{eq:Buch_comp_Sch} is approached, the central pressure diverges. The Buchdahl limit in this example is $\bet_{\text{B}}(1.5) \simeq 0.455661$. The dash-dotted curves correspond to configurations that violate the DEC.}
\label{fig:Schw_p_betas_chis}
\end{figure*}

To investigate the parametric regime of validity for the DEC, we will consider the ratio of the pressure over the energy density which is given by
\beq
\frac{p}{\rho}=\frac{\pi\left(\sqrt{1-2\beta x^2}-\sqrt{1-2\beta }\right)}{\left(\chi+3\pi \right)\sqrt{1-2\beta}-\pi\sqrt{1-2\beta x^2}}\,.
\label{eq:Sch_p_over_rho}
\eeq
The first derivative of~\eqref{eq:Sch_p_over_rho} with respect to $x$ is
\beq
\left(\frac{p}{\rho}\right)'=-\frac{2\pi\beta \sqrt{1-2\beta} \left(\chi+2\pi\right) x}{\sqrt{1-2\beta x^2} \left[\left(\chi+3\pi \right)\sqrt{1-2\beta}-\pi\sqrt{1-2\beta x^2}\right]^2}\,,
\eeq
and thus the ratio~\eqref{eq:Sch_p_over_rho} is a monotonically decreasing function of the compact radial coordinate a fact which entails that as long as the central value ($x=0$) of $p/\rho$ is below or equal to unity, the DEC is satisfied. It is then straightforward to determine via Eq.~\eqref{eq:Sch_p_over_rho} the maximum value of compactness for which the DEC is satisfied which turns out to be
\beq
\bet_{\text{DEC}}\left(\chi \right)=\frac{\chi^2+8\pi\chi+12\pi^2}{2(\chi+4\pi)^2}\,.
\label{eq:beta_DEC}
\eeq
For $\chi \gg 1$, $\bet_{\text{DEC}}$ asymptotes to the BH compactness limit, while for small deviations from GR ($\chi \ll 1$) one has that
\beq
\bet_{\text{DEC}}\left(\chi \right)=\frac{3}{8}+\frac{\chi}{16 \pi}+ \mathcal{O} \left(\chi^2 \right)\,.
\eeq
For any value of the MGT parameter $\chi$, the DEC-violation compactness limit~\eqref{eq:beta_DEC}, depicted with the red-dashed curve in Fig.~\ref{fig:fig01}, is less than the generalized Buchdahl limit $\bet_{\text{B}}\left(\chi\right)$~\eqref{eq:Buch_comp_Sch}. As a consequence, the latter bound further constrains the parametric regime for our solution to $\chi \geqslant 0$ and $\bet \leqslant \bet_{\text{DEC}}$. Notice that $\beta_{\text{DEC}}(2\pi)=4/9$ and thus, for $\chi \geqslant 2\pi$, one obtains configurations that can be more compact than the GR Buchdahl limit without violating any of the energy conditions, and at the same time exhibiting finite, positive and monotonically decreasing profiles for both the energy density and pressure everywhere in the interior. Finally, since both compactness limits ($\bet_{\text{B}}$ and $\bet_{\text{DEC}}$) asymptote to the BH compactness limit in the large $\chi$ regime, for sufficiently large values of the MGT parameter, the configuration approaches a BH mimicker state albeit as a causality-violating configuration. Nevertheless, this model constitutes a significant improvement over the GR variant of the model where a BH mimicker state can only be approached by going beyond the corresponding Buchdahl limit~\cite{Mazur:2015kia}. The consequences of crossing the generalized Buchdahl limit in our solution are briefly discussed in the appendix~\ref{Sec:Appx_T_III_Beyond_Buchdahl}.

\begin{figure*}[ht!]
\includegraphics[width=0.495\linewidth]{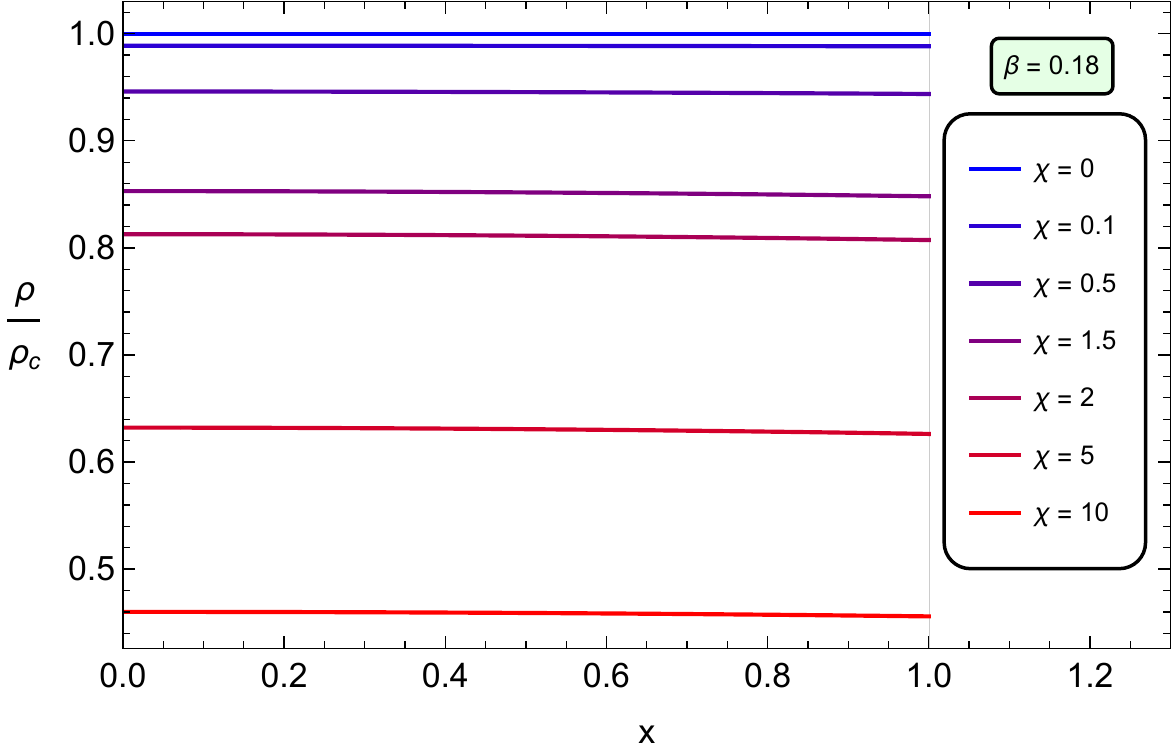}
\includegraphics[width=0.49\linewidth]{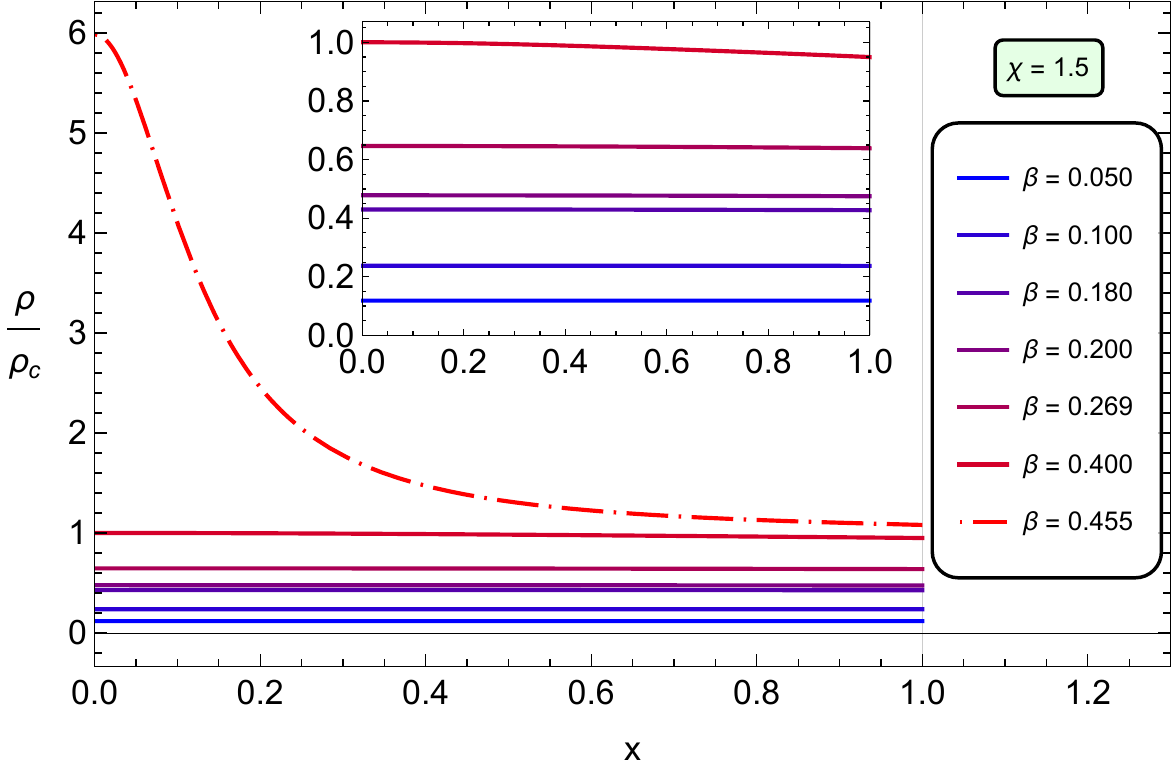}
\caption{The radial profile of the normalized energy density~\eqref{rhoFSs} for the extended T-III model, in the compact range $x \in [0,1]$. \textbf{Left panel:} for an indicative fixed value of the compactness $\beta=0.18$, and for various values of $\chi$. For the normalization we have used the central density~\eqref{eq:rho0} for $\chi=0$. The overall effect of $\chi$ is to suppress the energy density of the stellar configuration. \textbf{Right panel:} for an indicative fixed value of the MGT parameter $\chi=1.5$, and for various values of compactness. For the normalization we have used the central density~\eqref{eq:rho0} for $\bet=\bet_{\text{DEC}}(1.5)\simeq0.4$. The effect of $\beta$ is opposing the one of $\chi$ by increasing the matter energy density. The Buchdahl compactness limit~\eqref{eq:Buch_comp_Sch}, in this example, is $\beta_{\text{B}}(1.5) \simeq 0.455661$. The dash-dotted curves correspond to configurations which violate the DEC.}
\label{fig:Schw_rho_betas_chis}
\end{figure*}

Having constrained the parametric space of our solution to $\chi \geqslant 0$ and $\bet \leqslant \bet_{\text{DEC}}$ let us now turn to the effect of the free parameters $\chi$ and $\beta$ on the model. In the left panel of Fig.~\ref{fig:Schw_p_betas_chis} we plot the normalized pressure~\eqref{pFSs}, for various values of $\chi$, for $\beta=0.18$ corresponding to a configuration with the canonical-mass for a NS $M=1.4\,M_{\odot}$ and $R=11.4\,\text{km}$ ~\cite{Jiang:2019vmf}. We observe that the radial pressure decreases monotonically with $r$, with the central pressure corresponding to a global maximum. Note that the effect of $\chi$ is to suppress the pressure in the interior of the star, in this sense, the MGT effect can be thought of as a contribution of “tension" for this particular model. In the right panel of the same figure, we plot the pressure radial profile for fixed $\chi$ and varying compactness. As expected, we observe that as $\beta$ increases, the value of the central pressure also increases, up to the generalized Buchdahl limit $\beta_\text{B}(\chi)$~\eqref{eq:Buch_comp_Sch} where the central pressure diverges. However, for the particular indicative value of $\chi=1/2$ used in that figure, and according to the bound~\eqref{eq:beta_DEC}, the model violates the DEC for $\beta \geqslant 0.384384$.

\begin{figure}[ht!]
\includegraphics[width=\linewidth]{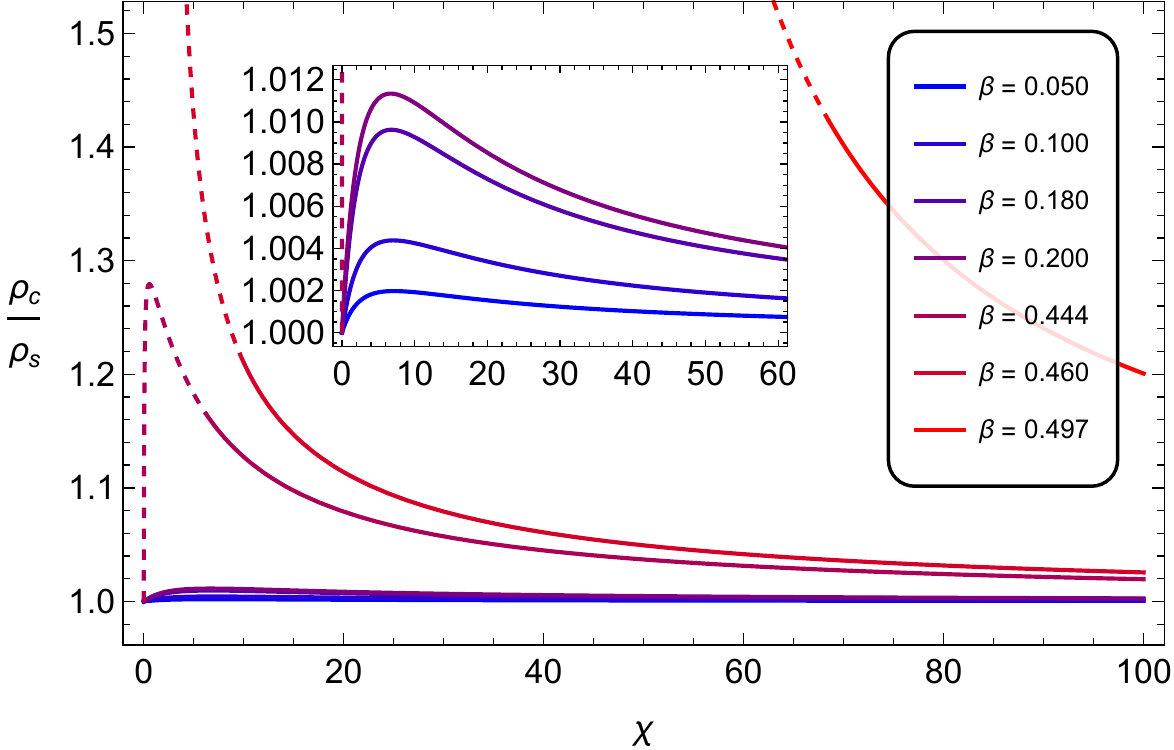}
\caption{The ratio of the central energy density~\eqref{eq:rho0}, over its surface value~\eqref{eq:T_III_rho_s}, for the extended T-III model, as a function of the parameter $\chi$, is depicted for various values of the compactness $\beta$. In the GR limit ($\chi=0$), the constant density (CD) configuration, where $\rho_c=\rho_s$, is reproduced exactly while for $\chi \gg 1$ it is recovered asymptotically for any $\beta$. The deviation from CD is significant only when $\bet$ is close to the generalized Buchdahl limit~\eqref{eq:Buch_comp_Sch}. The dashed part of the curves corresponds to configurations that violate the DEC according to the limit set by Eq.~\eqref{eq:beta_DEC}.  Notice that even “BH mimicker" states with $\beta\simeq0.5$ that comply with the DEC do not deviate drastically from the CD configuration at sufficiently large values of $\chi$.}
\label{fig:Schw_rho0_over_rho1}
\end{figure}

Turning now to the MGT effects on the energy density $\rho$~\eqref{rhoFSs}, in the left panel of Fig.~\ref{fig:Schw_rho_betas_chis}, for a fixed indicative value of compactness, we plot the radial profile of $\rho$ for a wide range of values of $\chi$. One can see that the effect of $\chi$ is to reduce the value of the energy density globally in the interior of the star. Another interesting feature of the solution is that $\rho$ remains nearly constant for all values of $\chi$ and so the solution can be characterized as a {\it quasiconstant density} interior solution, at least for values of compactness that are not close to the Buchdahl limit~\eqref{eq:Buch_comp_Sch}. More precisely, the central value of the energy density is given  by
\beq
\rho_c\equiv\rho(0)=\frac{6\beta\left[(3\pi+\chi)\sqrt{1-2\beta}-\pi\right]}{(\chi+2\pi)\left[(12\pi+3\chi)\sqrt{1-2\beta}-4\pi\right]}\,,
\label{eq:rho0}
\eeq
while the surface value is given in Eq.~\eqref{eq:T_III_rho_s}. In the GR limit ($\chi=0$), $\rho_c=\rho_s=3\bet/4\pi$ so the uniform density state is recovered, while for positive values of $\chi$, the condition $\rho_c>\rho_s$ is always satisfied. For large values of $\chi$, both $\rho_c$ and $\rho_s$ approach zero with the same dominant fall-off behavior $2\bet/\chi+\mathcal{O}\left(\chi^{-2} \right)$ and thus, for $\chi \gg 1$, the ratio $\rho_c/\rho_s = 1+ \mathcal{O}\left(\chi^{-1} \right)$ becomes independent of the compactness, to leading order in the expansion, and the constant density profile is recovered asymptotically. As illustrated in Fig.~\ref{fig:Schw_rho0_over_rho1}, for configurations at the typical NS compactness of $\bet \simeq 0.18$, the strongest deviations from constant density correspond to $\mathcal{O}\left(1\right)$ values of $\chi$ that maximize the deviation between central and surface energy density. The deviations from constant density are enhanced with compactness, however they are significant only close to the Buchdahl limit and thus the solution can indeed be characterized as a quasiconstant density for the majority of the parameter space.

\begin{figure*}[ht!]
\includegraphics[width=0.495\linewidth]{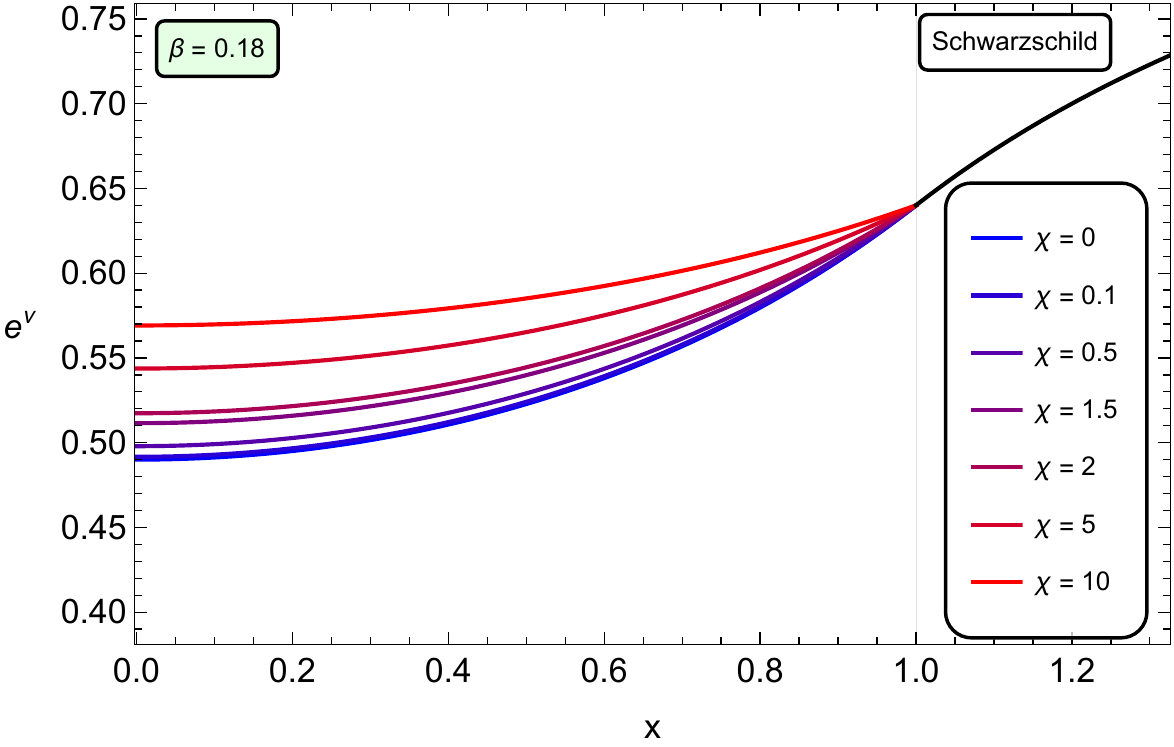}
\includegraphics[width=0.485\linewidth]{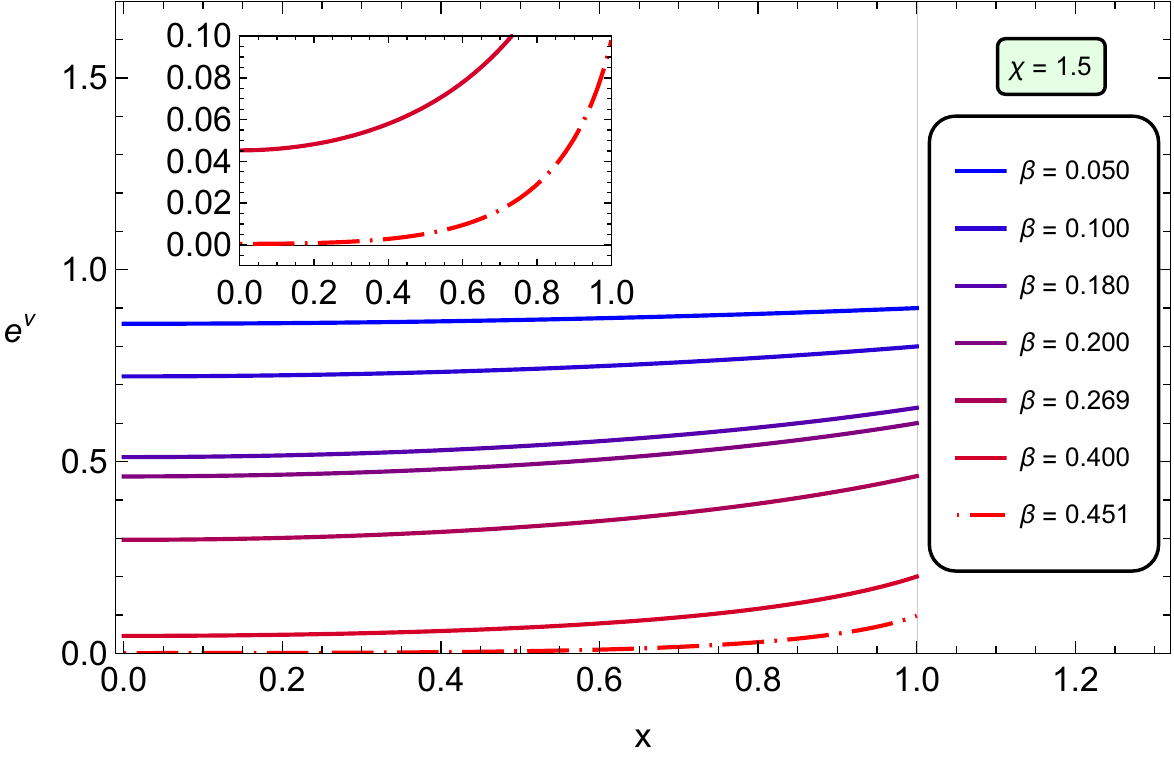}
\caption{The radial profile of the $g_{tt}(r)$ metric function~\eqref{gttFSs} for the extended T-III model, in the compact range $x \in [0,1]$. \textbf{Left panel:} for an indicative fixed value of the compactness $\beta=0.18$, and for various values of $\chi$. \textbf{Right panel:} For an indicative fixed value of the MGT parameter $\chi=1.5$, and for various values of compactness. In this example, the Buchdahl limit~\eqref{eq:Buch_comp_Sch} corresponds to $\bet_{\text{B}}(1.5)\simeq0.455661$. The dash-dotted curve corresponds to a configuration that violates the DEC bound~\eqref{eq:beta_DEC} of $\beta_{\text{DEC}}(1.5)\simeq 0.400238$. As the Buchdahl compactness is approached, the central value of $g_{tt}$ goes to zero, however for any $\beta <\bet_{\text{B}}$, it remains positive everywhere in the interior of the star.}
\label{fig:T_III_gtt_betas_chis}
\end{figure*}

For values of $\beta$ approaching $\beta_B$, the central value of the energy density increases monotonically as the right panel of Fig.~\ref{fig:Schw_rho_betas_chis} clearly illustrates. At the Buchdahl compactness, and in dire contrast with the GR variant of the model, not only the pressure but also the central energy density diverges. Note that this property of the solution does not require strong MGT effects as any nonzero $\chi$, however small in absolute value, is sufficient to drastically modify the behavior of the Schwarzschild star close to the Buchdahl limit, even though far from $\beta_B$ the solution is practically indistinguishable from the GR Schwarzschild star with its quasiconstant density nature.

The novel feature of the emergence of a pole in the energy density at the Buchdahl limit for the linear $f(\Ri,T)$ Schwarzschild star can be understood via the inspection of the $tt$ component~\eqref{EinsG00} of the field equations. The $G^0_{\,\,\,0}$ component of the Einstein tensor depends only on the $g_{rr}$ metric function for which we have adopted Tolman's ansatz~\eqref{eq:grr Sch.}. In GR, this choice for $g_{rr}$ generates the constant density interior solution i.e. we have the condition $G^0_{\,\,\,0}=8\pi \rho_c=6 \beta$. Thus, for our model, from~\eqref{EinsG00} we have the following constraint between energy density and pressure
\beq
\left(8\pi+3 \chi \right) \rho - \chi p \equiv 8 \pi \rho_{\text{eff}}=6 \beta\,.
\label{eq:T_III_constraint_p_rho}
\eeq
Since $\rho$ and $p$ enter the constraint equation~\eqref{eq:T_III_constraint_p_rho} with opposite signs, the divergence of $p$ at $\beta_B$ forces also the divergence of $\rho$ so that the effective energy density remains constant.

Let us finally turn to the effects of the interplay between $\chi$ and $\beta$ on the $g_{tt}$ metric function. We plot the dependence of the $g_{tt}(r)$ metric function on $\chi$ and $\beta$ in the left and right panel of Fig.~\ref{fig:T_III_gtt_betas_chis}  respectively. Once again, we have an opposing effect between the parameters $\chi$ and $\beta$, with the former causing an enhancement of the $g_{tt}$ metric function and the latter causing a suppression.

The surface value of $g_{tt}$ is of course independent of $\chi$ and is given by $g_{tt}(1)=1-2\beta$, see left panel in Fig.~\ref{fig:T_III_gtt_betas_chis}. This is a consequence of the boundary conditions that require the matching of the interior metric with the Schwarzschild exterior solution, which is independent of MGT effects given that it is a vacuum region $(T=0)$ in linear $f(\Ri,T)$ gravity.

\section{Tolman VII extended model in $f(\Ri,T)$ gravity}
\label{Sec:T_VII_analytic_extension}

Following Tolman's approach, as described in Sec.~\ref{Sec:T_III_analytic_extension}, we now proceed to find the extension of the so-called Tolman VII (T-VII) solution in linear $f(\Ri,T)$ gravity. We solve the modified field equations assuming the $g_{rr}$ metric element in the form~\cite{Tolman:1939jz}
\beq
e^{\lambda(r)}=\left(1-ax^2+bx^4\right)^{-1},
\label{eq:TVII_grr_ansatz}
\eeq
where $a$ and $b$ are integration constants, with appropriate units, and $x \equiv r/R$. Notice that for $b=0$ the T-III model is recovered from~\eqref{eq:TVII_grr_ansatz}, thus the T-VII model could be viewed, in the context of GR, as a generalization of the Schwarzschild's interior solution with uniform density. This association between the two models will of course extend, for the same reasons, in linear $f(\Ri,T)$ gravity since we are employing the same functional forms for $g_{rr}$ as in GR in order to obtain our extended solutions. Solving the system~\eqref{EinsG00}-\eqref{p isotropy Eq.}, with the ansatz~\eqref{eq:TVII_grr_ansatz}, requires four boundary conditions in total. To this end, in addition to~\eqref{bc1} and~\eqref{bc2}, we will further impose the physically sensible requirement that for ultracompact configurations, such as neutron and supermassive stars, the surface value of the energy density $\rho_s = \rho(R)$ is negligible compared to the central value $\rhoc = \rho(0)$. Thus, the final boundary condition will be taken to be
\beq
\rho(R)=0\,.
\eeq
We then obtain the following two-parametric family of solutions for the extended T-VII model in linear $f(\Ri,T)$ gravity in terms of the MGT parameter $\chi$ and compactness $\beta$
\begin{widetext}
\bea
g_{tt}(x) &=& \frac{1}{2}\left(1-\frac{5\beta}{3}\right)\left[1+\cos\left(C_{1}-\tanh^{-1}\theta(x)\right)\right] \,, \label{gttTVII}\\
g_{rr}(x) &=&e^{\lam(x)}= \left[1-\beta x^2 (5-3x^2)\right]^{-1}   \,, \label{grrTVII}\\
\rho(x) &=& \frac{60\pi\beta\left(1-x^2\right) + \chi\left\{\beta\left(20-21 x^2\right) + \sqrt{3 \beta 
e^{-\lambda}} \tan\left[\frac{1}{2} \left(C_{1}-\tanh^{-1}\theta(x)\right)\right]\right\}}{4(\chi+2\pi)(\chi+4\pi)}   \,, \label{rhoTVII}\\
p(x) &=& \frac{4\pi\beta\left(3 x^2-5\right) - 3\chi\beta x^2 + (8\pi+3\chi)\sqrt{3\beta e^{-\lambda}}\tan \left[\frac{1}{2}\left(C_{1}-\tanh ^{-1}\theta(x)\right)\right]}{4 (\chi +2 \pi ) (\chi +4 \pi )}        \,, \label{pTVII}
\eea
where
\beq
\theta(x) \equiv \sqrt{3\beta e^{\lambda}} \left(x^2-5/6\right), \quad
C_{1} = 2\tan^{-1}\left(\sqrt{\frac{\beta}{3(1-2\beta)}}\right)+\coth^{-1}\left(2\sqrt{\frac{3(1-2\beta)}{\beta}}\right).
\eeq
\end{widetext}

As we discussed in Sec.~\ref{Sec:T_III_analytic_extension},~\cite{Hansraj:2018jzb} proposed a generalization of the Tolman VII solution in linear $f(\Ri,T)$ gravity. They argue that their solution is the most general, considering that Tolman's was not, which reduces to theirs in the appropriate limit. Thus, they derived relations for $(g_{tt}, p, \rho)$ which are expressed in terms of certain integration constants. However, the authors did not carry out the appropriate matching with the exterior Schwarzschild solution; instead they assigned arbitrary values to these integration constants which produced erratic behaviors for the pressure and energy density profiles.

\begin{figure*}[ht!]
\centering
\includegraphics[width=0.49\linewidth]{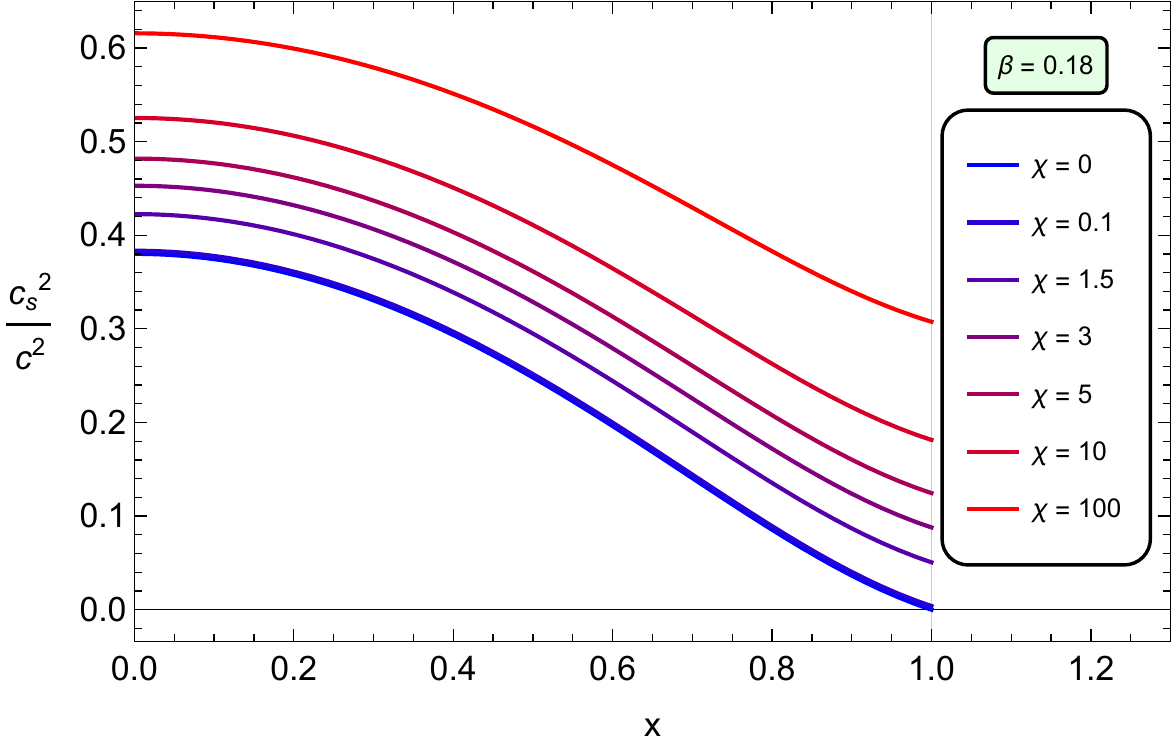}
\includegraphics[width=0.49\linewidth]{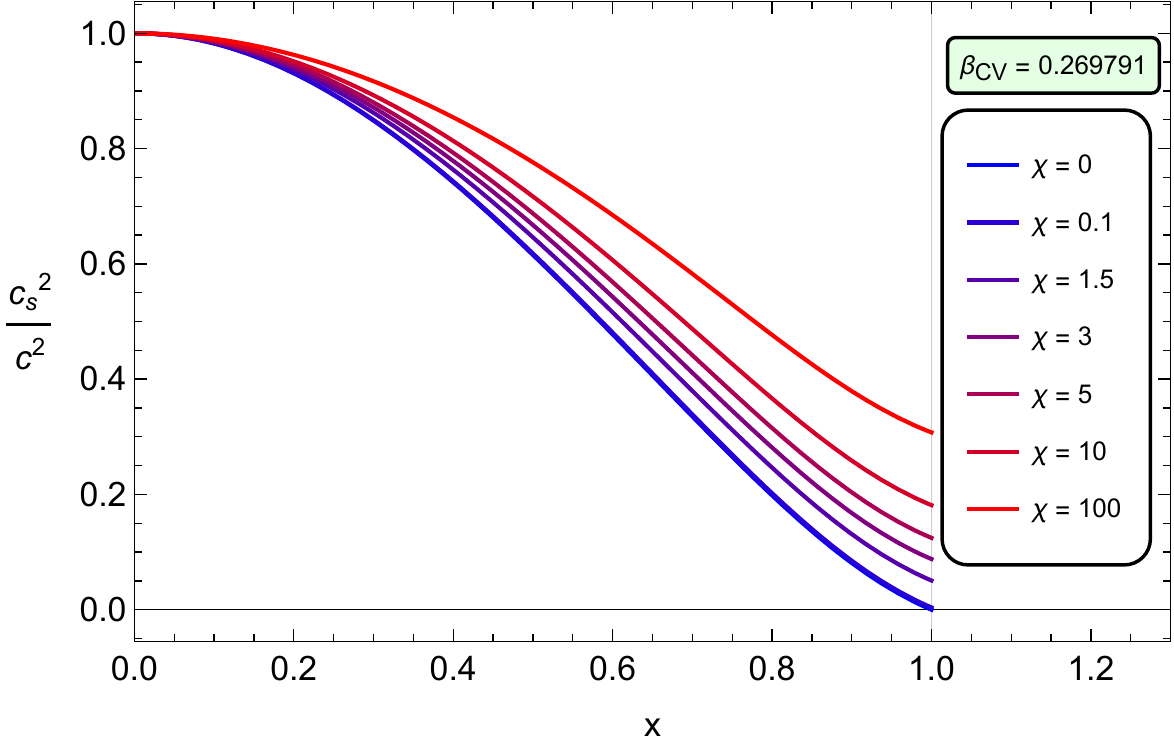}
\caption{The effect of the MGT parameter $\chi$ on the radial profile of the square of the speed of sound $c_s^2$ in the compact range $x \equiv r/R \in [0,1]$ for the extended T-VII model. \textbf{Left panel:} at the typical neutron star compactness of $\beta = 0.18$. \textbf{Right panel:} at the $\chi$-independent causality violation limit of $\beta_{\text{CV}} \simeq 0.269791$. The effect of $\chi$ is to universally enhance $c_s^2$ in the interior for any value of compactness that yields a causal configuration i.e. $\beta \leqslant \beta_{\text{CV}}$ and to induce a nonvanishing surface value $c_s^2(1)$~\eqref{eq:TVII_cs2_1}. In the GR limit ($\chi=0$), the profile of $c_s^2$ for the original T-VII model~\cite{Tolman:1939jz} is recovered.}
\label{fig:TVII_cs2_betas_effect}
\end{figure*}
\begin{figure*}[ht!]
\centering
\includegraphics[width=0.47\linewidth]{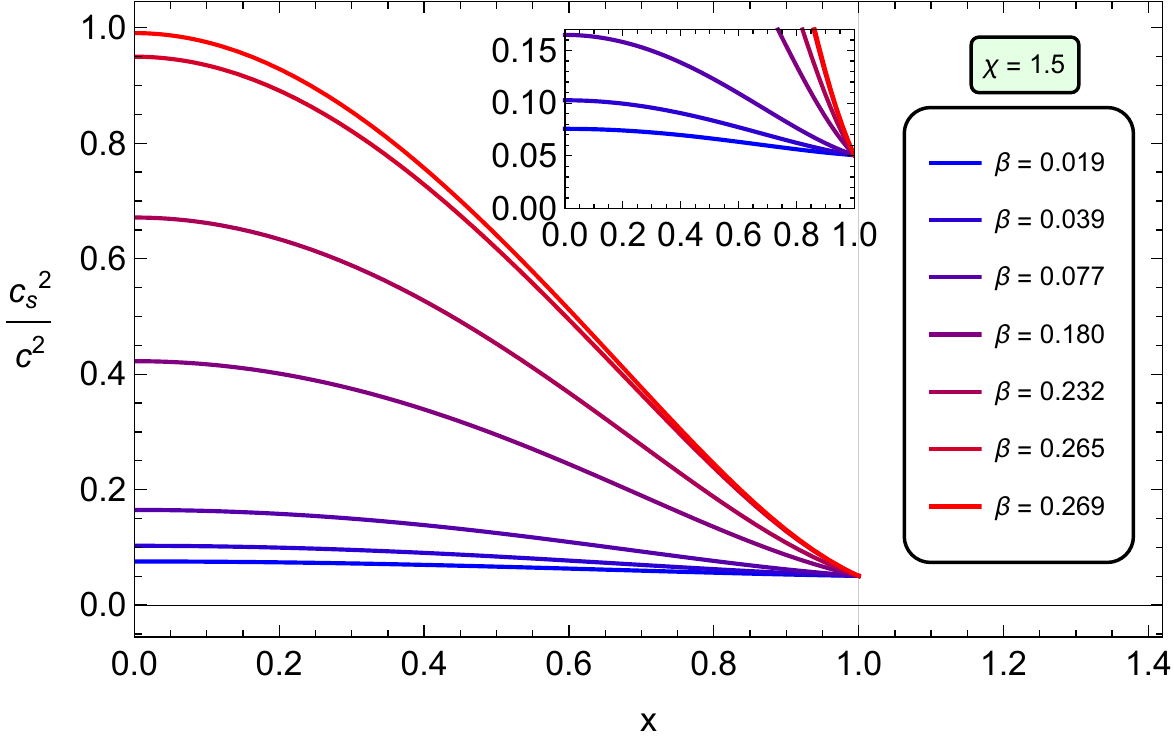}
\includegraphics[width=0.49\linewidth]{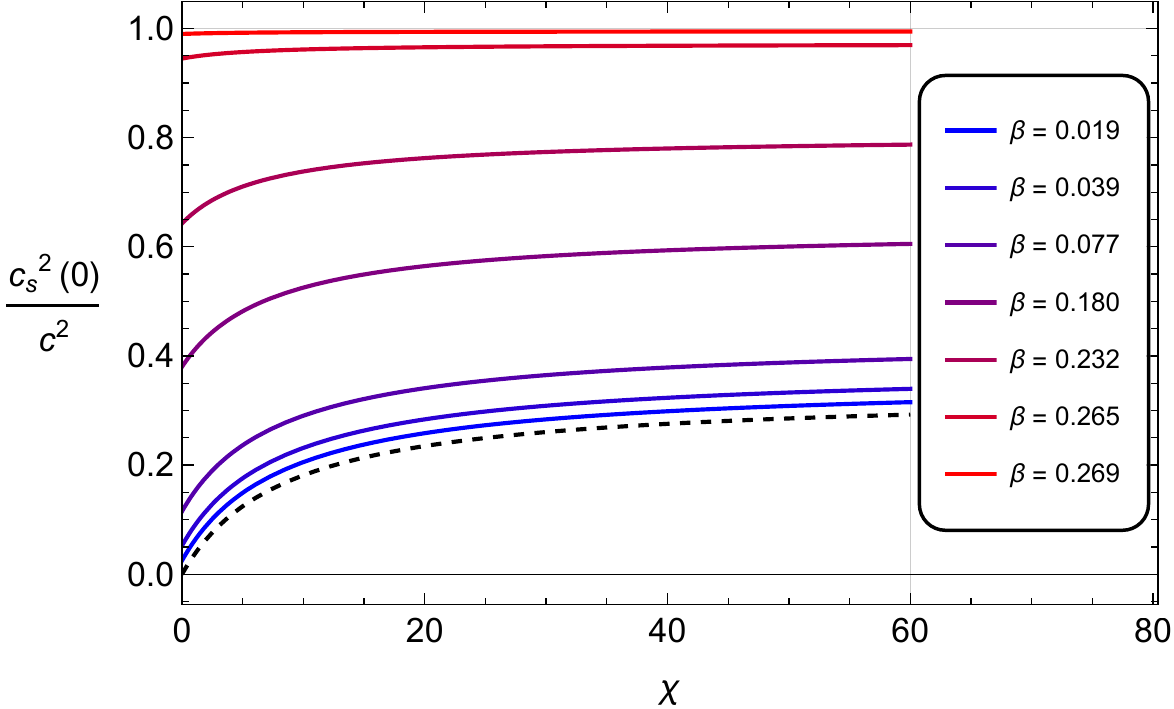}
\caption{For the extended T-VII model: \textbf{Left panel:} the effect of compactness on the radial profile of the square of the speed of sound $c_s^2$ in the compact range $x \equiv r/R \in [0,1]$ for an indicative value of the MGT parameter $\chi=1.5$. The surface value $c_s^2(1) \geqslant 0$ is independent of the compactness parameter $\beta$~\eqref{eq:TVII_cs2_1}. For the GR case ($\chi=0$), $c_s^2(1)=0$. \textbf{Right panel:} the effect of the compactness parameter $\beta$ on the central ($x=0$) value of $c_s^2$ as a function of the MGT parameter $\chi$. Causality is violated universally for any $\chi$ when $\beta > \beta_{\text{CV}} \simeq 0.269791$. For large values of $\chi$, $c_s^2(0)$ asymptotes to the $\beta$-dependent value of Eq.~\eqref{eq:TVII_cs2_0_max}, while, as $\beta_{\text{CV}}$ is approached, $c_s^2(0)\to c^2$ for any value of $\chi$. The black-dashed curve corresponds to the surface value of $c_s^2$~\eqref{eq:TVII_cs2_1} and is always less than its central value for any $\beta$ and $\chi$.}
\label{fig:TVII_chis_effect}
\end{figure*}

\begin{figure*}[ht!]
\centering
\includegraphics[width=0.49\linewidth]{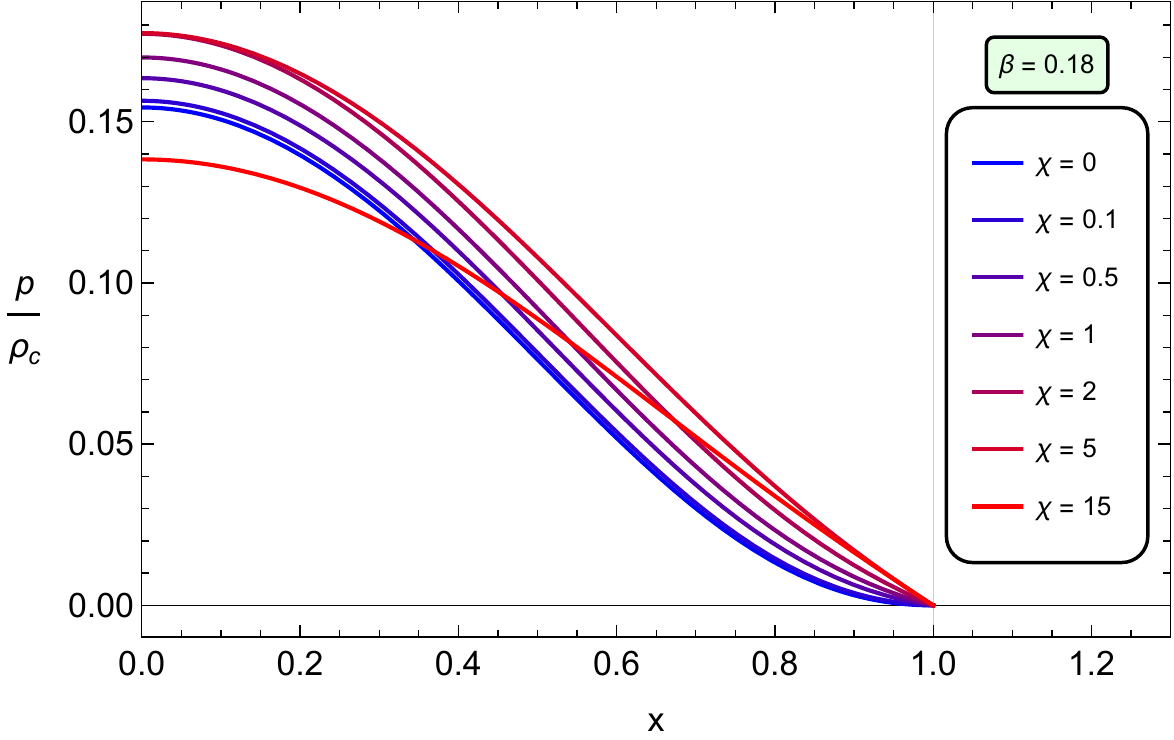}
\includegraphics[width=0.48\linewidth]{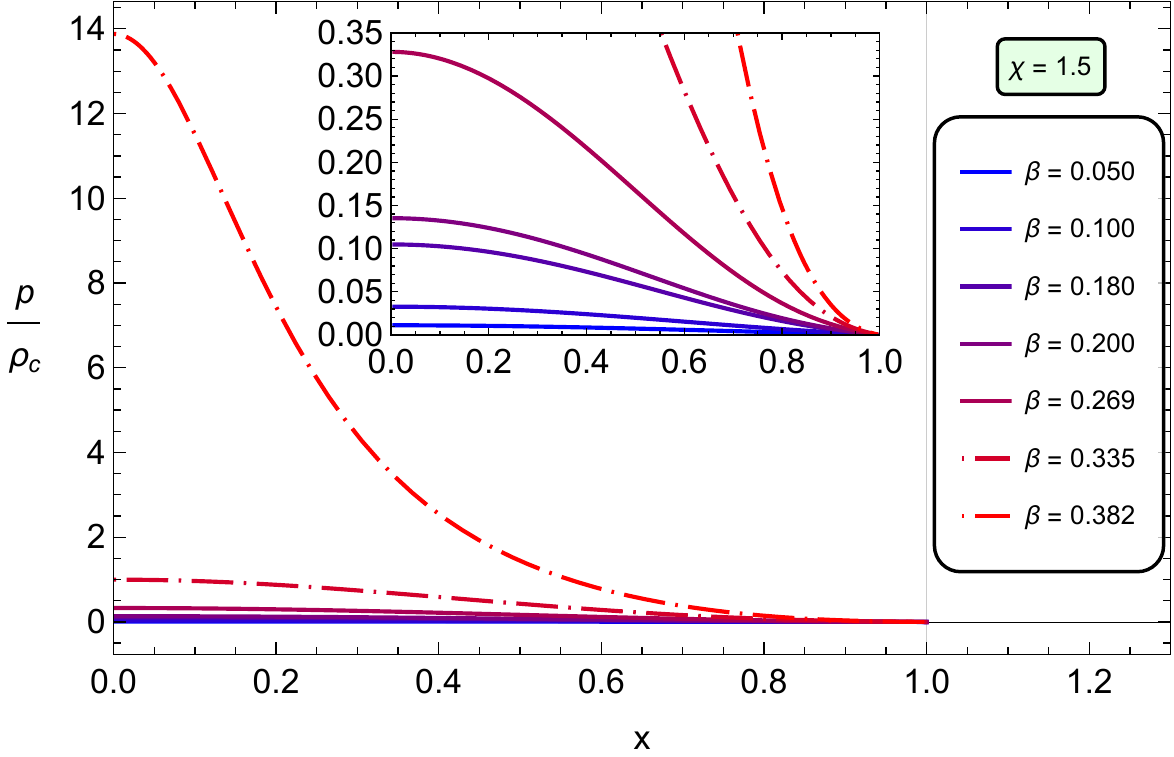}
\caption{Profile of the normalized radial pressure $p(r)$, as a function of $x \equiv r/R$, for the extended T-VII solution in linear $f(\Ri,T)$ gravity. \textbf{Left panel:} profiles for an indicative value of $\beta=0.18$, and for various values of $\chi$. For the normalization we have used the central density for $\chi=0$. \textbf{Right panel:} profiles for a fixed indicative value of $\chi=1.5$, and for different values of compactness. For the normalization we have used the central density for $\beta=\beta_{\text{DEC}}\simeq0.335$. The dash-dotted curves correspond to configurations that exhibit superluminal speed of sound. As the Buchdahl compactness limit $\beta_{\text{B}} \simeq 0.386$ is approached, the central pressure exhibits a sharp increase.}
\label{fig:p_TVII}
\end{figure*}
\begin{figure*}[ht!]
\centering
\includegraphics[width=0.49\linewidth]{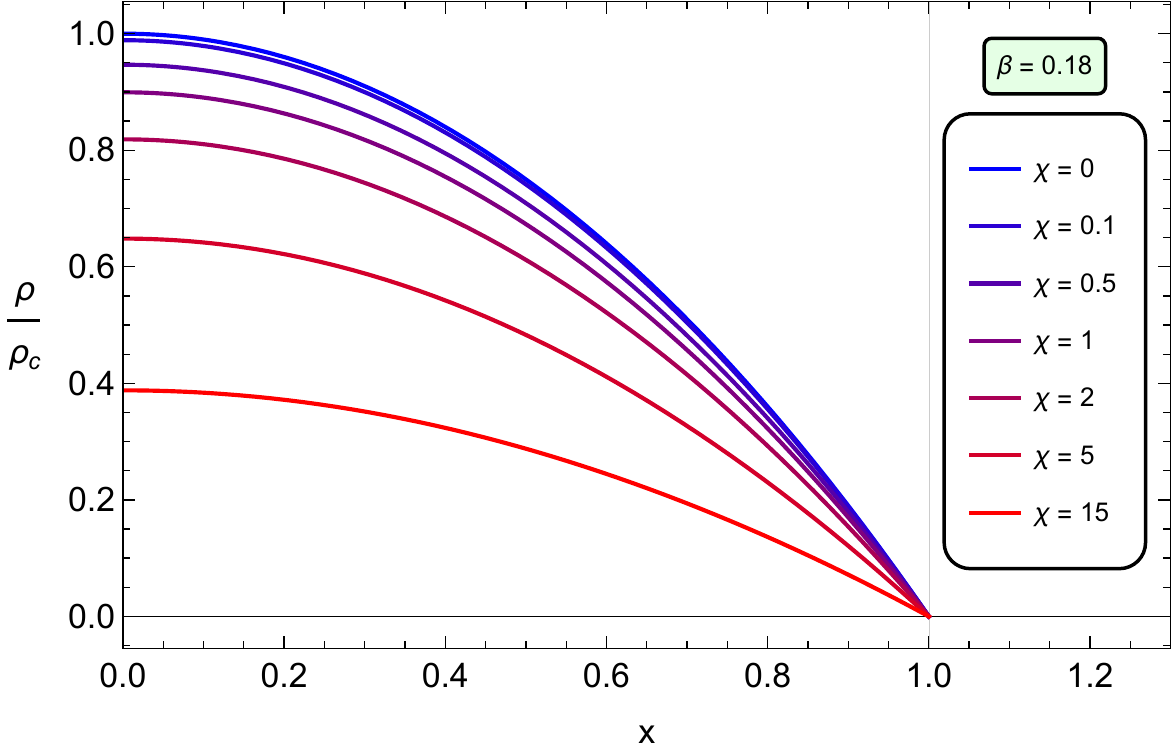}
\includegraphics[width=0.49\linewidth]{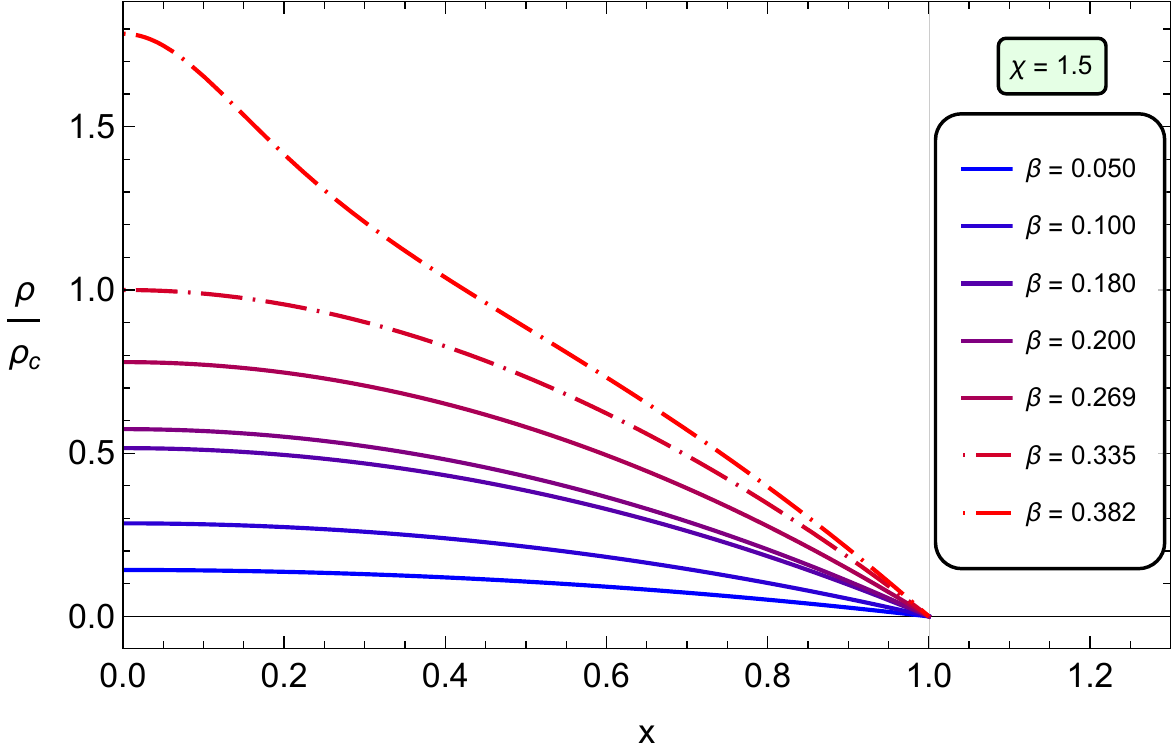}
\caption{Profile of the energy density $\rho(r)$, as a function of $x\equiv r/R$, for the extended T-VII solution in linear $f(\Ri,T)$ gravity. \textbf{Left panel:} profiles for an indicative value of $\beta=0.18$, and for various values of $\chi$. For the normalization we have used the central density for $\chi=0$. \textbf{Right panel:} profiles for a fixed indicative value of $\chi=1.5$, and for different values of compactness. For the normalization we have used the central density for $\beta=\beta_{\text{DEC}}\simeq0.335$. The dash-dotted curves correspond to configurations that exhibit superluminal speed of sound. As the Buchdahl compactness limit $\beta_{\text{B}} \simeq 0.386$ is approached, the central energy density exhibits a sharp increase.}
\label{fig:rho_TVII}
\end{figure*}

\begin{figure*}[ht!]
\centering
\includegraphics[width=0.49\linewidth]{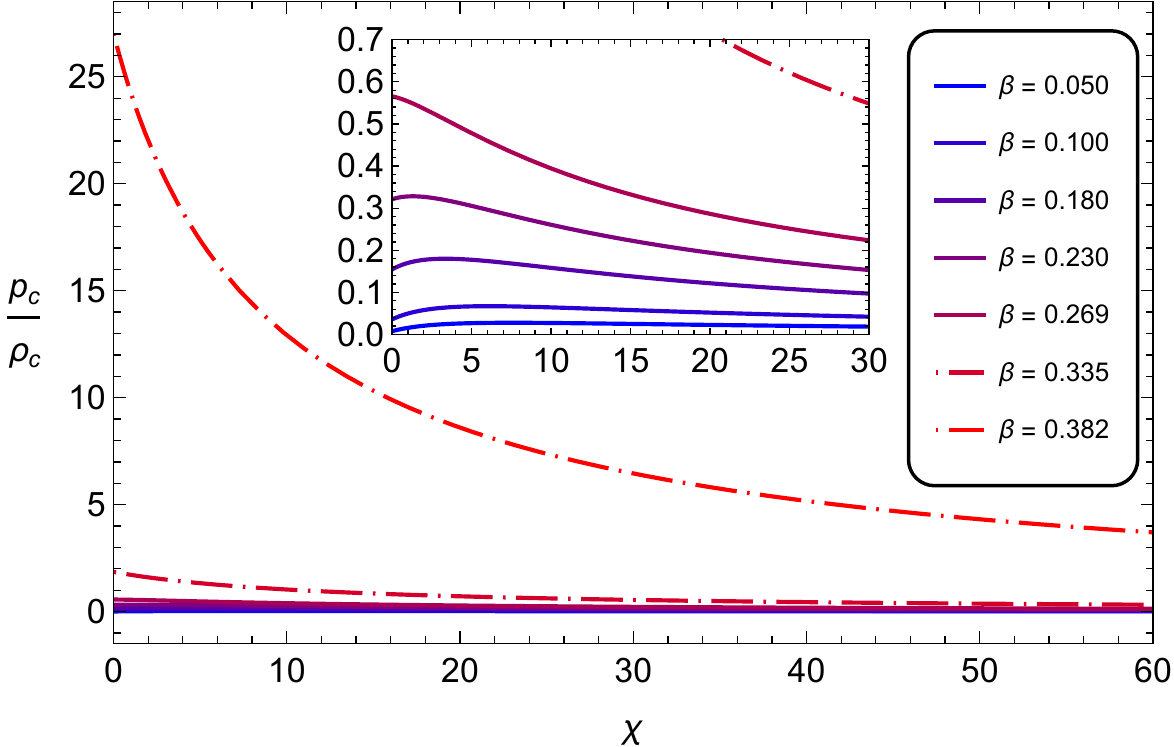}
\includegraphics[width=0.49\linewidth]{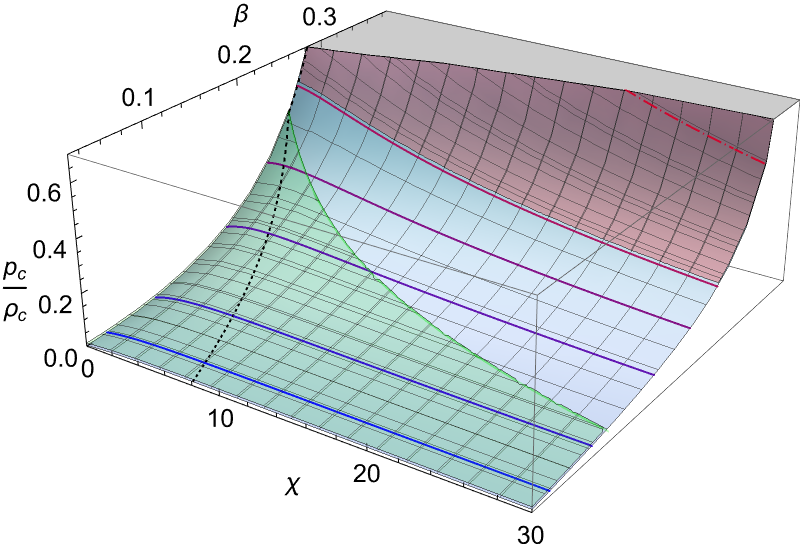}
\caption{The effect of $\chi$ and $\beta$ on the normalized central pressure $p_c\left(\chi,\beta\right)$, for the extended T-VII model. For the normalization, we have used the central density $\rho_c\left(\chi,\beta\right)$ evaluated for $\chi=0$ and $\bet=0.18$. \textbf{Left panel:} $p_c/\rho_c$ as a function of $\chi$ for various indicative values of $\beta$. The effect of $\beta$ is to always enhance the central pressure independently of $\chi$. Dash-dotted curves correspond to configurations that violate causality. \textbf{Right panel:} $p_c/\rho_c$ plotted on the $\chi-\beta$ parametric plane. The colored curves correspond to the values of compactness as displayed in the left panel with the same color coding. The red shaded regime corresponds to $\beta\geqslant \bet_{\text{CV}}$ and thus to configurations that violate causality. The green-shaded regime corresponds to configurations with $p_c\left(\chi,\bet\right)\geqslant p_c\left(0,\bet\right)$ i.e. to configurations with higher pressure than in GR for the same compactness. Everywhere else we have $p_c\left(\chi,\bet\right) <p_c\left(0,\bet\right)$. For each fixed value of compactness, the black dashed curve corresponds to the location $\chi_0\left(\bet\right)$~\eqref{eq:TVII_chi_0_max_p_0} of the maximum of $p_c \left(\chi\right)$. Notice that in the regime where $\beta \gtrapprox 0.257441$ the maximum is always located at $\chi=0$ and thus the effect of any $\chi \geqslant 0$ is to suppress the central pressure with respect to its GR value. Finally, in the bounded regime where $0<\chi\leqslant \chi_0\left(\bet\right)$, the effect of $\chi$ is to enhance $p_c$ while everywhere else its effect is to suppress $p_c$.}
\label{fig:p_0_TVII}
\end{figure*}

We now turn to the analysis of our solution. First of all, note that the metric elements $g_{tt}$ and $g_{rr}$ [Eqs.~\eqref{gttTVII},~\eqref{grrTVII}] do not depend on the parameter $\chi$, thus they preserve the same form as in GR. This behavior differs from the one we found for the T-III model in Sec.~\ref{Sec:T_III_analytic_extension}, and also from the one found in~\cite{Bhar:2021uqr} for the Tolman IV solution. On the other hand, we observe that the expressions for energy density and pressure [Eqs.~\eqref{rhoTVII} and~\eqref{pTVII}] have additional terms which are coupled to the parameter $\chi$. However, when $\chi=0$, we recover the T-VII solution in GR. An inspection of the expressions~\eqref{rhoTVII} and~\eqref{pTVII} reveals that they both involve the same $\chi$-independent quantity
\beq
\phi\left(x,\beta \right) \equiv \frac{1}{2} \left[C_{1}-\tanh^{-1}\theta(x)\right]\,,
\label{eq:TVII_angle}
\eeq
as the argument in the tangent function. Consequently, the central values of both the pressure and the energy density diverge when $\phi(0,\bet_{\text{B}})=\pi/2$ independently of $\chi$, at the same compactness as in the original T-VII model of GR, namely at $\bet_{\text{B}} \simeq 0.386195$~\cite{Lattimer:2000nx}. This property of the solution is in contrast to the quasiconstant density case studied in Sec.~\ref{Sec:T_III_analytic_extension}, where the Buchdahl limit is a function of $\chi$.

Let us now identify the regions in the $\chi-\beta$ parametric plane of our solution~\eqref{gttTVII}-\eqref{pTVII} where a configuration with subluminal speed of sound is obtained. In Fig.~\ref{fig:TVII_cs2_betas_effect} we plot the radial profile of the squared speed of sound $c_s^2\equiv dp/d\rho$ in terms of the compact radial coordinate $x\equiv r/R$, for various values of the MGT parameter $\chi$. Notice that the effect of $\chi$ is to induce a nonvanishing surface value for $c_s^2(x)$ which is given by the simple expression
\beq
c_s^2(1)= \frac{\chi}{8 \pi+3\chi}\,,
\label{eq:TVII_cs2_1}
\eeq
and is thus independent of the compactness (see also the left panel of Fig.~\ref{fig:TVII_chis_effect}). In the GR limit $c_s^2(1)=0$ as it should, since the original T-VII model is recovered. One of the necessary conditions for a causal configuration is $c_s^2(1) \in [0,1]$ and from Eq.~\eqref{eq:TVII_cs2_1} one can see that the aforementioned condition is realized for values of $\chi \in (-\infty,-4\pi] \cup [0,\infty)$. However, from the expression for the energy density~\eqref{rhoTVII} one can further show that for any value of $\chi <-4\pi$ the central energy density is negative, while for $\chi=-4\pi$ it exhibits a pole. Thus, we are constrained to $\chi \geqslant 0$ in order to have a physically plausible configuration. Finally, notice that for large values of $\chi$ we have the asymptotic limit of $c_s^2(1)_{\chi \gg 1}=1/3$ which is the maximum value that can be attained by $c_s^2(1)$.

The central value of the square of the speed of sound is given by the following expression
\beq
c_s^2(0)= 3+\frac{8\pi}{\chi}-\frac{120(2\pi+\chi)(4\pi+\chi)}{120\pi\chi+\chi^2\left(5\sqrt{3\beta}\tan\phi_0+3\sec^2\phi_0+42\right)} \,,
\label{eq:TVII_cs2_0}
\eeq
with $\phi_0 \equiv \phi\left(0,\beta \right)$ as defined in Eq.~\eqref{eq:TVII_angle}. In the GR limit, the $\beta$-dependent value for $c_s^2(0)$ for the original T-VII model is recovered. For large values of $\chi$, $c_s^2(0)$ asymptotes to the maximum value
\beq
c_s^2(0)_{\text{max}}=3-\frac{120}{5\sqrt{3\beta}\tan\phi_0+3\sec^2\phi_0+42}\,,
\label{eq:TVII_cs2_0_max}
\eeq
which is uniquely specified by the compactness of the configuration, (see also right panel of Fig.~\ref{fig:TVII_chis_effect}). Furthermore notice that for any $\chi \geqslant 0$ the central value of $c_s^2$ is always larger than its surface value and thus causality is violated whenever $c_s^2(0)>0$. As it can be seen from the rhs panels of Figs.~\ref{fig:TVII_cs2_betas_effect} and~\ref{fig:TVII_chis_effect}, the critical value of compactness for which causality is violated, $\beta_{\text{CV}} \simeq 0.269791$, is independent of $\chi$ and thus identical with the corresponding GR value of the original T-VII model. In summary, we have shown that causality considerations constrain the parametric space of the solution to $\chi \geqslant 0$ and $\beta \leqslant \bet_{\text{CV}}$.

Let us now turn to the constraints from the ECs. Once again, since both $\rho(x) \geqslant 0$ and $p(x) \geqslant 0$ everywhere in the interior for any value of $\chi \geqslant 0$ we need to consider only the DEC. Along the lines of the analysis we performed for the T-III extended solution, we find that the monotonically decreasing ratio of $p/\rho$ has a central value that becomes equal to unity for all values of $\chi$ at the DEC violation compactness, $\bet_{\text{DEC}}\simeq 0.335119$. Thus we have that for the T-VII extended model $\beta_{\text{CV}}<\beta_{\text{DEC}}<\beta_{\text{B}}$. Therefore, physically relevant configurations are obtained in the constrained parametric regime $\chi \geqslant 0$ and $\beta \leqslant \beta_{\text{CV}}\simeq 0.269791$. In the remaining of this section we will consider the effects of $\chi$ and $\beta$ on the extended T-VII model in the aforementioned parametric regime.

In Figs.~\ref{fig:p_TVII} and~\ref{fig:rho_TVII} we present the radial profiles of the energy density and pressure for the extended T-VII analytical solution in linear $f(\Ri,T)$ gravity. As we mentioned above, when $\chi=0$ we recover the T-VII solution in GR~\cite{Posada:2021zxk}. In the left panel of Fig.~\ref{fig:p_TVII}, we see that the pressure is always maximized at the center of the configuration and exhibits a monotonically decreasing radial profile. However, the effect of $\chi$ on the pressure is nontrivial. We observe that for the indicative fixed value of compactness, small values of $\chi$ cause an enhancement of pressure with respect to its GR value ($\chi=0$) while above some critical $\beta$-dependent value $\chi_0$, the role of $\chi$ is reversed and it results in a suppression of the pressure. We demonstrate in detail the aforementioned dual role of $\chi$ in Fig.~\ref{fig:p_0_TVII}.

The critical value of $\chi_0$, as a function of $\beta$, for which $p_c(\chi)$ is maximized is given by
\bea
\chi_0\left(\beta\right)&=&\frac{2\pi\cot(\phi_0)\sqrt{2\beta\left[50\beta+5\sqrt{3\beta}\tan(\phi_0)-3\tan^2(\phi_0)\right]}}{3\sqrt{3\beta}}\nonumber\\
&&+\frac{20\pi\beta\cot(\phi_0)}{3\sqrt{3\beta}}-\frac{8\pi}{3}\,,
\label{eq:TVII_chi_0_max_p_0}
\eea
with $\phi_0 \equiv \phi\left(0,\beta \right)$ as defined in Eq.~\eqref{eq:TVII_angle}. For $\beta \simeq 0.257441$, $\chi_0\left(\beta\right)$ vanishes and this implies that for more compact configurations the effect of any $\chi \geqslant 0$ is to suppress the central pressure $p_c$ with respect to its GR value.

In the right panel of Fig.~\ref{fig:p_TVII} we see that the effect of compactness on the configuration is to globally enhance the pressure, while as the Buchdahl limit $\beta_{\text{B}}\simeq 0.386195$ is approached, the central pressure exhibits a rapid increase. Finally in Fig.~\ref{fig:rho_TVII} we see that the effect of $\chi$ is to suppress the energy density, while the compactness always has the opposite effect of enhancing $\rho$. Notice that as the Buchdahl limit is approached, the central value of the energy density quickly diverges.

\section{A parametric deformation of the extended T-VII model}
\label{Sec:rho_s_extended_T_VII}

In Sec.~\ref{Sec:T_III_analytic_extension} we have obtained the extension of the Schwarzschild's constant-density (CD) interior solution (also known as T-III model) of GR in the context of linear $f(\Ri,T)$ gravity. As we have shown, the T-III extended exact analytic solution does not correspond to a CD configuration when the MGT effects are taken into account but rather to a quasiconstant density one. In order to  obtain the CD configuration in  linear $f(\Ri,T)$ one has to solve the field equations~\eqref{EinsG00}-\eqref{p isotropy Eq.} under the ansatz of $\rho(r)=\rho_c=\text{const.}$, for the energy density. However, the aforementioned system, in contrast to GR where an analytic solution can be readily obtained, is not easily solvable in $f(\Ri,T)$ gravity and thus the CD configuration has been missing from the literature. In this section, in an attempt to fill this gap, we construct a parametric deformation (PD) of the extended T-VII solution that interpolates between the T-III and T-VII models for any value of $\chi$. We show that via this PD solution, we are also able to obtain highly-accurate analytic approximations for the CD configuration in linear $f(\Ri,T)$ gravity. In the subsequent section~\ref{Sec:General_TVII_numer_sol}, we tackle the problem numerically, and present the solution for an exact CD configuration. The latter solution has been used as a test for the accuracy of our here-presented analytic approximation.

In the process of implementing the boundary conditions for the T-VII extended solution, we have required that the surface value of the energy density is zero. For realistic ultracompact configurations this is a reasonable assumption, but in principle, the general solution to the system of field equations admits an arbitrary constant value $\rho_s$ for the surface energy density. In this case, one ends up with the following three-parametric family of solutions in terms of $\rho_s,\beta,$ and $\chi$
\begin{widetext}
\bea
g_{tt}(x)&=&\left\{1-\frac{5\beta}{3}-\frac{\rho_s}{3}\left[\frac{\chi}{2}-\frac{4\pi}{3} \left(1+\frac{8\pi\rho_s}{6\beta-\rho_s(8\pi+3\chi)}\right)\right]\right\}\cos^2\left[\tan^{-1}(\phi_1)+\frac{\tanh^{-1}(\phi_2)-\tanh^{-1}(\phi_3(x))}{2}\right]\,,\label{eq:gtt(x)_generalized_TVII}\\
g_{rr}(x)&=&e^{\lam(x)}=\frac{2}{2\left[1-\beta x^2\left(5-3x^2\right)\right]+\rho_s(8\pi+3\chi)\left(x^2-x^4\right)}\,,\label{eq:grr(x)_generalized_TVII}\\
\rho(x)&=&\frac{40\beta(3\pi+\chi)-\rho_s\left(96\pi^2+68\pi\chi+12\chi^2\right)+x^2\left[\rho_s\left(160\pi^2+116\pi\chi+21\chi^2\right)-\beta(120\pi+42\chi)\right]}{8(2\pi+\chi)(4\pi+\chi)}+\nonumber \\
&&\frac{e^{-\lam(x)/2}\chi\sqrt{6\beta-\rho_s(8\pi+3\chi)}}{\sqrt{32}(2\pi+\chi)(4\pi+\chi)}\tan\left[\tan^{-1}(\phi_1)+\frac{\tanh^{-1}(\phi_2)-\tanh^{-1}(\phi_3(x))}{2}\right]\,,\label{eq:rho(x)_generalized_TVII}\\
p(x)&=&\frac{\pi[\rho_s(32\pi+12\chi)-40\beta]+x^2\left[6\beta(4\pi-\chi)-\rho_s\left(32\pi^2+4\pi\chi-3\chi^2\right)\right]}{8(2\pi+\chi)(4\pi+\chi)}+\nonumber\\
&&\frac{e^{-\lam(x)/2}(8\pi+3\chi)\sqrt{6\beta-\rho_s(8\pi+3\chi)}}{\sqrt{32}(2\pi+\chi)(4\pi+\chi)}\tan\left[\tan^{-1}(\phi_1)+\frac{\tanh^{-1}(\phi_2)-\tanh^{-1}(\phi_3(x))}{2}\right]\,,\label{eq:p(x)_generalized_TVII}
\eea
\end{widetext}
where
\beq
\phi_1\equiv\frac{2\beta-\rho_s\chi}{\sqrt{(1-2\beta)\left[12\beta-2\rho_s\left(8\pi+3\chi\right)\right]}}\,,
\eeq
\beq
\phi_2\equiv\frac{2\beta-\rho_s(8\pi+3\chi)}{2\sqrt{(1-2\beta)\left[12\beta-2\rho_s\left(8\pi+3\chi\right)\right]}}\,,
\eeq
\beq
\phi_3(x)\equiv\frac{\rho_s(8\pi+3\chi)-10\beta+2x^2\left[6\beta-\rho_s(8\pi+3\chi)\right]}{2\sqrt{12\beta-2\rho_s(8\pi+3\chi)}}e^{\lam(x)/2}\,.
\eeq

\begin{figure*}[ht!]
\centering
\includegraphics[width=0.49\linewidth]{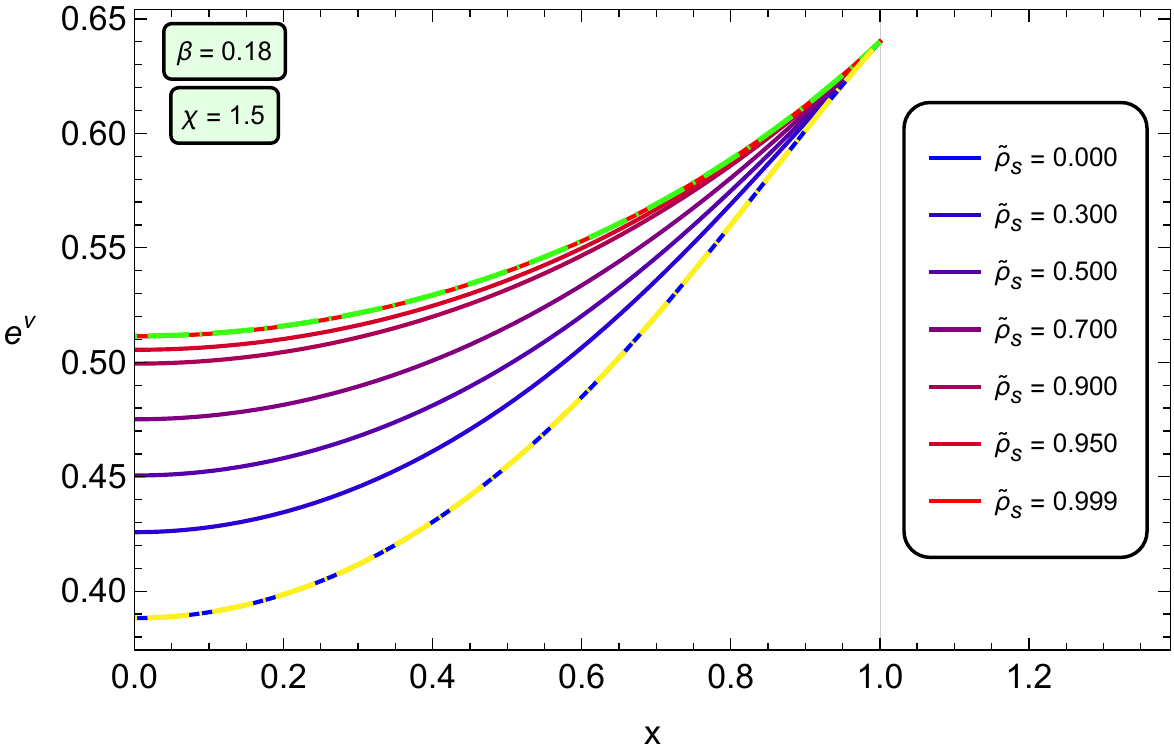}
\includegraphics[width=0.48\linewidth]{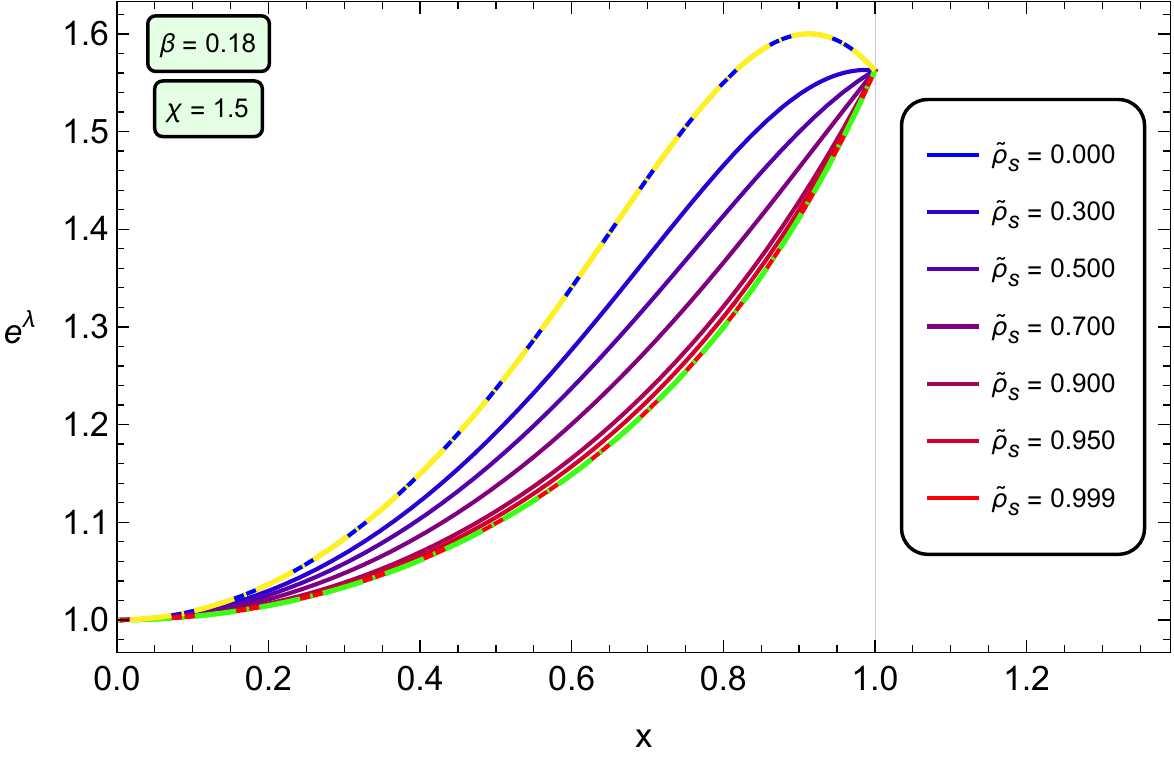}
\caption{The radial profiles of the $g_{tt}$~\eqref{eq:gtt(x)_generalized_TVII} and $g_{rr}$~\eqref{eq:grr(x)_generalized_TVII} metric functions of the parametrically deformed T-VII model, as the value of the surface energy density parameter ranges from $\rho_s=0$ where the T-VII model of  Sec.~\ref{Sec:T_VII_analytic_extension} is recovered (yellow dash-dotted curve), to the value given by Eq.~\eqref{eq:T_III_rho_s} for the T-III model of Sec.~\ref{Sec:T_III_analytic_extension} (green dash-dotted curve). To demonstrate the transition between the two solutions we use the parameter $\tilde{\rho}_s $ that corresponds to the free parameter $\rho_s$ normalized by the surface value of the energy density for the T-III extended analytic model~\eqref{eq:T_III_rho_s}.}
\label{fig:TVII_to_TIII_gtt_and_grr}
\end{figure*}
\begin{figure*}[ht!]
\centering
\includegraphics[width=0.49\linewidth]{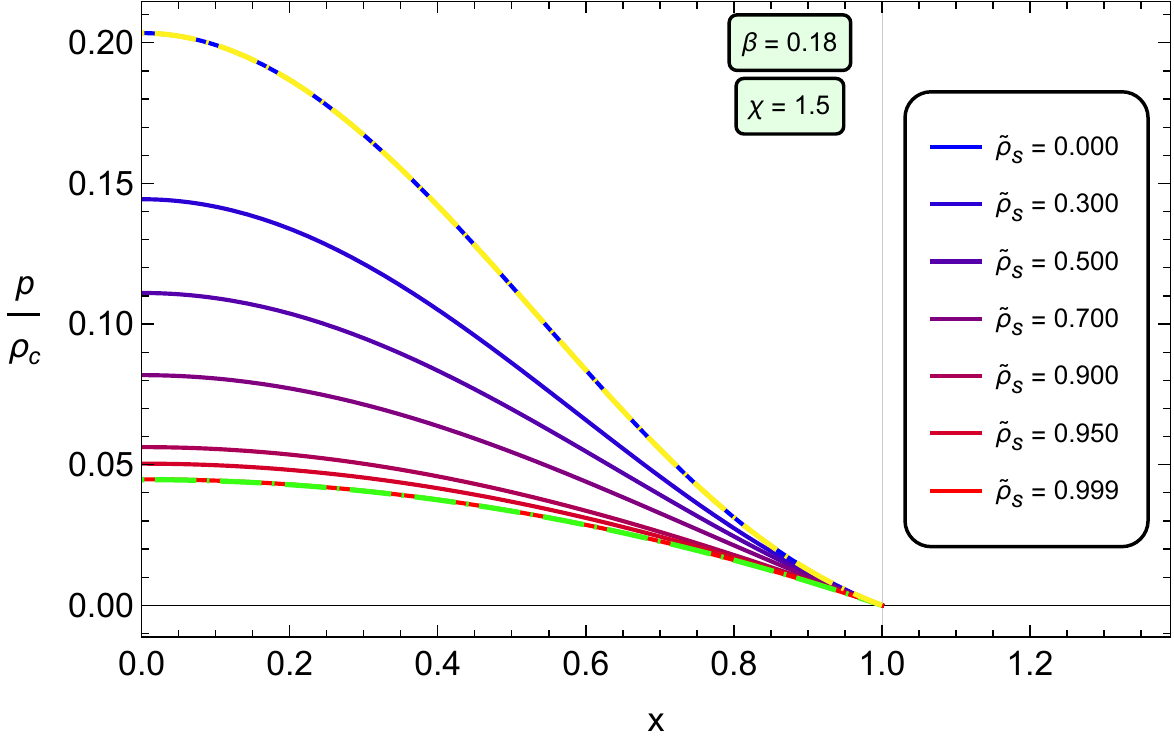}
\includegraphics[width=0.48\linewidth]{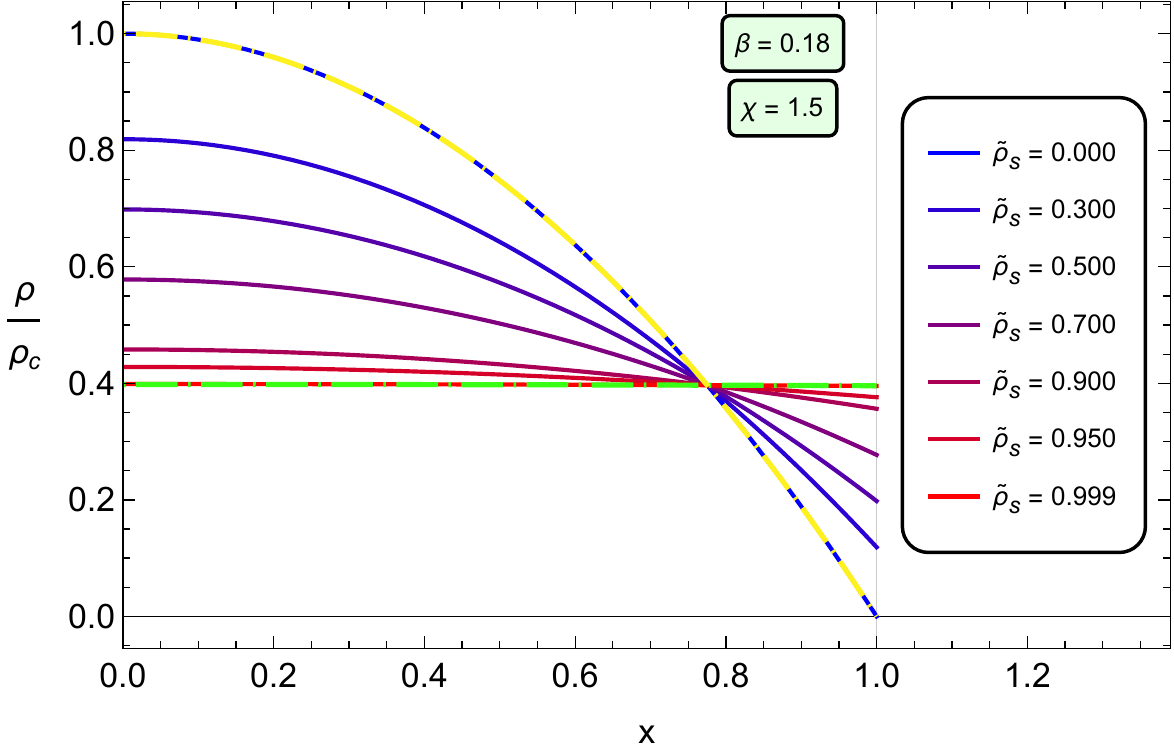}
\caption{The radial profiles for the normalized pressure~\eqref{eq:p(x)_generalized_TVII} and normalized energy density~\eqref{eq:rho(x)_generalized_TVII} of the parametrically deformed T-VII model, as the value of the surface energy density parameter ranges from $\rho_s=0$ where the T-VII model of  Sec.~\ref{Sec:T_VII_analytic_extension} is recovered (yellow dash-dotted curve), to the value given by Eq.~\eqref{eq:T_III_rho_s} for the T-III model of Sec.~\ref{Sec:T_III_analytic_extension} (green dash-dotted curve). For the normalization we have used in all cases the value of the central density for $\rho_s=0$. To demonstrate the transition between the two solutions we use the parameter $\tilde{\rho}_s $ that corresponds to the free parameter $\rho_s$ normalized by the surface value of the energy density for the T-III extended analytic model~\eqref{eq:T_III_rho_s}.}
\label{fig:TVII_to_TIII_p_and_rho}
\end{figure*}

This class of solutions can be thought of as a parametric deformation of the T-VII model in analogy to e.g.~\cite{Sotani:2018aiz}, where the T-VII energy-density is deformed by the introduction of a free parameter. However, with the approach we follow here, the extra parameter enters in all of the functions of the solution i.e. $g_{tt},g_{rr},\rho$ and $p$ allowing us to have an exact analytic deformation for the whole model. Upon fixing $\rho_s=0$ in the above exact analytic solution, one recovers the T-VII extended model~\eqref{grrTVII}-\eqref{pTVII} that has been studied in Sec.~\ref{Sec:T_VII_analytic_extension}. Furthermore, in the limit where $\rho_s$ approaches the surface value of the energy density given in Eq.~\eqref{eq:T_III_rho_s}, we recover asymptotically, the T-III extended solution of Sec.~\ref{Sec:T_III_analytic_extension}. Thus, the solution~\eqref{eq:gtt(x)_generalized_TVII}-\eqref{eq:p(x)_generalized_TVII} provides in a sense, a connection between, and a generalization of, our previous two exact analytic solutions. Notice that the parameter $\rho_s$ induces a dependence on $\chi$ for both metric functions~\eqref{eq:gtt(x)_generalized_TVII}-\eqref{eq:grr(x)_generalized_TVII}, a dependence that was absent for the extended T-VII model of Sec.~\ref{Sec:T_VII_analytic_extension} and it is exactly this property that allows us to interpolate between the T-III and T-VII models for any value of $\chi$ (including the GR value of $\chi=0$) by varying the value of the free parameter $\rho_s$. In Figs.~\ref{fig:TVII_to_TIII_gtt_and_grr} and~\ref{fig:TVII_to_TIII_p_and_rho} we illustrate the transition between the two solutions in terms of the parameter $\rho_s$. The central value of the energy density~\eqref{eq:rho(x)_generalized_TVII} is given in terms of $\chi, \beta$ and $\rho_s$ by
\begin{widetext}
\beq
\rhoc=\frac{40\beta (3\pi+\chi)-\rho_s\left(96\pi^2+68\pi\chi+12\chi^2\right)}{8(2\pi+\chi)(\chi+4\pi)}+\frac{\chi\sqrt{6\beta-\rho_s(8\pi+3\chi)}}{\sqrt{32}(2\pi+\chi)(4\pi+\chi)}\tan\left[\tan^{-1}(\phi_1)+\frac{\tanh^{-1}(\phi_2)-\tanh^{-1}(\phi_3(0))}{2}\right]\,.
\label{eq:rho(0)_generalized_TVII}
\eeq
\end{widetext}

\begin{figure*}[ht]
\centering
\includegraphics[width=0.49\linewidth]{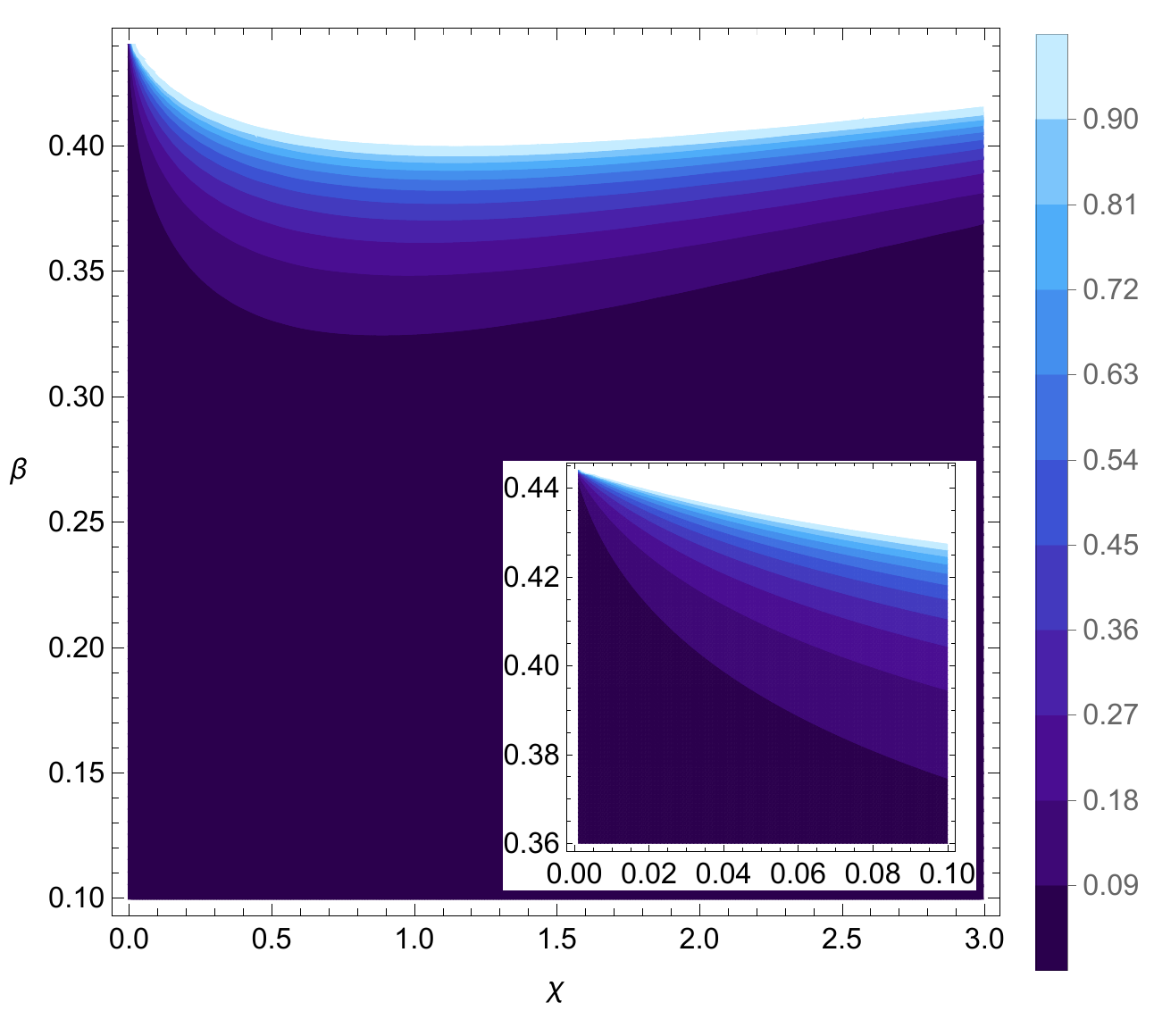}
\includegraphics[width=0.49\linewidth]{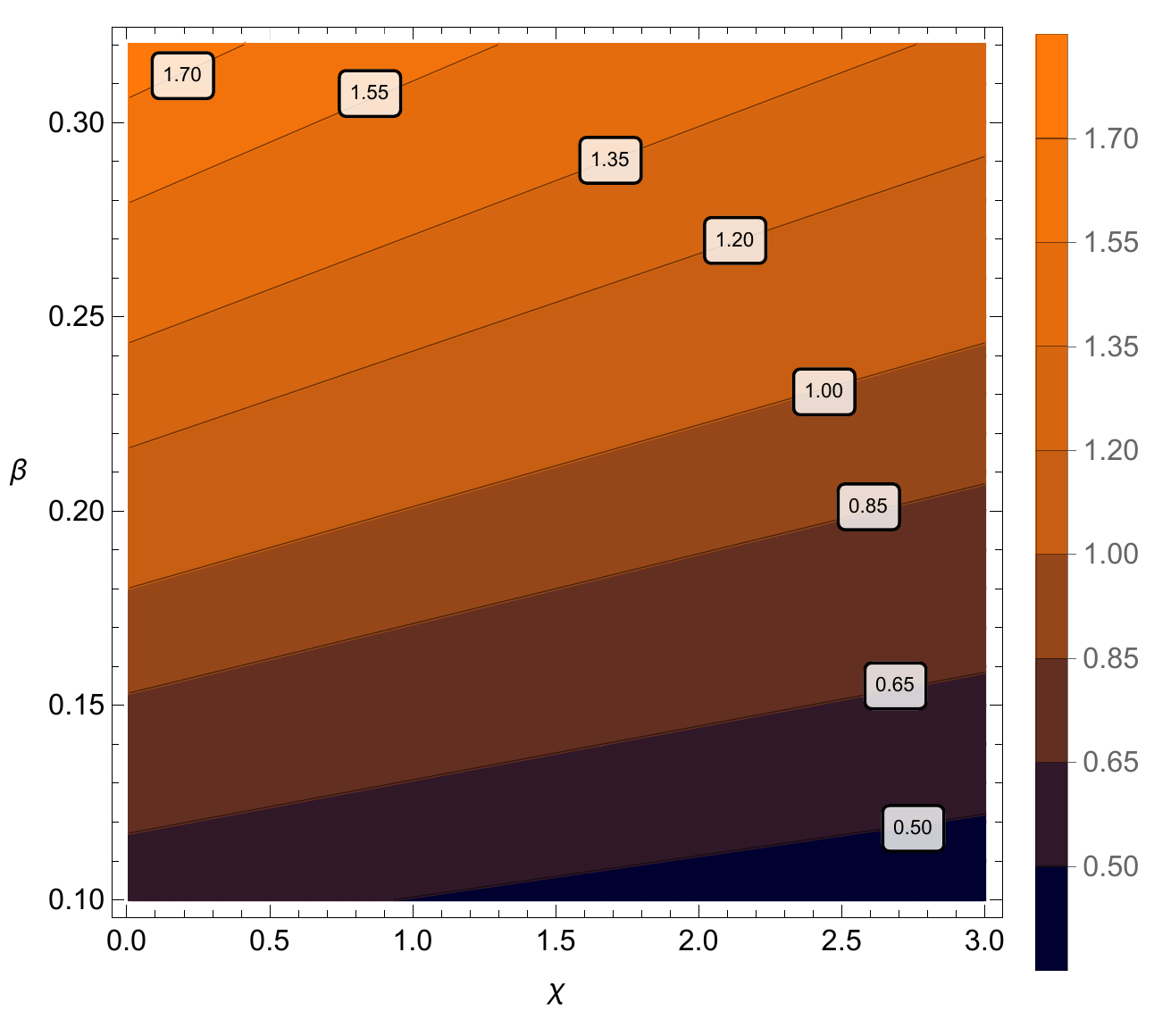}
\caption{The $\chi-\beta$ parametric plane, for the PD solution~\eqref{eq:grr(x)_generalized_TVII}-\eqref{eq:p(x)_generalized_TVII} when $\rho_s$ is given by Eq.~\eqref{eq:rho_s_analytic}. \textbf{Left panel:} the percentage of the maximum absolute relative error (MARE) in the compact radial range $x \in [0,1]$, between the energy density~\eqref{eq:rho(x)_generalized_TVII} and its central value $\rho_c$  This serves as an indication of the accuracy with which the PD solution approximates the CD configuration. The MARE is less than $1\%$ for any $\beta\leqslant0.4$. For small values of $\chi$, the approximation is very accurate even for larger values of compactness, as it can be seen in the inset. \textbf{Right panel:} the approximately uniform energy density normalized by the GR value of $\rho_c=3\beta/(4\pi)$ for $\beta=0.18$. Each line corresponds to a configuration with a fixed value of energy density and the whole parametric plane is covered continuously by such lines. The effect of $\chi$ is to suppress the density, while the effect of $\beta$ is to enhance it.}
\label{fig:fig13a}
\end{figure*}

Let us now discuss the adequacy of the PD solution in providing an analytic solution for the CD configuration in linear $f(\Ri,T)$ gravity. A necessary, albeit insufficient, condition for any solution to describe a uniform density configuration is that the central and surface values of the energy density must be equal. The latter condition amounts to solving the constraint equation~\eqref{eq:rho(0)_generalized_TVII} for $\rhoc=\rho_\text{s}$ thus obtaining the expression $\rho_\text{s}=\rho_\text{s}(\chi,\beta)$. This effectively amounts to removing the dependence of the solution on the extra free parameter $\rho_s$ and reducing it to the usual two-parametric family in terms of $\chi$ and $\beta$. Since the constraint equation cannot be solved analytically for arbitrary values of $\chi$ and $\beta$, one has to specify two out of the three free parameters ($\chi,\beta,\rho_s$) and then solve numerically Eq.~\eqref{eq:rho(0)_generalized_TVII} in order to determine the value of the third parameter to the desired accuracy. Subsequently, by substituting the obtained values for $\chi, \beta$ and $\rho_\text{s}$ into Eqs.~\eqref{eq:gtt(x)_generalized_TVII}-\eqref{eq:rho(x)_generalized_TVII}, we obtain a semianalytic solution for the CD configuration in linear $f(\Ri,T)$ gravity.

A second approach, stemming once again from a necessary but insufficient condition to be required from a CD configuration, is that all the derivatives of the energy density with respect to the radial coordinate should vanish at the surface of the star i.e.
\beq
\left.\frac{d^n\rho(x)}{dx^n}\right\vert_{x=1}=0\,,
\label{eq:drho_n_condition_for_CD}
\eeq
for any natural number $n$. Of particular importance here is the case of the first derivative, since then, Eq.~\eqref{eq:drho_n_condition_for_CD} corresponds to a constraint equation which is exactly solvable, yielding the following analytic expression for $\rho_s$ in terms of $\chi$ and $\beta$

\bea
\rho_s\left(\chi,\beta\right)&=&\frac{\sqrt{C_1-C_2}}{2\chi^2(4\pi+\chi)}+\frac{41\beta-20}{\chi}+\nonumber\\
&&5(2\beta-1)\left[\frac{8\pi}{\chi^2}+
\frac{1}{2(4\pi+\chi)}\right],
\label{eq:rho_s_analytic}
\eea
where
\beq
C_1 \equiv \left[(92\beta-45)\chi^2+8\pi \chi(61\beta-30)+320\pi^2(2\beta-1)\right]^2,
\eeq
and
\beq
C_2 \equiv 120\beta \chi^2(4\pi+\chi)(8\pi+3\chi)(2\beta-1)\,.
\eeq

Notice that in contrast to the first (semianalytic) approach that is valid for any $\chi \geqslant 0$, the second approach, even though it yields an exact analytic solution, cannot be considered for $\chi=0$. However, since we are here interested in obtaining the linear $f(\Ri,T)$ extension of the CD configuration, it is exactly the parametric regime for $\chi \neq 0$ that we wish to have a solution for.

It should be emphasized however, that in both of the aforementioned analytic approaches, the solutions thus obtained describe only approximately a CD state. The reason for this is that with the method we have developed here, the CD configuration is obtained via a parametric deformation of a non-CD solution. The boundary conditions are imposed on the energy density, and its first derivative at the surface of the configuration, and as a consequence, the intermediate values of the energy density i.e. in the regime $x\in(0,1)$ do not necessarily correspond to an exact CD configuration. Nevertheless, these solutions provide a very accurate approximation for the CD configuration with a maximum absolute relative error (MARE) of less than $0.1 \%$ for the majority of the $\chi-\beta$ parametric space, including the regime very close to $\chi=0$ for the second approach where $\rho_s$ is given by Eq.~\eqref{eq:rho_s_analytic} (see left panel of Fig.~\ref{fig:fig13a}). We have confirmed the accuracy of our analytic approximations by comparing them with the numerical solution for the exact CD configuration of the subsequent section.

In the right panel of Fig.~\ref{fig:fig13a},  on the $\chi-\bet$ parametric plane, we plot the central value of the normalized energy density of the PD solution~\eqref{eq:rho(x)_generalized_TVII} with $\rho_s$ given by Eq.~\eqref{eq:rho_s_analytic} (notice that according to the left panel of the same figure, in this parametric regime, the PD solution approximates the uniform density configuration with MARE less than $0.1\%$). Configurations with a fixed value of the energy density correspond to lines on the $\chi-\bet$ plane and this implies an approximately linear relationship between the two parameters for the CD configuration in linear $f(\Ri,T)$ gravity. This observation is in close agreement with the findings for the quasiconstant density configurations of Sec.~\ref{Sec:T_III_analytic_extension} (see Eqs.~\eqref{eq:T_III_rho_s}-\eqref{eq:rho0}). Furthermore, we see that the effect of $\chi$ is to suppress the energy density while $\bet$ has the opposite effect of causing an enhancement.

\section{Uniform density configuration in $f(\Ri,T)$ gravity}
\label{Sec:General_TVII_numer_sol}

\begin{figure*}[ht]
\centering
\includegraphics[width=0.5\linewidth]{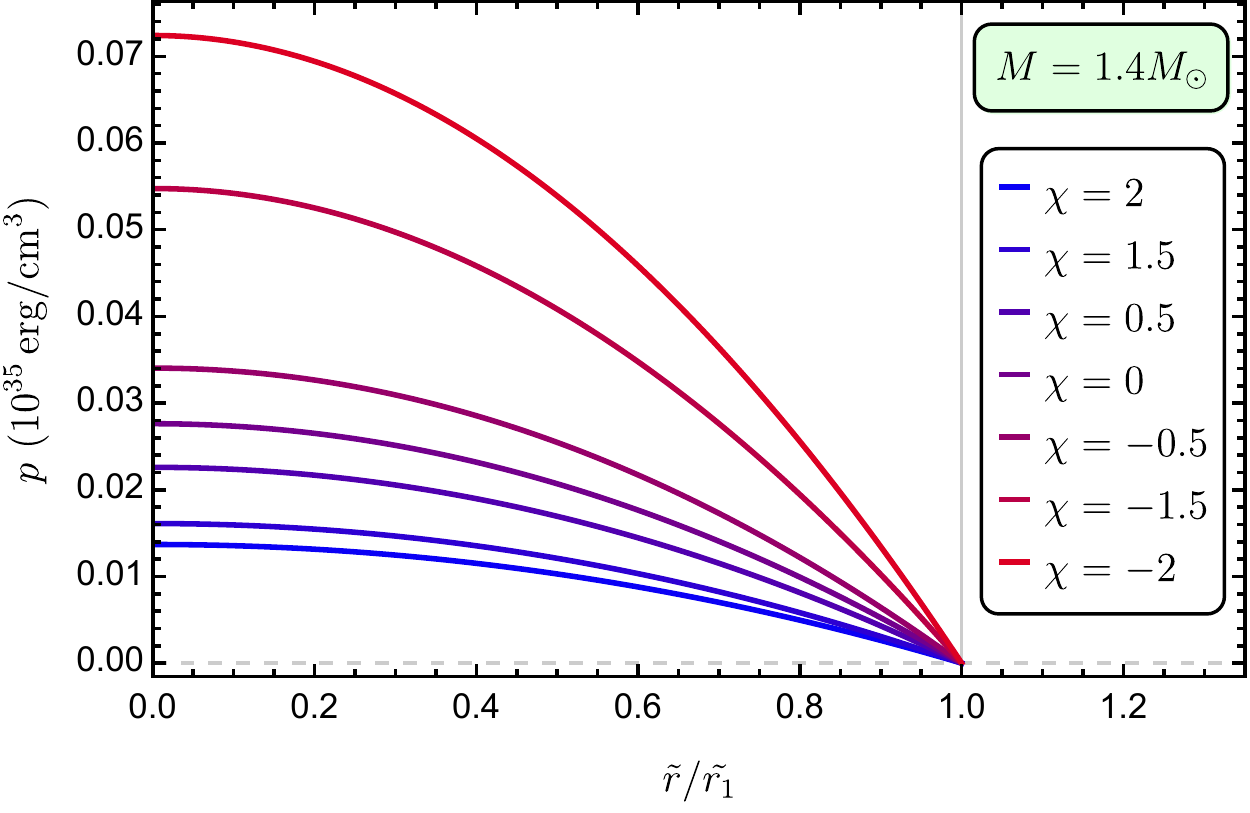}
\includegraphics[width=0.48\linewidth]{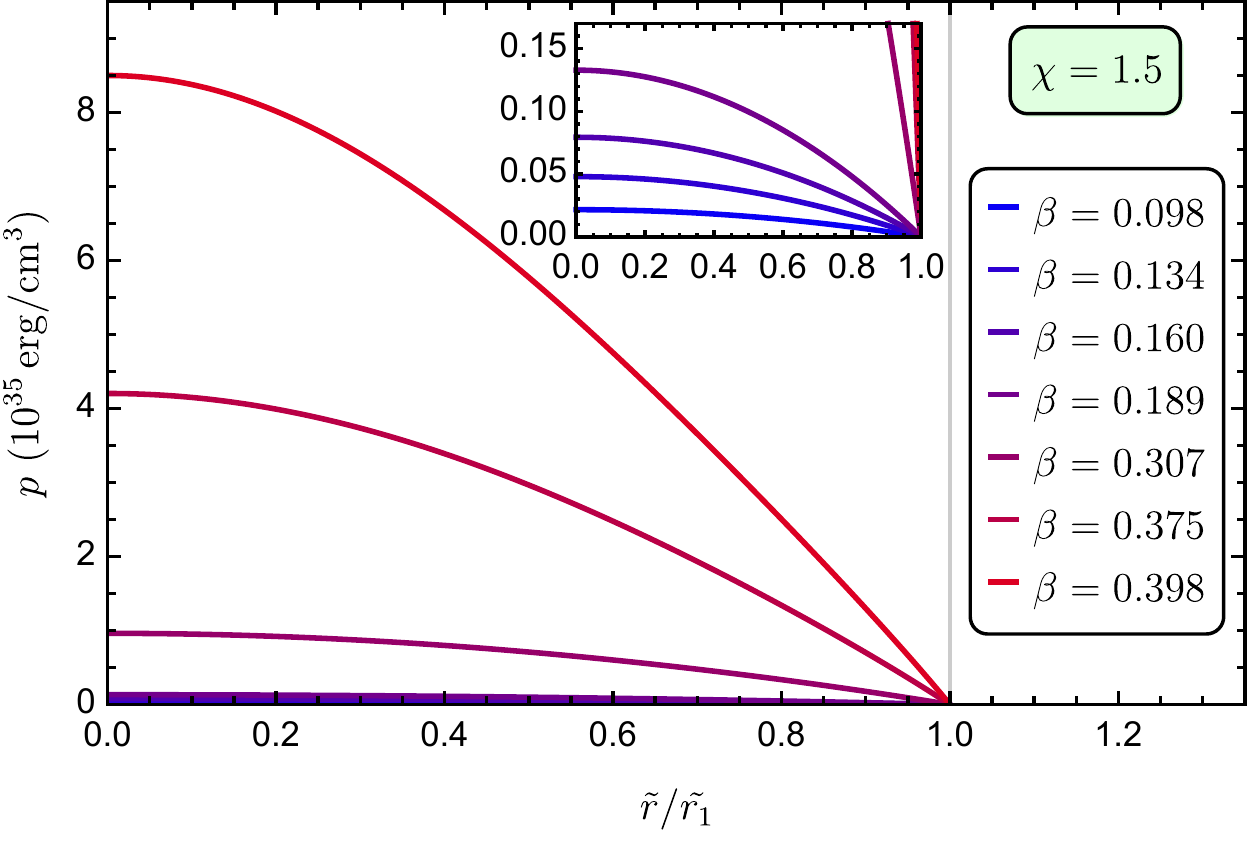}
\caption{Pressure, as a function of the dimensionless radial coordinate $\tilde{r}$ (in units of the dimensionless radius $\tilde{r}_1$), for a constant density configuration in linear $f(\Ri,T)$ gravity. \textbf{Left panel:} profiles for a configuration with fixed mass, $M=1.4~M_{\odot}$, for various values of $\chi$.\textbf{ Right panel:} profiles for a fixed value of $\chi=1.5$, and for various values of compactness.}
\label{fig:pres_CD_numer}
\end{figure*}

\begin{figure*}[ht]
\centering
\includegraphics[width=0.49\linewidth]{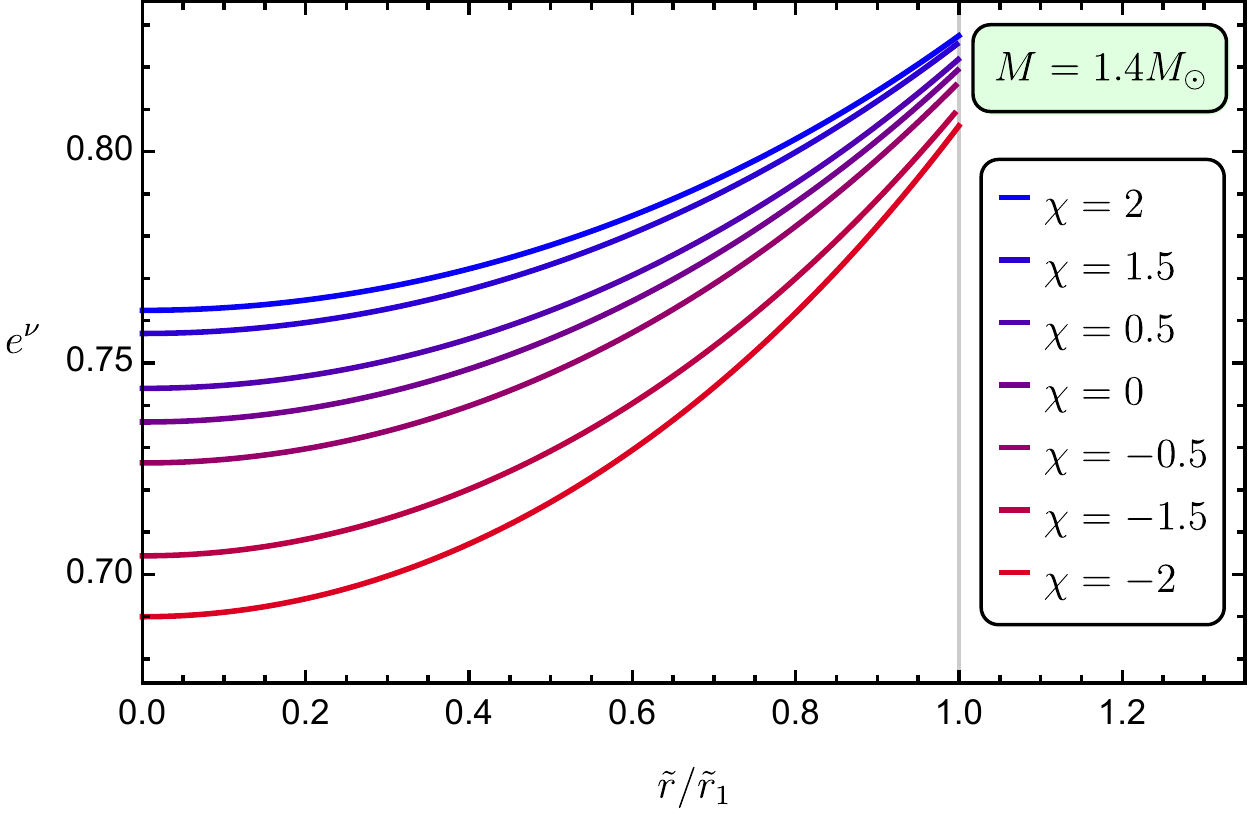}
\includegraphics[width=0.48\linewidth]{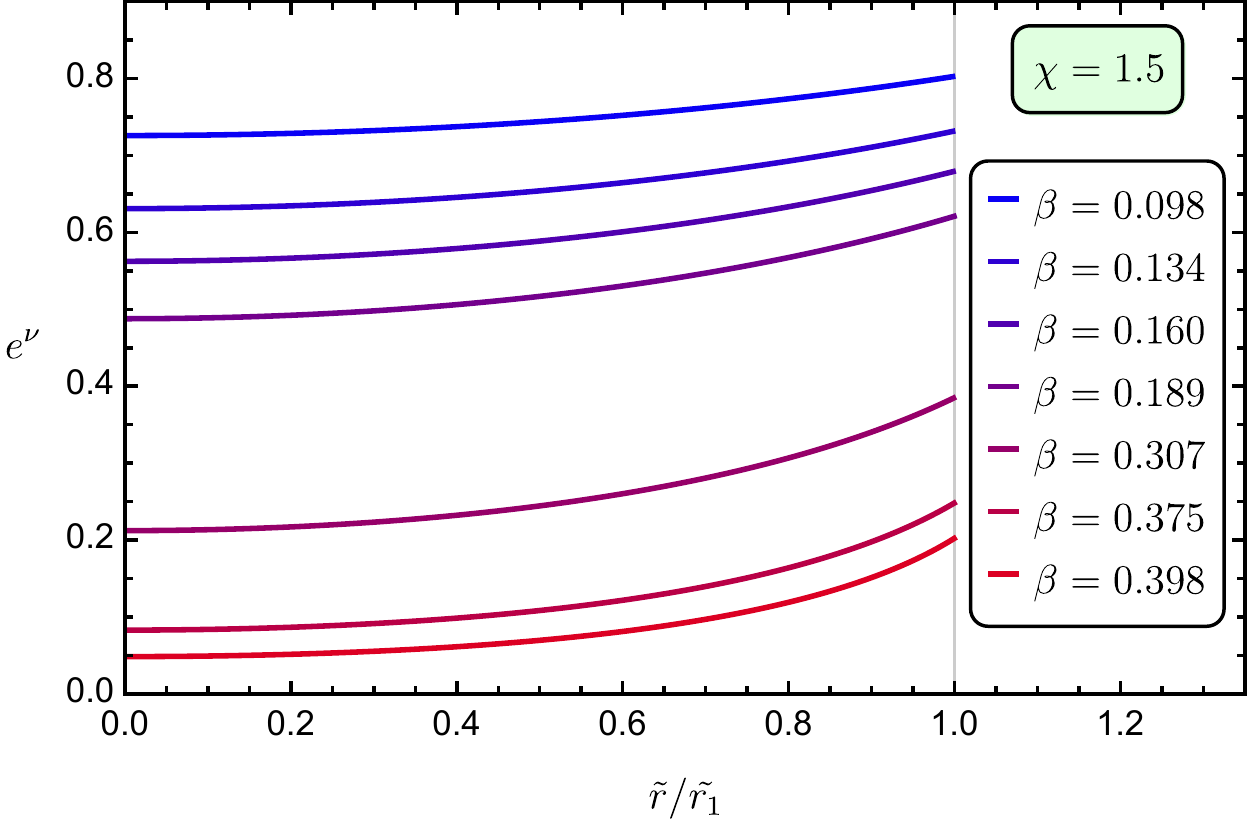}
\caption{Radial profile of the metric element $g_{tt}$ for a constant density configuration in linear $f(\Ri,T)$ gravity. \textbf{Left panel:} profiles for a star with a fixed mass, $M=1.4~M_{\odot}$, for various values of $\chi$. Although the mass is fixed, the radius of the star varies with $\chi$ and thus, so does its compactness. The latter fact is reflected at the surface value of $g_{tt}(1)$ that matches with  the corresponding Schwarzschild exterior $g_{tt}(1)=1-2\beta$ in all cases. \textbf{Right panel:} profiles for a fixed value of $\chi=1.5$ for various compactness.}
\label{fig:gtt_CD_numer}
\end{figure*}

\begin{figure*}[ht!]
\centering
\includegraphics[width=0.49\linewidth]{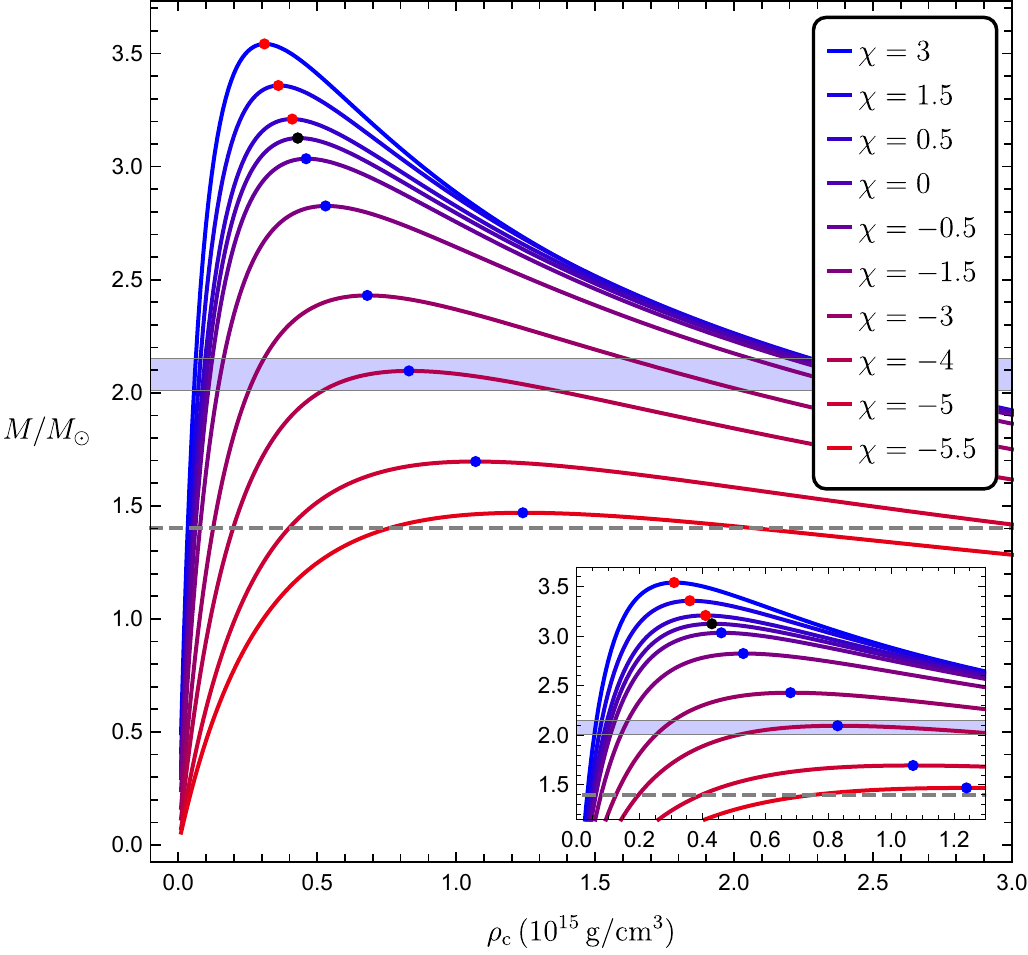}
\includegraphics[width=0.48\linewidth]{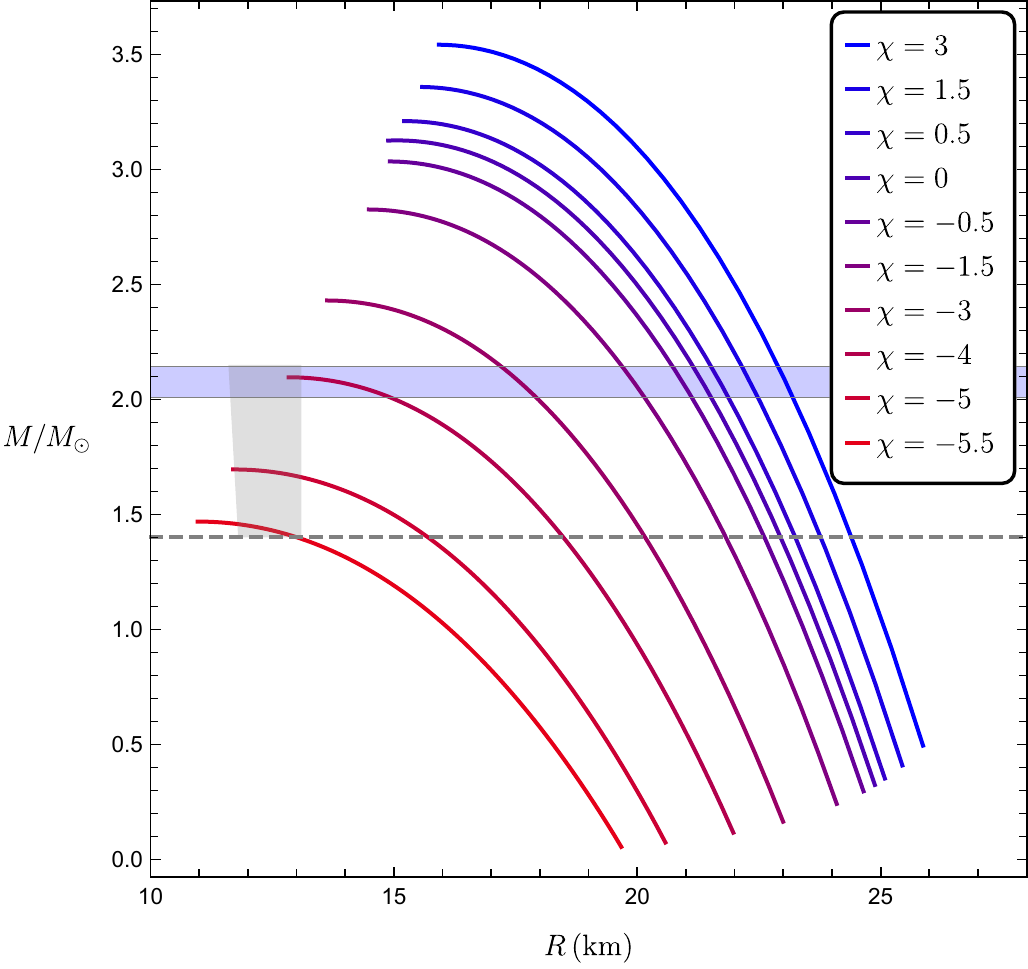}
\caption{\textbf{Left panel:} total mass $M$ (in solar-mass units), as a function of the central energy density $\rhoc$, for a CD configuration in linear $f(\Ri,T)$ gravity. In the inset we magnify the region between $\rhoc = (0.01 - 1.2)\times 10^{15}~\text{g}/\text{cm}^3$. The shaded horizontal blue stripe corresponds to the mass of $\text{PSR}~\text{J}0740+6620$, $M=2.08 \pm 0.07~M_{\odot}$. The dashed line indicates the value $M=1.4~M_{\odot}$. \textbf{Right panel:} total mass (in solar-mass units) as a function of the stellar radius for stable configurations. The shaded gray region corresponds to the estimates of radius for the canonical-mass, $M=1.4~M_{\odot}$, $R=12.45\pm 0.65~km$; and for a NS with mass $M=2.08~M_{\odot}$, $R=12.35\pm 0.75~km$~\cite{Miller:2021qha}. Note that the compactness $M/R$ increases with the mass.}
\label{fig:fig16a}
\end{figure*}

In contrast to GR where the constant energy density profile allows to obtain an exact analytical solution (T-III model), in linear $f(\Ri,T)$ gravity it is not possible to obtain the corresponding solution in analytical form. Therefore, in this context, to construct equilibrium configurations we must solve numerically the TOV system, Eqs.~\eqref{mass_Mod} and~\eqref{modTOV}. In order to do this, following~\cite{Schaffner-Bielich:2020psc}, we start by introducing dimensionless quantities in the form
\beq
p=\rhoc~\tilde{p},\quad \rho=\rhoc~\tilde{\rho}\,, 
\eeq
which we denote with a tilde. Here, $\rhoc$ corresponds to the central energy density (note that in natural units this $\rhoc$ has units of pressure). We will look for solutions to the TOV system in terms of $\rhoc$ only. Likewise, we introduce dimensionless quantities for $r$ and $m$ as follows
\beq\label{Eq:dimless_rm}
r=L\,\tilde{r}\,,\quad m=\mu\tilde{m}\,,
\eeq
where $L$ and $\mu$, with the corresponding units, are to be determined. In terms of the dimensionless quantities, the TOV system takes the form
\beq\label{mass_MOD2}
\frac{d\tilde{m}}{d\tilde{r}}=4\pi \tilde{r}^2 \tilde{\rho} + \frac{\chi}{2}\left(3\tilde{\rho}-\tilde{p}\right)\tilde{r}^2\,,
\eeq
\bea
\frac{d\tilde{p}}{d\tilde{r}}&=&-\frac{(\tilde{\rho}+\tilde{p})}{(1+\sigma)}\frac{\left[\tilde{m}+4\pi\,\tilde{p}\,\tilde{r}^3 - \frac{1}{2}\chi\,\tilde{r}^3\,(\tilde{\rho}-3\tilde{p})\right]}{\tilde{r}\left(\tilde{r}-2\tilde{m}\right)}+\nonumber\\
&&\frac{\sigma}{(1+\sigma)}\frac{d\tilde{\rho}}{d\tilde{r}}\,.
\label{modTOV2}
\eea

\noindent The dimensionful coefficients $L$ and $\mu$ satisfy the following scaling conditions
\beq\label{Eq:dimful_coeff}
L=\frac{1}{\sqrt{G~\rhoc}},\quad \mu=\frac{1}{\sqrt{G^3~\rhoc}}\,,
\eeq
\noindent where we have reintroduced the gravitational constant $G$, to show the explicit dependence. In this scenario, we have two free parameters in the system, namely, ($\rhoc$, $\chi$). Thus, for fixed values of the parameter $\chi$, we integrate the dimensionless TOV system [Eqs.~\eqref{mass_MOD2}-\eqref{modTOV2}] for the constant energy density profile, starting at the origin (or, rather at a very small cutoff value $\tilde{r}_0=10^{-7}$) where we impose the conditions: $\tilde{m}(\tilde{r}_0)=0$, and $\tilde{p}(\tilde{r}_0)=\tilde{\rho}_c$, with $\tilde{\rho}_c$ in this case denoting the dimensionless free parameter of the system which, appropriately rescaled, gives the physical energy density (see left panel of Fig.~\ref{fig:fig16a}). To determine the radius of the configuration, we look for the value $\tilde{r}=\tilde{r_\text{b}}$ where the pressure vanishes, $\tilde{p}(\tilde{r}_\text{b})=0$ [the dimensionful radius can be recovered from Eqs.~\eqref{Eq:dimless_rm} and~\eqref{Eq:dimful_coeff}]. Following this procedure, we can model different configurations by varying $\rhoc$, and then determine their corresponding radii. Once we know the stellar radius, and only then, we can determine the structure of the star in hydrostatic equilibrium and construct profiles for radial pressure, and mass-radius diagrams. Incidentally, the metric component $g_{tt}$ can be obtained from Eq.~\eqref{DivEMT_mod}, together with the boundary condition given by Eq.~\eqref{bc1}.

Following the method described above, here we present our results for a configuration with uniform density in the context of linear $f(\Ri,T)$ gravity. For the extended T-III and T-VII analytical models of the previous sections, following various constraints that yield physically plausible configurations, we have restricted $\chi$ to be non-negative. However, keeping in mind that the constraints on $\chi$ are model dependent, there is no reason \emph{a priori} to discard negative values of $\chi$ for our analysis in this section. For all the configurations considered here, the pressure is a monotonically decreasing function of the radial coordinate and all of the energy conditions are satisfied.

In Fig.~\ref{fig:pres_CD_numer} we show the radial pressure profiles, as a function of the dimensionless radial coordinate $\tilde{r}$ (in units of the radius $\tilde{r_1}$). In the left panel, we present the radial pressure, for a configuration with $M=1.4~M_{\odot}$, for different values of the parameter $\chi$. We observe that the pressure is maximum at the center, and it decreases monotonically with $\tilde{r}$. Similar to what we found for the quasiconstant density configuration in Sec.~\ref{Sec:T_III_analytic_extension}, we observe that the effect of positive $\chi$ is to quench the interior pressure, with respect to GR, which could be interpreted once again as a “tension." On the other hand, we observe that negative values of $\chi$ produce the opposing effect, namely, they enhance the interior pressure with respect to the GR value. In general, the overall effect of $\chi$ is to suppress the pressure. In the right panel of the same figure, we plot the radial pressure for a fixed value of $\chi$ with varying compactness. We also observe that as $\beta$ increases, the value of the central pressure also increases.

In Fig.~\ref{fig:gtt_CD_numer} we display the radial profile of the $g_{tt}$ metric component. In the left panel, we present the results for a configuration with fixed mass, $M=1.4~M_{\odot}$, for different values of $\chi$. We observe that the metric is continuous and well behaved in the interior of the star. In contrast to the pressure profile of Fig.~\ref{fig:pres_CD_numer}, we observe an inverted role of the parameter $\chi$, i.e., for positive values of $\chi$ the $g_{tt}$ function is enhanced, with respect to the GR profile ($\chi=0$), while for negative $\chi$-values we observe that $g_{tt}$ decreases, for a given $r$. Note that as we fix the total mass, different values of $\chi$ will give different radii, therefore different compactness. In other words, as we increase (or decrease) $\chi$ we must adjust the value of the central energy density (the free parameter of the solution), in order to have a total fixed mass. Thus, although we have $M$ fixed, we obtain different radii, and as a consequence, different compactness. The right panel of the same figure shows the profile for $g_{tt}$, now for a fixed value of $\chi$, for different compactness. Note that this profile shows the same general features of the $g_{tt}$ metric for the quasiconstant density solution found in Sec.~\ref{Sec:T_III_analytic_extension} (see Fig.~\ref{fig:T_III_gtt_betas_chis}). Moreover, we obtained excellent agreement between this numerical solution, and the analytical one obtained through the parametric deformation [Eq.~\eqref{eq:gtt(x)_generalized_TVII}], thus confirming our results.

The total mass, in units of the solar mass, for sequences of stellar models with uniform density and different values of the parameter $\chi$, is shown in the left panel of Fig.~\ref{fig:fig16a}, as a function of the central energy density $\rhoc$. The range of values of $\rhoc$ is typical for realistic NSs~\cite{Sotani:2018aiz,Capozziello:2015yza}. The inset shows a magnification of the region between $\rhoc\sim(0.01 - 1.2)\times 10^{15}~\text{g}/\text{cm}^3$. The horizontal shaded blue region indicates the values of mass of $\text{PSR}~\text{J}0740+6620$, $M=2.08\pm 0.07~M_{\odot}$, corresponding to the highest reliably determined mass of a NS; and the shaded gray region corresponds to the estimates of radius for the canonical-mass, $M=1.4~M_{\odot}$, $R=12.45\pm 0.65~km$; and for a NS with mass $M=2.08~M_{\odot}$, $R=12.35\pm 0.75~km$ reported by~\cite{Miller:2021qha}.

First of all, all of the curves exhibit a local maximum where $dM/d\rhoc=0$. This value of central density indicates the corresponding limit where the star becomes dynamically unstable. We observe that for increasing positive values of the parameter $\chi$, the maximum mass allowed for stability increases, while the limiting central energy density decreases. \emph{Thus, a positive value of the modified gravity parameter $\chi$ tends to destabilize the configurations}. On the other hand, we observe that negative values of $\chi$ quench the mass-radius profile giving a lower maximum mass, with respect to the GR case $(\chi=0)$, while the limiting central energy density, for stability, increases. Therefore, \emph{negative values of $\chi$ tend to stabilize the configurations}. It is noteworthy that for the values $\chi\in(-4,-5.5)$, the CD model, in linear $f(\Ri,T)$ gravity, predicts a maximum mass, and corresponding radii, which are within the accepted range of values of mass and radius for realistic NSs. For instance, for $\chi=-4$, the CD model predicts a maximum mass $M_{\text{max}}\sim 2.1~M_{\odot}$ with radius $R\sim 12.8~\text{km}$ (see the rhs panel), while a CD configuration with $\chi=-5.5$ and central density $\rhoc\simeq 0.77\times 10^{15}~\text{g}/\text{cm}^3$, predicts a mass $M\sim 1.40~M_{\odot}$ with a corresponding radius $R\simeq 13~\text{km}$. These values are in very good agreement with the typical values for NSs, namely, $M=1.4~M_{\odot}$ and $R\simeq 12.5~\text{km}$~\cite{Miller:2021qha}. For lower values of $\chi$ the maximum mass is below $1.4~M_{\odot}$, and for $\chi\leq-8.4$ we obtained pressure profiles which grow monotonically with $r$, which is unphysical. Thus, in contrast to the GR case where a CD model, besides being unrealistic, provides values of maximum mass and radius well above the accepted bounds for realistic NSs, a CD star in $f(\Ri,T)$ seems to provide a better model, for $\chi$ in the approximate range $\chi\in(-4,-6)$.

\section{Conclusions}
\label{Sec:Conclusions}

In this work, we have obtained analytic and numerical solutions corresponding to perfect-fluid spheres in hydrostatic equilibrium in the context of linear $f(\Ri,T)=\Ri+\chi T$ gravity, where $\Ri$ is the Ricci scalar, $T$ is the trace of the energy-momentum tensor of matter, and $\chi$ is a dimensionless free parameter of the theory. To obtain exact analytic solutions, we have followed the “mathematical approach" pioneered by Tolman~\cite{Tolman:1939jz} in the late 1930s where, in principle, solutions to the field equations may be obtained upon appropriately choosing the functional form for one of the functions in such a way that the field equations simplify. Once the complete system is solved and the metric functions $g_{tt},\,g_{rr}$ as well as the energy density $\rho$ and pressure $p$ are known, the physical plausibility of the obtained solutions is then investigated.

One of the advantages of considering the linear model of $f(\Ri,T)$ gravity, is that the field equations have the same structure as in GR with the only modification being encoded in the effective energy momentum tensor, which does not depend on the metric. Thus, various ans\"atze made for the metric functions in order to simplify the Einstein equations in GR may be equally well utilized in this theory. Let us point out that in GR, the T-III (T-III) and T-VII (T-VII) solutions can be obtained analytically either by assuming the $g_{rr}$ metric function, or the energy density profile. However the induced mixing between energy density and pressure with respect to the EMT part of the modified field equations, no longer allows us to obtain analytic solutions given an energy density profile, thus requiring numerical solutions.

Therefore, in order to obtain analytical solutions, following Tolman, we have considered as ans\"atze the functional forms for the $g_{rr}$ metric functions of the T-III and T-VII models and have obtained their extensions in linear $f(\Ri,T)$ gravity. Upon imposing the appropriate boundary conditions on our solutions, such that we have a smooth matching of the interior region with the Schwarzschild exterior, we obtained two-parametric families of solutions in terms of the parameter $\chi$ and the compactness $\beta$. All the solutions reduce to the corresponding GR models for $\chi=0$.

Subsequently, we have restricted the $\chi-\beta$ parametric space of the solutions by imposing some fundamental physical requirements, e.g., energy density and pressure radial profiles that are monotonically decreasing. Furthermore, we have derived the constraints imposed on the solutions by analyzing the speed of sound profiles and the compliance with respect to the energy conditions. Once we have identified the restricted parametric regime in each case we then proceeded with the investigation of the effects of the parameters $\chi$ and $\beta$ on our solutions.

As the analysis for our extended T-III model in Sec.~\ref{Sec:T_III_analytic_extension} has revealed, even though the original GR model corresponds to a uniform density configuration, its extension to linear $f(\Ri,T)$ gravity no longer shares the same feature and instead corresponds to a quasiconstant density configuration. The uniform density state is however recovered asymptotically for $\chi \to 0$ and $\chi \gg 1$. In the parametric regime $\chi \geqslant 0$ and $\bet \leqslant \beta_{\text{DEC}} \left( \chi \right)$, where $\beta_{\text{DEC}}$~\eqref{eq:beta_DEC} corresponds to the $\chi$-dependent value of compactness for which the DEC is violated; the quasiconstant density configuration exhibits monotonically decreasing profiles for both the energy density and pressure, and satisfies the null, weak, strong and dominant energy conditions. However, we have found that causality is violated in this model as the speed of sound is superluminal. Nevertheless, the model provides an improvement over the uniform density configuration of GR where the speed of sound is undefined. 

Furthermore, in the extended T-III model we have found that for any value of $\chi$, the energy density and pressure simultaneously diverge at the generalized Buchdahl bound $\beta_{\text{B}}\left(\chi\right)$~\eqref{eq:Buch_comp_Sch}. Quite interestingly, both $\beta_{\text{DEC}}$ and $\beta_{\text{B}}$ asymptote to the black-hole (BH) compactness limit for sufficiently large values of $\chi$ (see Fig.~\ref{fig:fig01}) and as a consequence, “BH-mimicker" states are obtained in this parametric regime, which do not exhibit divergence of physical quantities nor violate any of the energy conditions.

In the appendix~\ref{Sec:Appx_T_III_Beyond_Buchdahl}, we have also considered the quasiconstant density solution beyond the Buchdahl bound, up to the BH compactness limit, i.e. for $\beta\in(\beta_{\text{B}}\left(\chi\right),1/2]$, as a first investigation of the relevance of our extended T-III solution in connection with gravastars. We found that even though the linear-$f(\Ri,T)$ extension shares many similarities with the GR version of the model, such as a single zero of $g_{tt}$ which occurs at the same radius where the pressure diverges and shifts to the surface of the configuration for $\beta \to 1/2$, leaving behind a constant negative pressure interior; it is nevertheless plagued by the emergence of shells of negative energy density during the transition from $\beta_{\text{B}}\left(\chi\right)$ to $\beta \to 1/2$. The final state, however, is characterized by a strictly positive, uniform energy density $\rho \left(\chi,\beta\right)$ with pressure given by $p=-\rho$.

For our more realistic T-VII extended model studied in Sec.~\ref{Sec:T_VII_analytic_extension}, we have obtained three bounds for compactness which are universal for any value of $\chi$ and are thus invariant from the original model of GR. In particular, we have found that the Buchdahl compactness limit is given by $\beta_{\text{B}} \simeq 0.386195$, the dominant energy condition violation limit corresponds to compactness $\beta_{\text{DEC}} \simeq 0.335119$, while causality is violated for even less-compact configurations at about $\beta_{\text{CV}} \simeq 0.269791$. To ensure that the energy density and pressure are everywhere non-negative and monotonically decreasing functions of the radial coordinate, the MGT parameter $\chi$ must be non-negative for this model. In summary, we have physically acceptable configurations in the parametric regime $\chi \geqslant 0$ and $\beta \leqslant \bet_{\text{CV}}$.

In Sec.~\ref{Sec:rho_s_extended_T_VII}, we have constructed a parametric deformation (PD) for the extended T-VII solution (Sec.~\ref{Sec:T_VII_analytic_extension}), by allowing the surface value of the energy density $\rho_s$ to be a free parameter of the model during the implementation of the boundary conditions. When $\rho_s=0$, the PD solution reduces exactly to the model of Sec.~\ref{Sec:T_VII_analytic_extension} while for $\rho_s$ approaching the surface value of the energy density of the quasiconstant density configuration~\eqref{eq:T_III_rho_s}, the PD solution asymptotes to the extended T-III model. This correspondence between the T-III and T-VII models via the PD solution extends to all values of $\chi \geqslant 0$ and thus holds even for the original models of GR. Furthermore, in terms of the PD solution, we have shown that it is possible to obtain highly-accurate (with maximum absolute relative error less than $0.1 \%$ for the majority of the parametric space) analytic approximations for the uniform density configuration in linear $f(\Ri,T)$, an exact analytic solution to which is still missing from the literature.

Besides the analytical extensions of the Tolman III and VII solutions, in Sec.~\ref{Sec:General_TVII_numer_sol} we considered a constant density configuration. For this ansatz, it is not possible to find analytical solutions to the modified field equations, therefore we solved numerically the TOV system, thus obtaining relations for mass, pressure and $g_{tt}$. We also provided mass-radius relations for this particular configuration. In this general approach, we considered positive and negative values of the parameter $\chi$. In general, we found that positive (negative) values of $\chi$, increase (decrease) the maximum mass of stable configurations, while the corresponding limiting central energy density $\rhoc$, for the onset of dynamical instability, decreases (increases). Thus, the main conclusion is that positive (negative) values of the modified gravity parameter $\chi$ destabilize (stabilize) the configurations. It is worthwhile to remark that these results are, in principle, an indication of the dynamical stability of the configuration. However, a thorough investigation of the stability would require a study of the time-dependent radial perturbations as developed by Chandrasekhar in GR~\cite{Chandrasekhar:1964ApJ}, and extended to linear $f(\Ri,T)$ by~\cite{Pretel:2021kgl}. Also, we found that positive values of $\chi$ predict masses and radii which are not in agreement with the current constraints for realistic NSs. However, for negative values of $\chi$, in the approximate range $(-4,-6)$, the predicted masses and radii lie within the limits of the current bounds for NSs. The case of a T-VII quadratic energy density profile is left for future work.

Finally, let us here summarize the effect of the parameters $\chi$ and $\beta$ on the pressure, and density profiles of our solutions. We found that in all cases, the compactness $\beta$ causes an enhancement of both, the pressure and the energy density, which is the anticipated effect. On the other hand, in all cases, we found that the effect of $\chi$ is opposite to the one of $\beta$, since it causes a suppression, except for the pressure of the T-VII model where we found that, for compactness $\bet \lessapprox 0.257$ and for $\mathcal{O}\left(1\right)$ values of $\chi$ (see Fig.~\ref{fig:p_0_TVII}), the effect of the latter is to cause an enhancement.

\begin{acknowledgments}
The authors acknowledge the support of the Research Centre for Theoretical Physics and Astrophysics of the Institute of Physics at the Silesian University in Opava.
\end{acknowledgments}

\appendix

\section{Beyond the generalized Buchdahl limit: The ultracompact Schwarzschild star in linear $f(\Ri,T)$ gravity}
\label{Sec:Appx_T_III_Beyond_Buchdahl}
Here, we consider the implications of going beyond the generalized Buchdahl limit~\eqref{eq:Buch_comp_Sch} for the quasiconstant density configuration model~\eqref{gttFSs}-\eqref{pFSs} in linear $f(\Ri,T)$ gravity. The motivation to consider such ultracompact configurations stems from the already existing analyses performed in the context of GR for the Schwarzschild's interior solution with constant density, or Schwarzschild star, where by introducing a pressure anisotropy, Mazur and Mottola~\cite{Mazur:2015kia} (MM-15) showed that the solution allows compactness beyond the Buchdahl limit. Moreover, in the BH limit, $\beta\to 1/2$, the Schwarzschild star becomes essentially the gravastar, hypothesized 20 years ago as a possible final state of gravitational collapse~\cite{Mazur:2001fv,Mazur:2004fk}. The gravastar of~\cite{Mazur:2015kia} may be considered as the universal limit of the original gravastar where a thin layer of ultrarelativistic fluid $p=\rho$ was interposed, near the would-be-horizon, to match the interior de Sitter with the exterior Schwarzschild spacetime. The universality may be seen in the sense that when the thickness of the intermediate layer goes to zero, the resulting configuration is independent of any equation of state one may introduce in the intervening layer. Thus, in the MM-15 model, the matching of the interior and exterior spacetimes occurs exactly at the \emph{null} surface $R=2M$ (up to possible Planckian corrections), endowed with an anisotropic stress tensor which produces a surface tension. The MM-15 has maximal compactness $\beta=1/2$, and similar to the model proposed in~\cite{Mazur:2001fv,Mazur:2004fk}, it is a cold condensate state with zero entropy, and no information paradox.

A crucial element in the construction of gravastars in the way envisaged by~\cite{Mazur:2015kia}, is that the pole in the pressure occurs at the same radius where the metric function $g_{tt}=0$, for any value of compactness beyond Buchdahl. Furthermore, in the BH compactness limit, the pole reaches the surface of the configuration $R=R_\text{S}=2M$, producing a $\delta$-function associated with an anisotropic stress-tensor contribution, leaving behind a constant negative pressure interior. Motivated by these results in the context of GR, we proceed here to examine the regime beyond the Buchdahl bound in our quasiconstant density model in linear $f(\Ri,T)$ gravity. It is worthwhile to remark that the following analysis is not supposed to be exhaustive. The main objective here is to explore some of the consequences when we go beyond the Buchdahl limit in our quasiconstant density model, and draw some general conclusions, inspired by the results in~\cite{Mazur:2015kia}.

\begin{figure}[ht!]
\includegraphics[width=\linewidth]{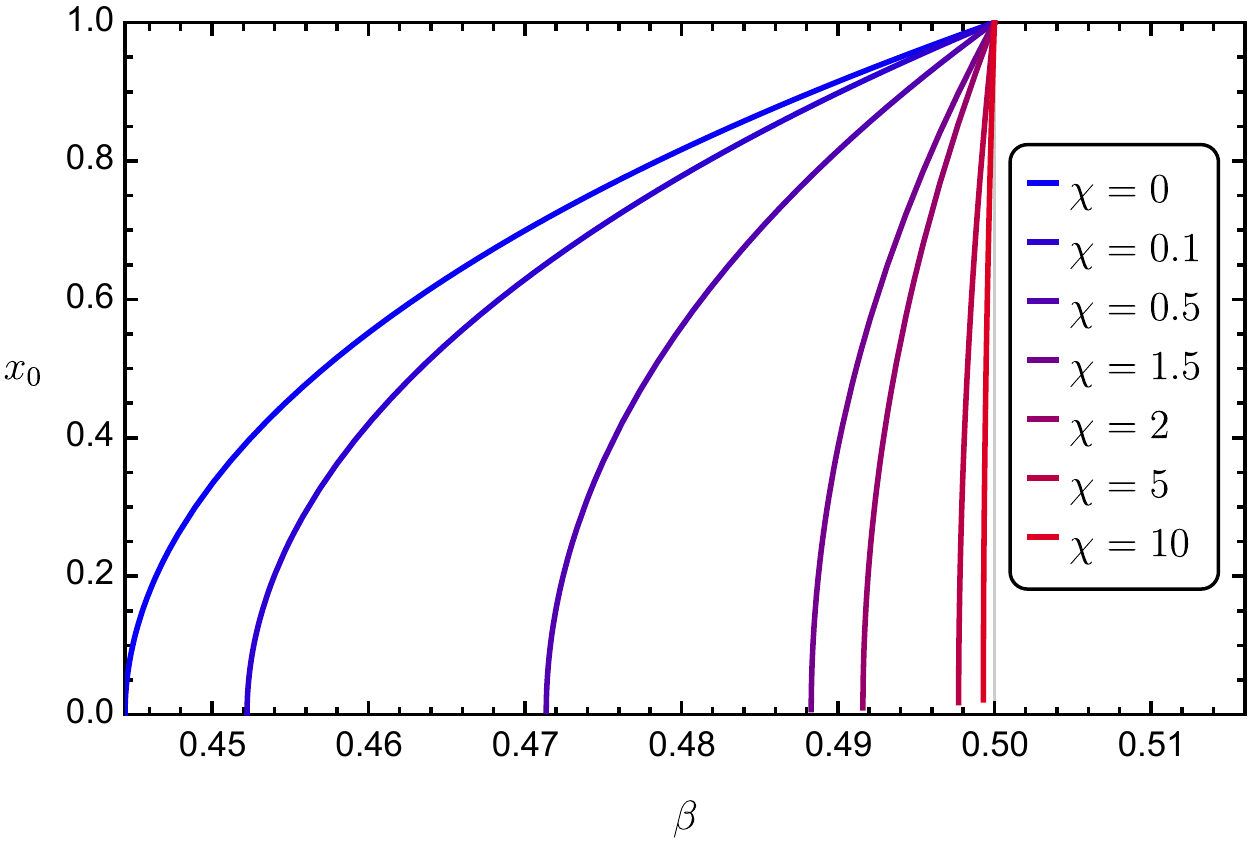}
\caption{The zero of the denominator $D$~\eqref{eq:D}, as a function of the compactness, for various values of $\chi$. Notice that, at the Buchdahl limit, $x_0$ appears first at $x=0$ and then, as $\beta\to 1/2$, it moves outwards up to the surface where $x=x_0=1$. The effect of $\chi$ is to shift the pole emergence for larger values of $\beta$, in agreement with the generalized Buchdahl limit~\eqref{eq:Buch_comp_Sch}.}
\label{fig:fig17}
\end{figure}

\begin{figure*}[ht!]
\centering
\includegraphics[width=0.49\linewidth]{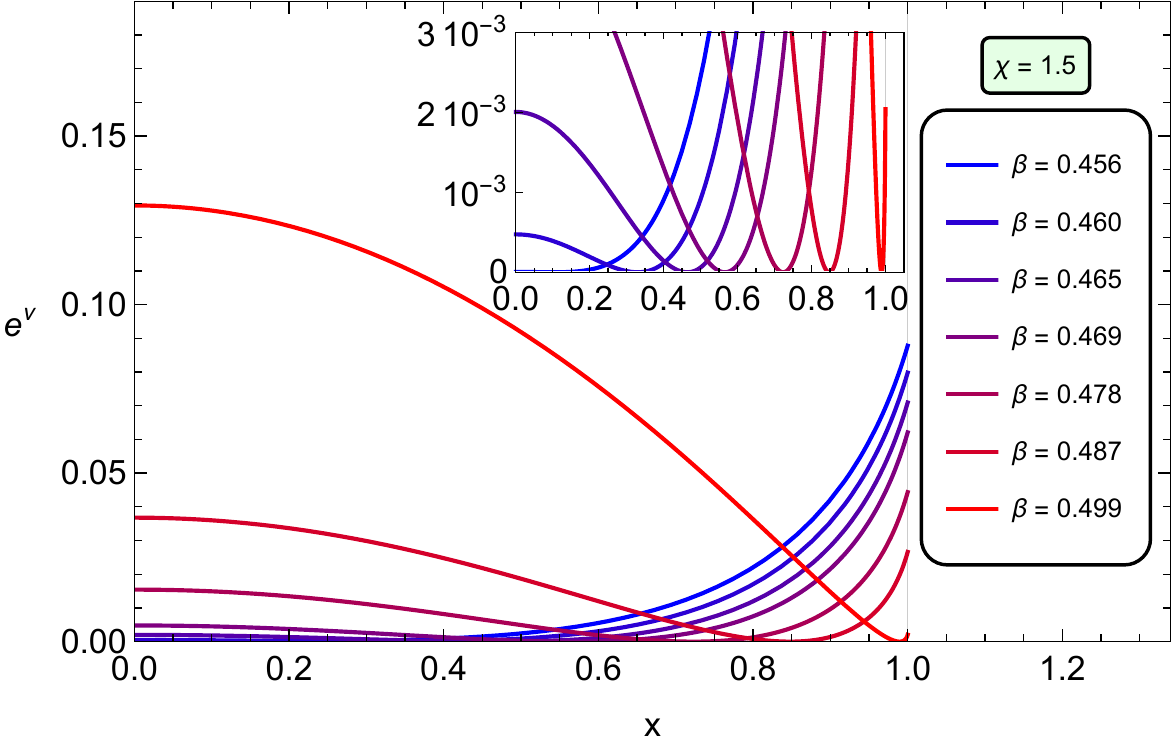}
\includegraphics[width=0.497\linewidth]{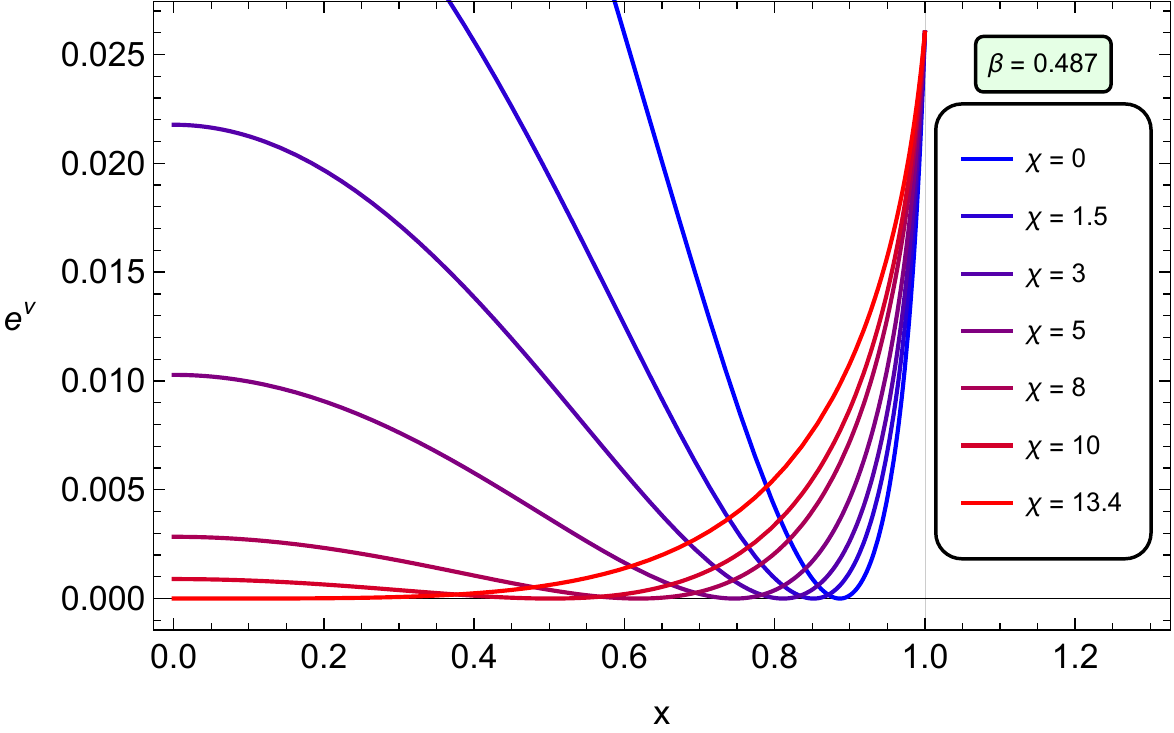}
\caption{The radial profile of the $g_{tt}(x)$ metric function~\eqref{gttFSs} for configurations beyond the generalized Buchdahl limit~\eqref{eq:Buch_comp_Sch}. \textbf{Left panel:} for an indicative fixed value of $\chi=1.5$ and for various values of compactness between the generalized Buchdahl limit $\bet_{\text{B}}(1.5)\simeq 0.455661$ and the black-hole compactness limit $\beta_{BH}=1/2$. \textbf{Right panel:} the effect of $\chi$ for configurations with compactness approaching the BH compactness limit.}
\label{fig:gtt_Τ_ΙΙΙ_beyond_Buchdahl}
\end{figure*}
A first inspection of Eqs.~\eqref{gttFSs}-\eqref{pFSs} immediately reveals some interesting features. For instance, we observe that the solution is regular everywhere in the interior, except at some value $x=x_0$ where the term
\beq
D\equiv 3\sqrt{1-2\beta}-\sqrt{1-2\beta x^2}+ \frac{3\chi}{4\pi}\sqrt{1-2\beta}\,,
\label{eq:D}
\eeq
vanishes in the range $x\in[0,1]$, for $\beta\geqslant \beta_{\text{B}}$~\eqref{eq:Buch_comp_Sch}. Note that $D$ appears in the denominator of both $\rho$ and $p$ as well as in the numerator of $g_{tt}$. As a consequence, remarkably, and in strict analogy with the situation in GR, the pressure, and in this case also the energy density, vanish at the same radius $\left(x=x_0\right)$ where also $g_{tt}\left(x_0\right)=0$. At any other radius $x \neq x_0$, the metric function $g_{tt}$ is non-negative and regular everywhere in the interior [see Fig.~\ref{fig:T_III_gtt_betas_chis}]. The location of the single root of the equation $D=0$ in $x \in [0,1]$ is given by $x=x_{0}$ where
\beq
x_0=\sqrt{1-\frac{(1-2\beta)\left(9\chi^2+72\pi\chi+128\pi^2\right)}{32\pi^2\beta}}\,,
\label{eq:T_III_root_of_gtt}
\eeq
which is a non-negative real number for $\beta \geqslant \beta_{\text{B}}$. Note that for $\chi=0$,~\eqref{eq:T_III_root_of_gtt} reduces to the GR value [see Eq. (2.22) in~\cite{Mazur:2015kia}]. In analogy to the GR case, at the generalized Buchdahl limit~\eqref{eq:Buch_comp_Sch}, we have $x_0=0$, and the single zero of $g_{tt}$ occurs at the center of the configuration (see Fig.~\ref{fig:gtt_Τ_ΙΙΙ_beyond_Buchdahl}).

\begin{figure*}[ht!]
\centering
\includegraphics[width=0.49\linewidth]{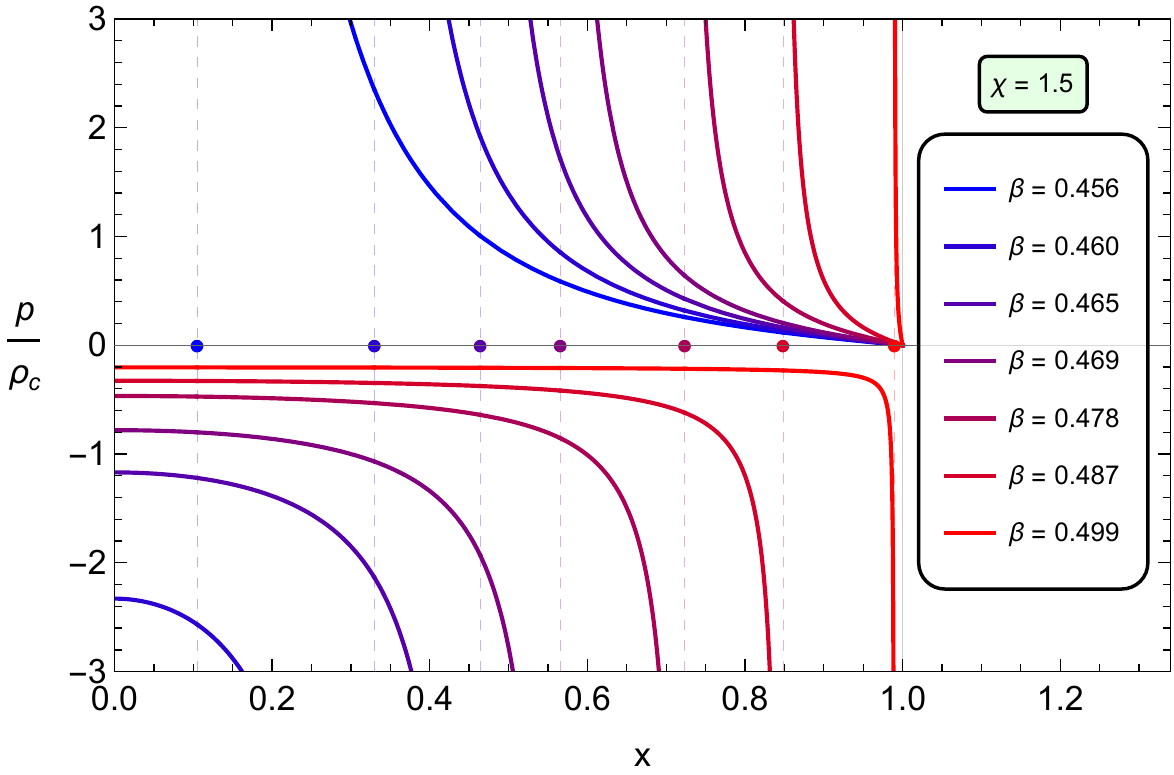}
\includegraphics[width=0.497\linewidth]{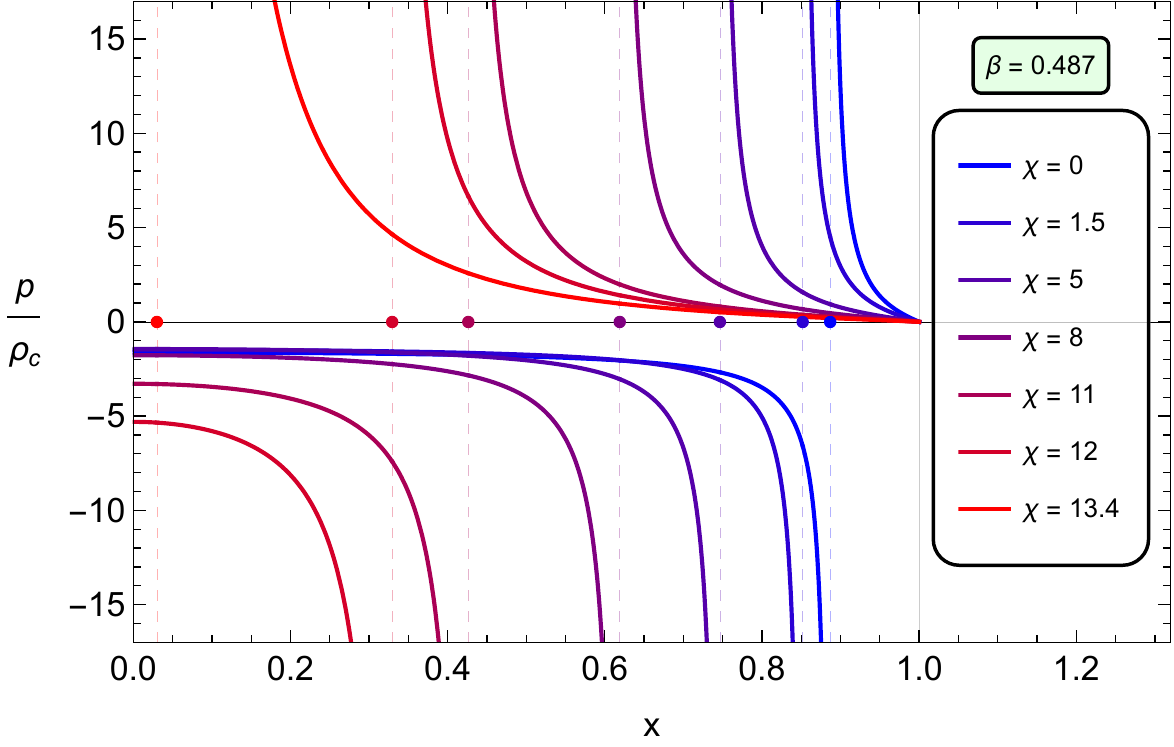}
\caption{The radial profile of the normalized pressure~\eqref{pFSs} for configurations beyond the generalized Buchdahl limit~\eqref{eq:Buch_comp_Sch}. The dots correspond to the locations of the poles~\eqref{eq:T_III_root_of_gtt} where both the pressure and the energy density diverge. \textbf{Left panel:} for an indicative fixed value of $\chi=1.5$ and for various values of compactness between the generalized Buchdahl limit $\bet_{\text{B}}(1.5)\simeq 0.455661$ and the black-hole compactness limit $\beta_{BH}=1/2$. For the normalization we have used the absolute value of the central density $\lvert\rho_c\rvert$ for $\bet=0.456$. \textbf{Right panel:} the effect of $\chi$ for configurations with compactness approaching the BH compactness limit. For the normalization we have used $\rho_c$ for $\chi=0$.}
\label{fig:p_Τ_ΙΙΙ_beyond_Buchdahl}
\end{figure*}

If we consider the solution in the regime where $\beta_{\text{B}}<\beta<1/2$, we observe that $x_{0}$ moves out from the origin to values $0<x_0<1$ (see Fig.~\ref{fig:fig17}). Thus, Eq.~\eqref{pFSs} shows that a region $0<x<x_{0}$ emerges naturally where $D<0$, therefore $p<0$, while the $g_{tt}$ metric component remains positive [see Figs.~\ref{fig:gtt_Τ_ΙΙΙ_beyond_Buchdahl},~\ref{fig:p_Τ_ΙΙΙ_beyond_Buchdahl}] . Meanwhile, the pressure remains positive in the region $x_{0}<x<1$. This behavior remarkably mirrors the one found for the Schwarzschild star in GR~\cite{Mazur:2015kia}.

As $\beta\to 1/2$, $x_0$ moves radially outward up to the black-hole compactness limit where $x_0 \to 1$ and the root of $g_{tt}$ reaches the surface of the configuration. In Fig.~\ref{fig:gtt_Τ_ΙΙΙ_beyond_Buchdahl} we plot the $g_{tt}$ metric function for various values of the parameter $\chi$ and for various compactness values beyond the generalized Buchdahl limit $\beta_{\text{B}}(\chi)$~\eqref{eq:Buch_comp_Sch}. We see that the transition from the Buchdahl limit to the BH compactness limit is qualitatively the same as in the GR variant of the model. The effect of modified gravity, on a configuration of a given compactness, is to reduce the value of $x_0$.

\begin{figure*}[ht!]
\centering
\includegraphics[width=0.495\linewidth]{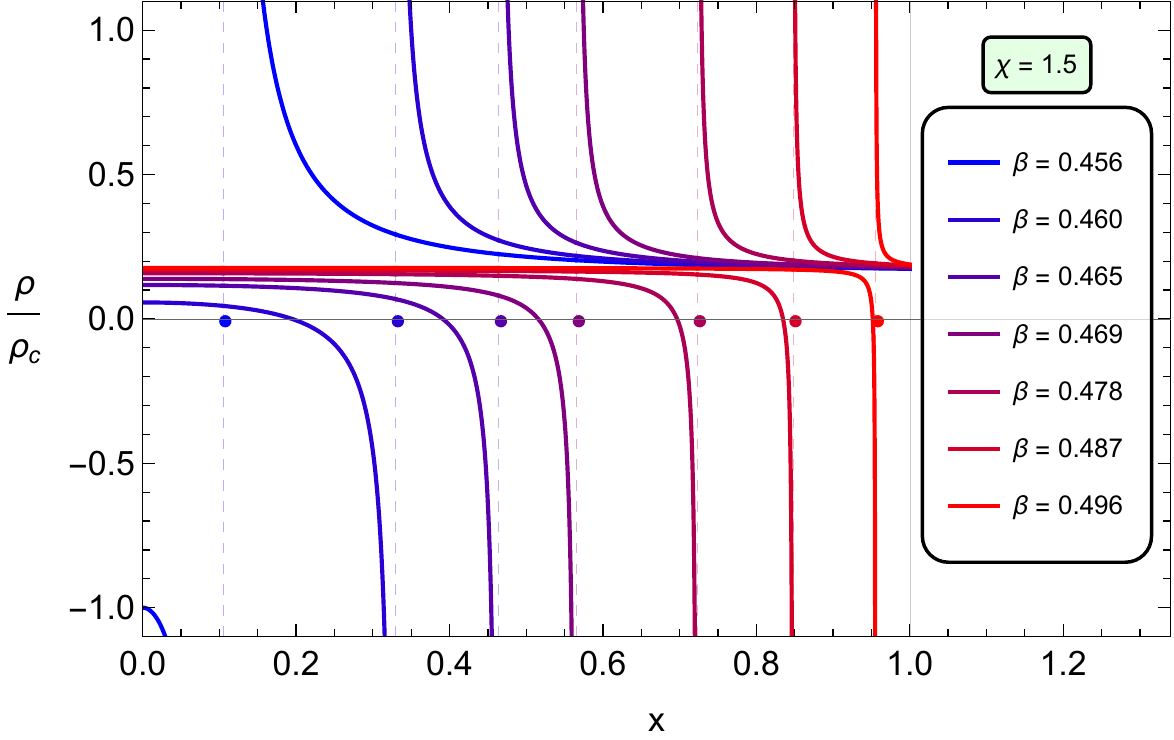}
\includegraphics[width=0.495\linewidth]{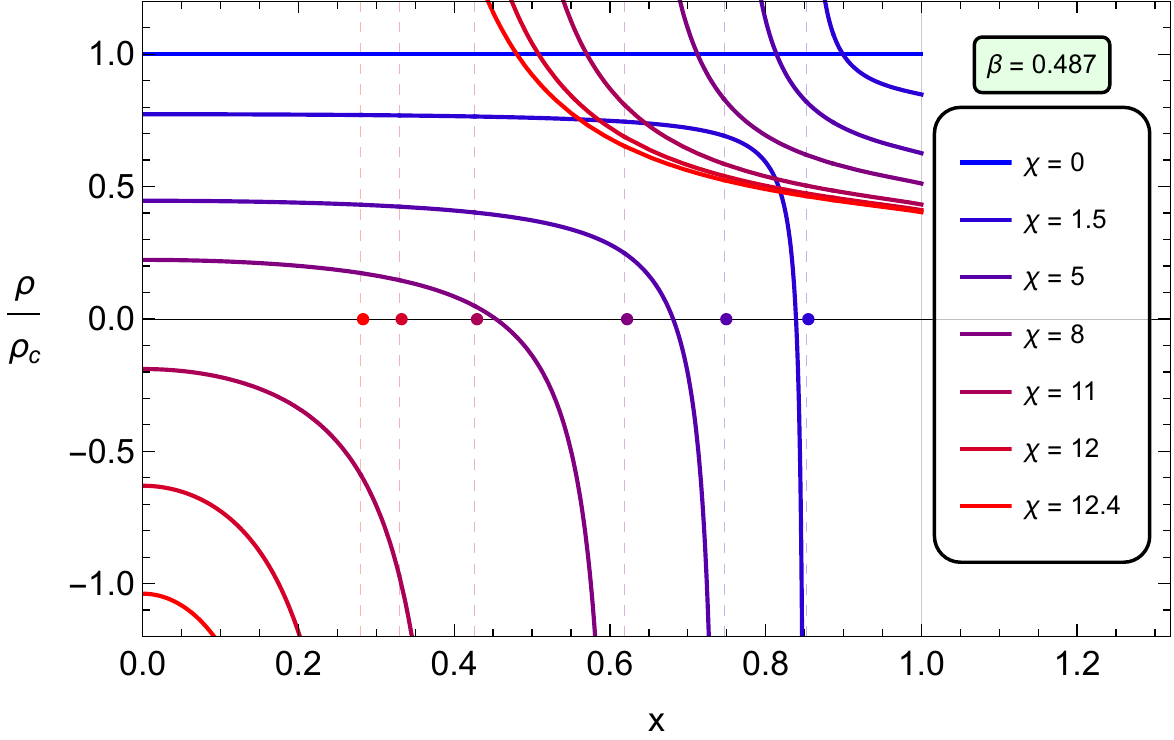}
\caption{The radial profile of the normalized energy density~\eqref{rhoFSs} for configurations beyond the generalized Buchdahl limit~\eqref{eq:Buch_comp_Sch}. The dots correspond to the locations of the poles~\eqref{eq:T_III_root_of_gtt} where both the energy density and the pressure diverge. \textbf{Left panel:} for an indicative fixed value of $\chi=1.5$ and for various values of compactness between the generalized Buchdahl limit $\bet_{\text{B}}(1.5)\simeq 0.455661$ and the black-hole compactness limit $\beta_{BH}=1/2$.  For the normalization we have used the absolute value of the central density $\lvert\rho_c\rvert$ for $\bet=0.456$. \textbf{Right panel:} the effect of $\chi$ for configurations with compactness approaching the BH compactness limit. For the normalization we have used $\rho_c$ for $\chi=0$.}
\label{fig:rho_Τ_ΙΙΙ_beyond_Buchdahl}
\end{figure*}
We see that as the BH limit is approached, the pressure in the interior region with $x \in [0,x_0)$ approaches a constant negative value. To find this value, we perform an expansion of~\eqref{pFSs} around the BH compactness that yields
\beq
p(x)=-\frac{3}{4\left(2\pi+\chi\right)}-\frac{3 \left(8\pi+3\chi \right) \sqrt{1-2\bet}}{16\pi \left(2\pi+\chi \right) \sqrt{1-x^2}}+\mathcal{O}\left(\frac{1}{2}-\beta\right).
\eeq
Similarly, the expansion around the BH compactness for the energy density gives
\beq
\rho(x)=\frac{3}{4 \left(2\pi+\chi\right)}-\frac{3 \chi \sqrt{1-2 \bet}}{16\pi \left(2\pi+\chi\right) \sqrt{1-x^2}}+\mathcal{O}\left(\frac{1}{2}-\beta\right).
\eeq
Thus, for $\bet \to 1/2$, a constant-density configuration is obtained with a constant negative pressure given by
\beq
p=-\rho=-\frac{3}{4\left(2\pi+\chi\right)}\,.
\label{eq:p_minus_rho}
\eeq
This result is in exact analogy with the MM-15 model and reduces to it when $\chi=0$. Finally, at the BH compactness the $g_{tt}$ metric function becomes
\beq
e^{\nu}=\frac{16\pi^2}{(8\pi+3\chi)^2}(1-x^2)\,,
\eeq
which corresponds to a patch of de Sitter spacetime, modified by the factor $16\pi^2/(8\pi+3\chi)^2$ from its standard value, which is essential for the proper matching at the boundary~\cite{Mazur:2015kia}. Note that for $\chi=0$, the $g_{tt}$ metric function of the MM-15 model is recovered.

However, there is a serious drawback of the $f(\Ri,T)$-gravity extended MM-15 model over the GR variant for configurations with compactness above the Buchdahl limit~\eqref{eq:Buch_comp_Sch} and this has to do with a negative energy density shell (NEDS) that emerges during the transition of the configuration from the Buchdahl compactness to the BH compactness (see Fig.~\ref{fig:rho_Τ_ΙΙΙ_beyond_Buchdahl}). The outer radius of the NEDS is identified with $x_0$ while the inner radius is given by
\beq
x_{-}=\sqrt{1-\frac{\left(1-2\bet\right)\left(\chi^2 +6\pi\chi+8\pi^2\right)}{2\bet \pi^2}}\,,
\eeq
and corresponds to the radius where the energy density becomes zero (see Fig.~\ref{fig:rho_Τ_ΙΙΙ_beyond_Buchdahl}). In the BH compactness limit, $x_{-}=x_0$ thus the configuration is free from a NEDS and is characterized by a strictly positive uniform energy density and a negative uniform pressure~\eqref{eq:p_minus_rho} as it can be seen in the left panels of Figs.~\ref{fig:rho_Τ_ΙΙΙ_beyond_Buchdahl} and~\ref{fig:p_Τ_ΙΙΙ_beyond_Buchdahl} respectively. In the GR limit, where $\chi=0$, one has that $x_{-}=x_0$ and so the solution is free of a NEDS for any $\bet \in (\bet_{\text{B}},1/2)$ as it should since the energy density is a positive constant. In the quasiconstant density solution~\eqref{gttFSs}-\eqref{pFSs} on the other hand, a NEDS unavoidably emerges in the interior when the generalized Buchdahl limit is crossed. The reason for this is the constraint~\eqref{eq:T_III_constraint_p_rho} which we have discussed in Sec.~\ref{Sec:T_III_analytic_extension}.

Nevertheless, despite the fact that when our solution is considered beyond the Buchdahl limit, it exhibits a NEDS from which the original MM-15 model is free from, below Buchdahl, our solution has important advantages over the latter for sufficiently large values of $\chi$. In particular, and as we have discussed in Sec.~\ref{Sec:T_III_analytic_extension}, with our solution, we may obtain configurations that are close to a BH mimicker state without crossing the Buchdahl limit (thus avoiding the divergence of physical quantities) and quite importantly, do so, while respecting the dominant energy condition violation compactness limit. Both of these limits have to be violated in the GR variant of the model in order to obtain a BH mimicker state.

It is worthwhile to remark that the quasistationary, and adiabatic, contraction shown above, taking the compactness beyond the Buchdahl bound, up to the BH limit, must be considered as an ideal artifact; although useful to show the geometrical properties of the solution, it could not be realized in a realistic scenario. Thus, the relevant configuration corresponds to the one obtained when $R=R_{S}=2M$, which corresponds to a fully formed gravastar.

Further extensions and modifications of this beyond-GR model that may result in configurations that do not exhibit NEDSs during the transition from Buchdahl to BH compactness is currently under investigation.

\clearpage

\bibliography{main}
\bibliographystyle{utphys}
\end{document}